%% file: main.tex
\documentclass[acmsmall]{acmart}
\newcommand{\revision}[1]{\textcolor{black}{#1}}

\AtBeginDocument{%
  \providecommand\BibTeX{{%
    \normalfont B\kern-0.5em{\scshape i\kern-0.25em b}\kern-0.8em\TeX}}}

\begin{document}

\title{Indexing Techniques for Graph Reachability Queries}


\author{Chao Zhang}
\affiliation{%
  \institution{University of Waterloo}
  \city{Waterloo}
  \country{Canada}}
\email{chao.zhang@uwaterloo.ca}

\author{Angela Bonifati}
\affiliation{%
  \institution{Lyon 1 University}
  \city{Lyon}
  \country{France}}
\email{angela.bonifati@univ-lyon1.fr}

\author{M. Tamer Özsu}
\affiliation{%
  \institution{University of Waterloo}
  \city{Waterloo}
  \country{Canada}}
\email{tamer.ozsu@uwaterloo.ca}


\begin{abstract}
We survey graph reachability indexing techniques for efficient processing of graph reachability queries in two types of popular graph models: \textit{plain graphs} and \textit{edge-labeled graphs}. 
\revision{Reachability queries are Boolean in nature, determining whether a directed path exists between a given source and target vertex.
They form a core class of navigational queries in graph processing. Reachability indexes are specialized data structures designed to accelerate reachability query processing.}
Work on this topic goes back four decades -- we include 33 of the proposed techniques.
Plain graphs contain only vertices and edges, with reachability queries checking path existence between a source and target vertex.
Edge-labeled graphs, in contrast, augment plain graphs by adding edge labels. Reachability queries in edge-labeled graphs incorporate path constraints based on edge labels, assessing both path existence and compliance with path constraints.

We categorize techniques in both plain and edge-labeled graphs and discuss the approaches according to this classification, using existing techniques as exemplars. 
We discuss the main challenges within each class and how these might be addressed in other approaches. 
We conclude with a discussion of the open research challenges and future research directions, along the lines of integrating reachability indexes into modern graph database management systems. 
This survey serves as a comprehensive resource for researchers and practitioners interested in the advancements, techniques, and challenges on reachability indexing in graph analytics.
\end{abstract}






\maketitle

\input{sections/introduction.tex}

\input{sections/background}

\input{sections/plain-reachability-indexes.tex}

\input{sections/path-constraint-reachability-index.tex}

\input{sections/challenge.tex}

\bibliographystyle{ACM-Reference-Format}
\bibliography{publications, bibliography}

\input{sections/appendix}

\end{document}

%% file: sections/introduction.tex
\section{Introduction}
Graphs play a pervasive role in modeling real-world data \cite{Newman2010,SakrBVIAAAABBDV21}, as they provide a unified paradigm to represent entities and relationships as first-class citizens. 
Vertices represent individual entities, while edges denote the relationships among them. 
Instances of graphs can be found in a variety of application domains such as financial networks \cite{financialg}, biological networks \cite{koutrouli2020guide},  social networks \cite{10.1145/2723372.2742786}, knowledge graphs \cite{10.1145/3418294}, property graphs \cite{10.5555/3307192} and transportation networks  \cite{barthelemy2011spatial}.
When dealing with data structured as graphs, one of the most intriguing queries involves determining whether a directed path exists between two vertices. This query examines the transitive relationship between entities within the network and is commonly referred to as a \textit{reachability query} ($Q_r$). 
Reachability queries serve as fundamental operators in graph data processing and have found extensive practical applications \cite{10.1145/3186728.3164139,sahu2020ubiquity}. 
They are often considered the most interesting queries pertaining to graph-oriented analytics \cite{SakrBVIAAAABBDV21} due to the functionality of  identification of transitive relationships and the assessment of how entities are connected in the network.

\textit{Scope}.
\revision{
This survey provides a comprehensive technical review and analysis of the primary indexing techniques developed for reachability queries. These queries return a binary answer indicating whether a path exists between a given source and target vertex in a graph.}
The purpose of reachability indexes is to process such queries with minimal or no graph traversals.
We focus on two distinct types of indexes according to graph types: indexes for plain graphs (focusing only on the structure of the graph) and indexes for edge-labeled graphs (additionally including labels on edges).
In plain graphs, reachability queries assess the existence of a path from a source to a target, while in edge-labeled graphs, they also consider the satisfaction of specified path constraints based on edge labels. This survey is structured following this categorization.

We identify the general indexing frameworks, such as tree cover, 2-hop labeling and approximate Transitive Closure (TC), and discuss individual techniques within these classes. 
This allows us to focus on specific problems and present corresponding techniques for index construction, query processing, and dynamic graph updates.

\textit{Survey structure}.
We begin by providing the necessary background on reachability queries in graphs (\S \ref{sec:background}). Next, we delve into the examination of indexes for plain reachability queries (\S \ref{sec:pr_indexes}) and path-constrained reachability queries (\S \ref{sec:pcr_indexes}). It is in the latter section that we consider edge labels and the evaluation of specified path constraints within the index structures.
Concluding the survey, we discuss the open challenges that remain in this field (\S \ref{sec:our_vision}). We provide research perspectives and insights towards the development of full-fledged indexes for graph database management systems.

\textit{Intended audience}.
The primary intention of the survey is to demystify the extensively studied indexing techniques for the communities working with graph data.
Reachability indexes have been studied for four decades, and abundant optimization techniques have been proposed.
However, the existing approaches have not been well characterized, and many studies are from the purely theoretical research perspective.
Our categorization of index classes and identification of the main challenges in each index class can help researchers understand the current state of the research in the field and
differentiate different techniques.
Besides the research community, engineers involved in developing graph database management systems or working with graph data analytics can also benefit from the survey, especially for those who are seeking for advanced technique related to traversing paths in graphs. To this end, we restrict formalization to the core notations and provide a consistent running example over the analysed indexing techniques.

\textit{Previous surveys}.
There \revision{exist} numerous surveys related to graph data management, which focus on different aspects of graph databases, including graph data models \cite{10.1145/1322432.1322433,6313676}, graph query languages from both a theoretical perspective \cite{10.1145/2206869.2206879,10.1145/2463664.2465216} and a practical one \cite{10.1145/3104031}, particular problems on graph pattern matching \cite{bunke00,gallagher2006matching,Riesen2010,10.1007/s10044-012-0284-8,10.1145/2911996.2912035}, RDF systems \cite{tamer16rdfsurvey}, the landscape of graph databases \cite{Angles2018}, system aspects of graph databases \cite{10.1145/3604932}, and a vision on graph processing systems \cite{SakrBVIAAAABBDV21}.
Although reachability queries are discussed in some of these, especially in the ones related to graph query languages, the focus is not on indexing techniques.
Fletcher et al. \cite{Fletcher2018} describe the indexes for graph query evaluation  and briefly mention reachability indexes for plain graphs.
Yu et al. \cite{yu2010graph} investigate the early indexing solutions for plain graphs while Bonifati et al. \cite{10.5555/3307192} discuss in detail a few representative indexing techniques for plain graphs.

The current survey is different in that it provides a characterization of the various indexing techniques and discusses the approaches according to this characterization (Tables \ref{table:pr_indexes} and \ref{table:pcr_indexes}). 
\revision{
We note that this is the first survey dedicated to indexing techniques for reachability queries with path constraints. These queries are becoming increasingly important due to their growing support in modern graph query language standards and the systems that implement them.}
We focus on both the basic techniques and the scalability of the approaches as graph sizes grow. 
\revision{Our survey targets a broader audience than previous ones as our discussion and presentation are aided by a consistent running example with judicious use of formal notation.}
We identify the main characteristics across a great variety and number of indexing techniques (we cover $33$ techniques developed over the last four decades). We also single out the fine-grained components and discuss in detail the problems and solutions within each.

A preliminary and short version of the survey was presented as a conference tutorial \cite{10.1145/3555041.3589408}, and the slides of our tutorial are available online\footnote{\textbf{Slides}: \url{https://github.com/dsg-uwaterloo/ozsu-grp/blob/main/An_Overview_of_Reachability_Indexes_on_Graphs.pdf}}. \revision{Due to space limit, additional materials and examples are provided in our online appendix \cite{zhang2023indexingtechniquesgraphreachability}.}

%% file: sections/background.tex
\section{Background}\label{sec:background}
\begin{figure}
\begin{minipage}{0.49\linewidth}
\centering
    \resizebox{\textwidth}{!}{\includegraphics{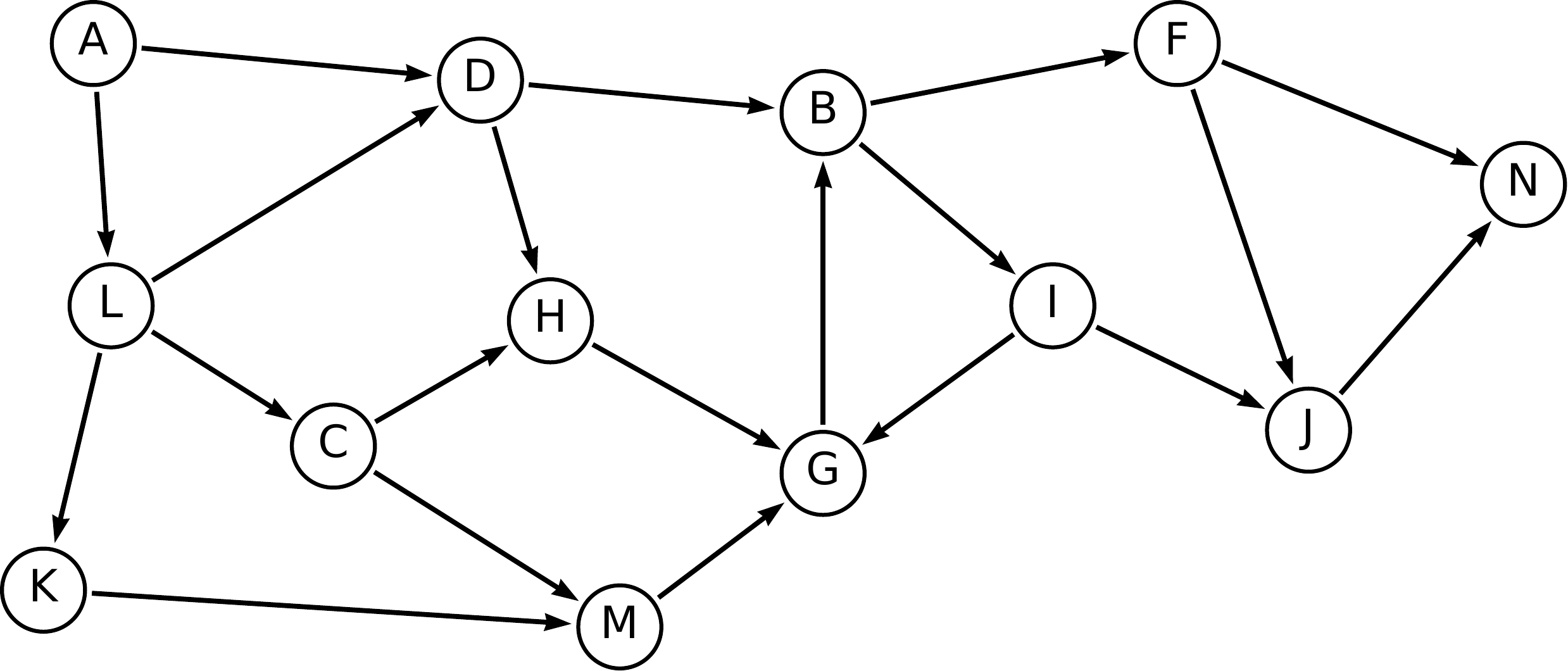}}    
    \textbf{(a)}
\end{minipage}
\hfill
\begin{minipage}{0.49\linewidth}
    \centering
    \resizebox{\linewidth}{!}{\includegraphics{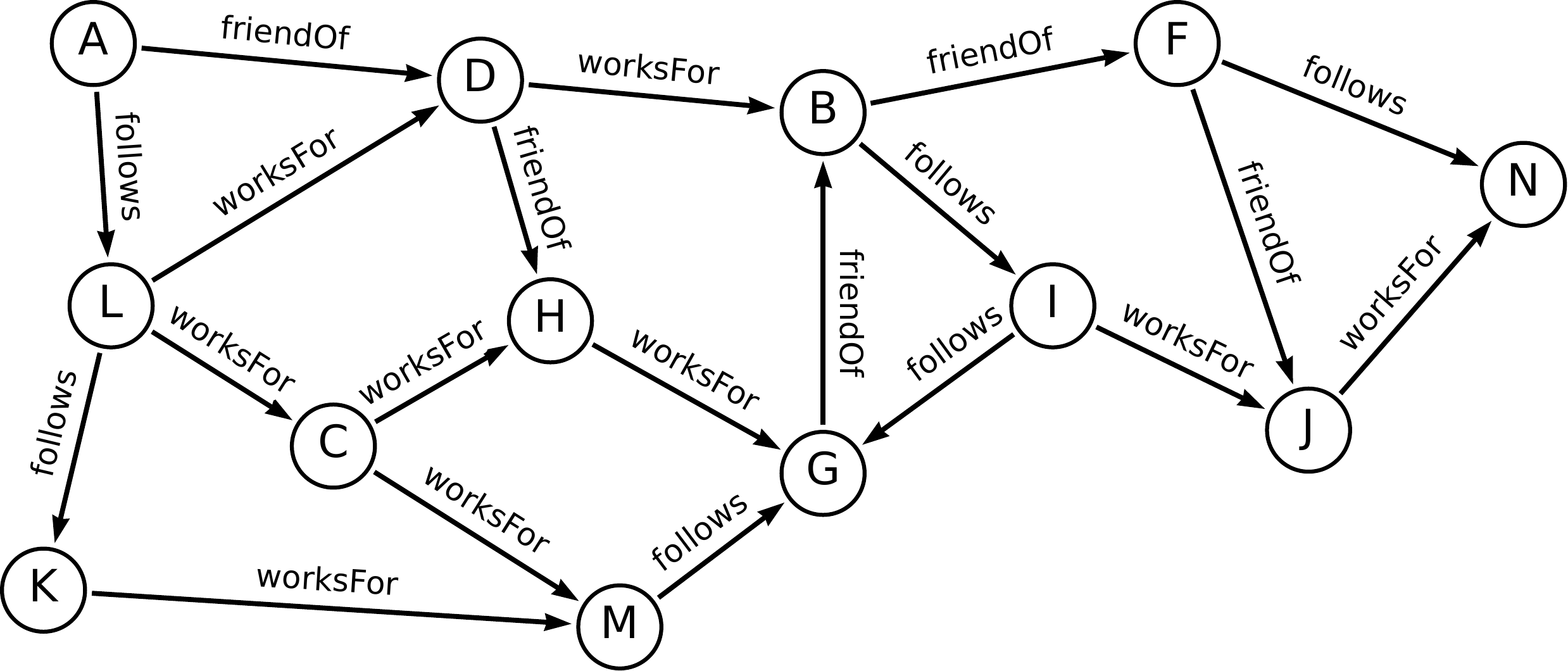}} 
    \textbf{(b)}
\end{minipage}
\caption{A plain graph (a), and an edge-labeled graph (b).}\label{fig:graphs}
\end{figure}

\textit{Plain graphs} are graphs with only vertices and edges, denoted as $\mathcal{G}=(V,E)$, where $V$ is a set of vertices and $E$ is a set of edges, \textit{e.g.}, Fig. \ref{fig:graphs}(a).
Queries in plain graphs include only topological predicates.
\textit{Edge-labeled graphs} \cite{10.5555/3307192,10.1145/3104031} augment plain graphs by assigning labels to edges, \textit{i.e.}, $\mathcal{G}=(V,E,\mathcal{L})$, where $\mathcal{L}$ is a set of labels and each $e \in E$ is assigned a label in $\mathcal{L}$, \textit{e.g.}, Fig. \ref{fig:graphs}(b).
Queries in edge-labeled graphs include predicates on edge labels in addition to topological predicates.
\textit{Property graphs} \cite{10.5555/3307192,10.1145/3104031}  extend edge-labeled graphs by adding attributes as key/value pairs to vertices 
and edges. \revision{An example of property graph is presented in our online appendix \cite{zhang2023indexingtechniquesgraphreachability}.}

\subsection{The landscape of graph queries}
Graph queries can be broadly categorized into two classes: \textit{pattern matching queries} and \textit{navigational queries}.
Pattern matching queries define a query graph that is matched against the input data graph.
Matching triangles is a notable instance of pattern matching.
An example query is to find vertices $\{x,y,z\}$ in the edge-labeled graph in Fig. \ref{fig:graphs} such that the following patterns can be satisfied: $(x,\texttt{follows}, y)$, $(x,\texttt{follows}, z)$, and $(y,\texttt{worksFor}, z)$.
$\{A,L,D\}$ is a triple of such vertices in Fig. \ref{fig:graphs}.
Navigational queries define a mechanism (usually a regular expression) that will be used to guide the navigation in the input graph. 
An example of navigational query is to find all the direct and indirect friends of person $A$ in Fig. \ref{fig:graphs}(b), where the query searches for friends of $A$ and also the friends of the friends of $A$, and so on. 
Navigational queries are clearly more expressive than pattern matching queries as they include recursion.
We note that both pattern matching and navigational queries can be defined on plain graphs, edge-labeled graphs, and property graphs. 
Reachability queries are subclasses of navigational queries that will be elaborated on. 
We refer readers to the recent survey \cite{10.5555/3307192} on graph query languages.

Since this survey is on reachability indexes, we focus on navigational queries. The most notable navigational queries are \textit{regular path queries} \cite{abiteboul1997regular,10.1145/3104031} on edge-labeled graphs.
Such queries inspect the input graph to retrieve paths of arbitrary length between two vertices, and the visited paths should satisfy the constraints specified by regular expressions based on edge labels. 
Regular path queries are a subclass of \textit{path queries}, and path queries can have more complex path constraints, \textit{e.g.}, checking the relationships between the sequences of edge labels in paths \cite{10.1145/2389241.2389250}. 
We note that there are  a few additional types of navigational queries, \textit{e.g.}, where the navigation is through repeated trees or graph patterns \cite{10.5555/3307192}. 
In addition, it is also possible to combine regular path queries with graph patterns to define a query graph including paths with regular expressions as path constraints, which is known as navigational graph pattern \cite{10.1145/3104031}.

Regular path queries can be classified according to query types and edge directions on paths \cite{10.1145/3104031}. 
\begin{itemize}
    \item \textit{Boolean}: These queries take a pair of source and target vertices and a regular expression as input and return only \textit{True} or \textit{False}, indicating the existence of the path between the source and the target, which can satisfy the constraint imposed by the regular expression. \textit{Boolean} queries are also \revision{known} as \textit{path-existence queries} \cite{10.1145/3104031} or \textit{path-finding queries} \cite{Wolde2023DuckPGQEP}.
    \item \textit{Nodes}: These queries take a regular expression as input and return pairs of source and target vertices such that the path between them  satisfies the query constraint. If the source (or the target) vertex is fixed in the queries, only a list of target (source) vertices are returned.
    \item \textit{Path}: These queries are similar to the node queries, but also return the full paths. 
\end{itemize}

Regular path queries are usually defined on edge-labeled graphs that are directed graphs (each edge has a specific direction).
Practical graph query languages such as SQL/PGQ \cite{10.1145/3514221.3526057}, GQL \cite{10.1145/3514221.3526057}, and openCypher \cite{openCypher} allow specifying the directions of edges in paths that the queries \revision{need}  to retrieve. \textit{Connectivity} queries \cite{Foulds1992} ignore the direction of the edges in \revision{a} path. \textit{Reachability} queries \cite{yu2010graph,10.1145/3104031} require each edge in the path to have the same direction. Finally, in \textit{two-way} queries \cite{CALVANESE2002443,10.1145/959060.959076,10.1145/3335409.3335411,10.1145/3104031} every two contiguous edges in the path have the opposite directions.
In this survey, we focus on reachability queries that return Boolean results, which are recognized as the most fundamental type of regular path queries \cite{10.1145/3104031}.

\subsection{Plain reachability}
Reachability queries in plain graphs are known as \textit{plain reachability queries}.
Plain reachability query $Q_r(s,t)$ has source vertex $s\in V$ and target vertex $t\in V$ as input arguments and checks whether there exists a path from $s$ to $t$ in $\mathcal{G}$.
For instance, in Fig. \ref{fig:graphs}(a), $Q_r(A,G)=True$ because of the path $(A,D,H,G)$.

A naive way to process plain reachability queries is to use a graph traversal method, \textit{e.g.}, breadth-first traversal (BFS), depth-first traversal (DFS), or bidirectional breadth-first traversal (BiBFS). 
Such traversal methods do not require any offline computation cost. However, their online query processing cost might be high as the traversal can visit a large portion of the input graph.
A naive form of indexing for plain reachability queries is the \textit{transitive closure} (TC) \cite{10.1145/67544.66950}, that is costly to compute and materialize. 
Even though the online query processing cost with a TC is negligible, it is unfeasible in practice due to high time complexity.

\subsection{Path-constrained reachability}\label{sec:pcr-queries}
Reachability queries with path constraints in edge-labeled graphs are referred to as \textit{path-constrained reachability queries}.
Compared to plain reachability queries, a path-constrained reachability query $Q_r(s,t,\alpha)$ has an additional argument $\alpha$, which is a regular expression based on edge labels.
We consider the basic regular expression in this survey.
More precisely, $\alpha$ has edge labels as literal characters, and concatenation `$\cdot$', alternation `$\cup$', and the Kleene operators (star `$*$' or plus `$+$') as meta-characters.
The grammar of $\alpha$ is $\alpha:: = l| \alpha\cdot\alpha| \alpha \cup \alpha | \alpha^+ | \alpha^*$. 
 $Q_r(s,t,\alpha)$ checks not only whether there exists a path from $s$ to $t$, but also whether the path can satisfy the path constraint specified by $\alpha$, \textit{i.e.}, whether the sequence of edge labels in the path from $s$ to $t$ form a word in the language of the regular expression $\alpha$.
For instance, path constraint $\alpha=(\texttt{friendOf}\cup\texttt{follows})^*$ enforces that the path from $s$ to $t$ can only contain edges of labels  \texttt{friendOf} or \texttt{follows}, and $Q_r(A,G,\alpha)=False$ in Fig. \ref{fig:graphs}(b) because every path from $A$ to $G$ must include \texttt{worksFor}. 
Path-constrained reachability queries are a subclass of regular path queries \cite{5767858}, as they only determine the existence of a path that satisfies a given constraint, whereas regular path queries can additionally return the source and target nodes of such paths.

The approaches for processing regular path queries can be used to process path-constrained reachability queries, \textit{e.g.}, building a finite automata according to $\alpha$ and using the finite automata to guide an online traversal. 
We note that the path semantic needs to be specified for processing such queries. 
In practice, the arbitrary path semantic has been widely adopted, \textit{e.g.}, SPARQL 1.1 \cite{sparql}.
Due to the presence of the Kleene operators in $\alpha$, path-constrained reachability queries are inherently recursive, which are known to be computationally expensive.
Another naive approach for processing path-constrained reachability queries is precomputing a \textit{generalized transitive closure} (GTC) \cite{10.1145/1807167.1807183,10.1109/ICDE55515.2023.00013}. It records for any pair of vertices $(s,t)$, not only whether $t$ is reachable from $s$, but also the information related to edge labels of the paths from $s$ to $t$, which is necessary for evaluating path constraint $\alpha$.
However, GTCs are even more expensive to compute than TCs that are already known to be computationally expensive.
Thus, precomputing GTCs is not feasible in practice.

\begin{figure}
    \centering
    \includegraphics[width=0.8\linewidth]{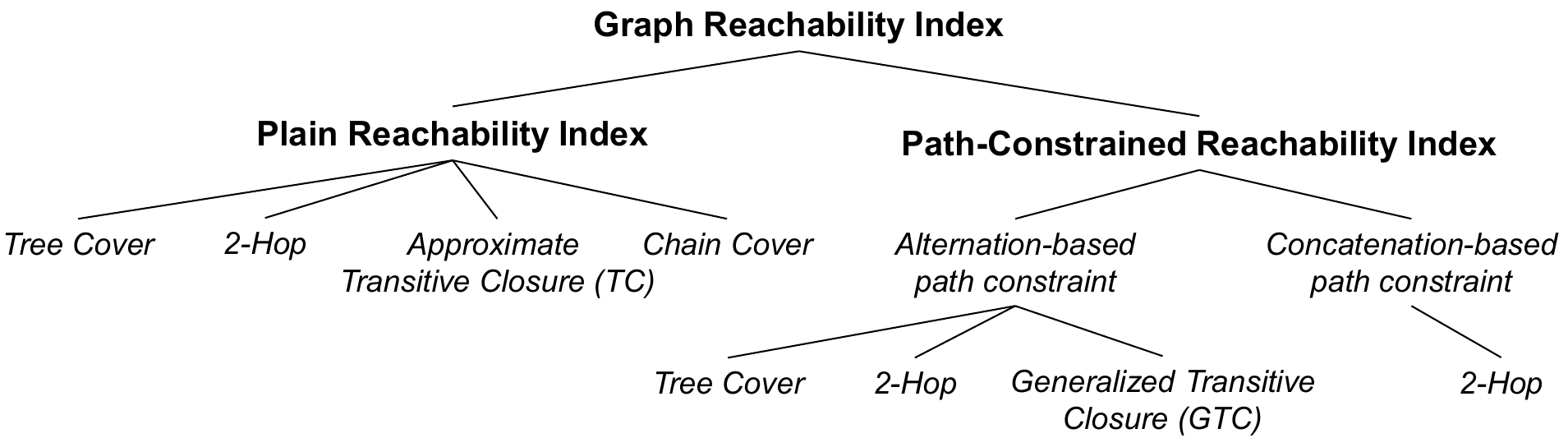}
    \caption{\revision{Taxonomy of reachability problems and their corresponding indexing classes. The details of each reachability indexes are presented in Table \ref{table:pr_indexes} and Table \ref{table:pcr_indexes}, respectively.}}
    \label{fig:taxonomy_index_class}
\end{figure}

\subsection{Reachability indexes}\label{sec:reachability_indexes}
Reachability indexes are non-trivial data structures that can effectively compress the naive indexes, \textit{i.e.}, TCs or GTCs, and can also be efficiently computed. 
In addition, query processing using reachability indexes can be more efficient than the approaches based on graph traversals.
The core intuition behind designing reachability indexes is to strike a balance between offline index construction and online query processing, aiming to optimize the overall cost of answering reachability queries.
According to the types of reachability queries, there exist \textit{plain reachability indexes} \cite{10.1145/67544.66950,10.1145/99935.99944,10.5555/545381.545503,1617443,4497498,10.1145/1376616.1376677,10.1145/1559845.1559930,10.1145/2463676.2465286,10.14778/2556549.2556578,10.5555/1083592.1083651,10.1145/1247480.1247573,10.14778/1920841.1920879,6544893,Veloso2014ReachabilityQI,10.14778/2732977.2732992,7750623, 10.1145/2588555.2612181, 10.1145/1871437.1871457,10.1145/2505515.2505724,10.1145/3556540,merz2014preach,4912201,1410143,492668,10.5555/2627817.2627899,10.1145/1007352.1007387,10.1007/978-3-030-73197-7_52,Yildirim2013DAGGERAS} and \textit{path-constrained reachability indexes} \cite{10.1145/1807167.1807183,ZOU201447,10.1145/3035918.3035955,10.14778/3380750.3380753,10.1145/3451159,10.14778/3529337.3529348,10.1109/ICDE55515.2023.00013}. 
\revision{A taxonomy of reachability problems and the corresponding index classes is presented in Fig. \ref{fig:taxonomy_index_class}.}
We further categorize plain reachability indexes and path-constrained reachability indexes in Tables \ref{table:pr_indexes} and \ref{table:pcr_indexes}, respectively.

%% file: sections/plain-reachability-indexes.tex
\section{Plain Reachability Indexes}\label{sec:pr_indexes}
Most of the indexes can be grouped into index classes according to the underlying classes with a few additional techniques.
The three main classes are: \textbf{Tree cover} \cite{10.1145/67544.66950}; \textbf{2-Hop labeling} \cite{10.5555/545381.545503}; \textbf{Approximate transitive closure} \cite{10.14778/2732977.2732992}.
These indexes can be further characterized according to $4$ metrics (Table \ref{table:pr_indexes}). We provide a detailed explanation of each column in the table.

The index class column indicates the main class of the indexing approach: tree cover, 2-hop, approximate transitive closure, and chain cover. A few other approaches also exist, which have specific designs that are different from these main classes. 
    
The index type column indicates whether an indexing approach is a \textit{complete index} or a \textit{partial index}. A complete index records all the reachability information in the graph such that queries can be processed by using only index lookups. In contrast, a partial index  records only partial information such that query processing might need to traverse the input graph. 
Partial indexes can be further classified into  \textit{partial indexes without false positives} and \textit{partial indexes without false negatives}, indicated by no FP and no FN, respectively. 
In the first subclass, if index lookups return $True$, query result $True$ can be immediately returned as the index does not contain false positives. Otherwise, graph traversals are needed for query processing. 
Techniques in the second subclass work in the opposite way, \textit{i.e.}, graph traversals can be avoided if the index lookups return $False$.
    
The input column indicates whether an indexing approach assumes a directed acyclic graph (DAG) or a general graph as input. Assuming DAGs as input is not a major issue in terms of query processing in static graphs as an efficient reduction can be used \revision{(see details in Section \ref{sec:tree_cover_pipline})}. 
    
The dynamic column indicates whether an indexing approach can handle dynamic graphs with updates. There exist two types of dynamic graphs: \textit{fully dynamic graphs} with both edge insertions and deletions indicated by I\&D, and \textit{insertion-only graphs} with only edge insertions indicated by I. 
We note that although assuming DAG as input is not a major issue for query processing on static graphs, that assumption becomes a major bottleneck in maintaining indexes on dynamic graphs. This is because indexing techniques designed with the DAG assumption have to deal with the problem of maintaining strongly connected components, which can be computationally expensive.

\revision{The construction time, index size, and query time columns reflect the complexity of each approach. Due to space constraints, a detailed discussion is provided in the online appendix \cite{zhang2023indexingtechniquesgraphreachability}.}

\begin{table}
    \centering
\caption{\revision{Overview of plain reachability indexes.
    Notations: $|NTE|$: \#non-tree edges;
$k$: \#spanning trees;
$|DP|$: \#disjoint paths;
$|HDV|$: \#high-degree vertices;
$|MC|$: min\#chains;
$|C|$: \#chains;
$T$: characteristic of the  DAG (see TFL \cite{10.1145/2463676.2465286});
$|L|$: size of labeling; 
$|H|$: factor related to \#backbones (see HL \cite{10.14778/2556549.2556578});
$|BF|$: size of Bloom filer (bit);
$|ME|$: \#minimum elements in MinHash;
$|HVL|$: size of a huge-vertex label (see IP \cite{10.14778/2732977.2732992})
$r$: reachability ratio;
$Con(\mathcal{G})$: transitive closure contour \cite{10.1145/1559845.1559930};
$|HS|$: size of hash set;
$|LV|$: \#landmark vertices;
$|TO|$: \#topological orderings;
$|SV|$: \#supportive vertices;
$|CSV|$: \#candidates per supportive vertex;
$|V_i|$/$|E_i|$: \#vertices or \#edges visited in the $i$-th iteration (see TOL \cite{10.1145/2588555.2612181});
$D$: width of the DAG (see Optimal Chain Cover \cite{4497498}).
    }} \label{table:pr_indexes}
    \resizebox{\linewidth}{!}{
    \begin{tabular}{|c|c|c|c|c|p{3cm}|c|p{3cm}|} \hline
        \textbf{Indexing Technique}                     & \textbf{Index Class}&\textbf{Index Type} &\textbf{Input} & \textbf{Dynamic}  & \textbf{Construction Time}  & \textbf{Index size}  & \textbf{Query Time}\\ \hline \hline
        \textbf{Tree Cover} \cite{10.1145/67544.66950}  & Tree Cover &   Complete        & DAG &     No              & $O(|V||E|)$& $O(|V|^2)$& $O(\log|V|)$     \\  \hline
           Tree+SSPI \cite{10.5555/1083592.1083651}     & Tree Cover   & \begin{tabular}{@{}c@{}}Partial \\ (no FP)\end{tabular} & DAG & No & $O(|E|+|V|)$& $O(|E|+|V|)$& $O(|E|-|V|)$ \\ \hline
           Dual labeling \cite{1617443} & Tree Cover & Complete & DAG & No &$O(|V|+|E|+|NTE|^3)$  &  $O(|V|+|E|+|NTE|^2)$& $O(1)$/$O(\log |NTE|)$ \\ \hline
           GRIPP \cite{10.1145/1247480.1247573} & Tree Cover & \begin{tabular}{@{}c@{}}Partial \\ (no FP)\end{tabular} & General  & No & $O(|E|+|V|)$& $O(|E|+|V|)$& $O(|E|-|V|)$ \\ \hline
           Path-tree \cite{10.1145/1376616.1376677,10.1145/1929934.1929941}  & Tree Cover & Complete & DAG & \begin{tabular}{@{}c@{}}Yes \\ (I\&D)\end{tabular} & $O(|DP||E|)$/$O(|V||E|)$ & $O(|DP||V|)$&  $O((\log |DP|)^2)$\\ \hline
           GRAIL \cite{10.14778/1920841.1920879} & Tree Cover & \begin{tabular}{@{}c@{}}Partial \\ (no FN)\end{tabular} & DAG & No &$O(k(|V|+|E|))$& $O(k|V|)$& $O(k)$/$O(|V|+|E|)$\\ \hline
           Ferrari \cite{6544893} & Tree Cover & \begin{tabular}{@{}c@{}}Partial \\ (no FN)\end{tabular} & DAG & No &$O(k^2|E|)$&$O((k+|S|)|V|)$& $O(k)$/$O(|V|+|E|)$\\ \hline 
           DAGGER \cite{Yildirim2013DAGGERAS} & Tree Cover & \begin{tabular}{@{}c@{}}Partial \\ (no FN)\end{tabular} & DAG & \begin{tabular}{@{}c@{}}Yes \\ (I\&D)\end{tabular} &$O(k(|V|+|E|))$& $O(k|V|)$& $O(k)$/$O(|V|+|E|)$ \\ \hline \hline
           \textbf{2-Hop} \cite{10.5555/545381.545503} & 2-Hop & Complete & General & No &$O(|V|^4) $&$O(|V||E|^{1/2})$&$O(|E|^{1/2})$ \\ \hline
           Ralf et al. \cite{1410143} & 2-Hop & Complete & General & \begin{tabular}{@{}c@{}}Yes \\ (I\&D)\end{tabular} & - & - & -  \\ \hline
           3-Hop \cite{10.1145/1559845.1559930} & 2-Hop & Complete & DAG & No &$O(|C||V|^2|Con(\mathcal{G})|)$&$O(|C||V|)$&$O(\log|V|+|C|)$  \\ \hline
           U2-hop \cite{4912201} & 2-Hop & Complete & DAG & \begin{tabular}{@{}c@{}}Yes \\ (I\&D)\end{tabular} &-& -& - \\ \hline           
           Path-hop \cite{10.1145/1871437.1871457} & 2-Hop & Complete & DAG  & No &-&-& -\\ \hline
           TFL \cite{10.1145/2463676.2465286} & 2-Hop & Complete & DAG & No &$O(T)$ &-& $O(|L_{out}(s)| + |L_{in}(t)|)$\\ \hline
           DL \cite{10.14778/2556549.2556578} & 2-Hop & Complete & General & No &$O(|V|(|V|+|E|)|L|)$&-&$O(|L_{out}(s)| + |L_{in}(t)|)$ \\ \hline
           HL \cite{10.14778/2556549.2556578}      &   2-Hop   &   Complete    &   DAG &   No &$O(H)$& - &$O(|L_{out}(s)| + |L_{in}(t)|)$ \\ \hline 
           PLL \cite{10.1145/2505515.2505724} & 2-Hop & Complete & General & No &$O(|V||E|)$&$O(|V|\log|V|)$&$O(\log|V|)$ \\ \hline
           TOL \cite{10.1145/2588555.2612181} & 2-Hop & Complete & DAG & \begin{tabular}{@{}c@{}}Yes \\ (I\&D)\end{tabular} &$O(\sum_{i}(|E_i|+i|V_i|))$&$O(\sum_{i}|V_i|)$& $O(|L_{out}(s)| + |L_{in}(t)|)$ \\  \hline 
           DBL \cite{10.1007/978-3-030-73197-7_52} & 2-Hop & \begin{tabular}{@{}c@{}}Partial \\ (mixed)\end{tabular}& General & \begin{tabular}{@{}c@{}}Yes \\ (I)\end{tabular} &$O((|HS| + |LV|)(|V|+|E|))$&$O((|HS| + |LV|)|V|)$&$O(|HS| + |LV|)$/$O((|HS| + |LV|)(|V|+|E|))$ \\ \hline
           O'Reach \cite{10.1145/3556540}     &   2-Hop   &   \begin{tabular}{@{}c@{}}Partial \\ (mixed)\end{tabular}     &   DAG &   No &$O((|TO|+|SV||CSV|)(|V|+|E|))$&$O(|V|)$& $O(|SV| + |TO|)$/$O(|V|+|E|)$ \\ \hline \hline
           \textbf{IP} \cite{10.14778/2732977.2732992,10.1007/s00778-017-0468-3} & Approximate TC & \begin{tabular}{@{}c@{}}Partial \\ (no FN)\end{tabular} & DAG & \begin{tabular}{@{}c@{}}Yes \\ (I\&D)\end{tabular} & $O((|ME|+|HVL|)(|V|+|E|))$&$O((|ME|+|HVL|)|V|)$&$O(|ME|)$/$O(|ME||V|r^2)$\\ \hline
           BFL \cite{7750623} & Approximate TC & \begin{tabular}{@{}c@{}}Partial \\ (no FN)\end{tabular} & DAG & No &$O(|BF|(|V|+|E|))$&$O(|BF||V|)$& $O(|BF|)$/$O(|BF||V|+|E|)$ \\ \hline \hline
           \textbf{Chain Cover} \cite{10.1145/99935.99944} & Chain Cover& Complete & DAG & \begin{tabular}{@{}c@{}}Yes \\ (I\&D)\end{tabular} &$O(|V|^3)$&$O(|MC||V|)$&$O(\log |MC|)$ \\ \hline
           Optimal Chain Cover \cite{4497498} & Chain Cover & Complete & DAG & \begin{tabular}{@{}c@{}}Yes \\ (I\&D)\end{tabular} &$O(|V|^2+D^{3/2}|V|)$&$O(D|V|)$&$O(\log D)$  \\ \hline \hline           
           Feline \cite{Veloso2014ReachabilityQI}  &   -   &   \begin{tabular}{@{}c@{}}Partial \\ (no FN)\end{tabular}    &   DAG &   No &$O(|V|\log|V|+|E|)$&$O(|V|)$& $O(1)$/$O(|V|+|E|)$ \\ \hline
           Preach \cite{merz2014preach}            &   -   &   \begin{tabular}{@{}c@{}}Partial \\ (no FP)\end{tabular}    &   DAG &   No &$O(|V|\log|V|+|E|)$&$O(|V|)$& $O(1)$/$O(|V|+|E|)$\\ \hline           
    \end{tabular}}
\end{table}


\subsection{Tree-cover-based indexes}\label{sec:tree-cover} 
In the  tree cover index (original paper: \cite{10.1145/67544.66950}), each vertex is assigned one or more intervals, and queries are processed by using intervals of the source and target vertices.
The intervals are obtained based on spanning trees of the input graph, followed by specifically addressing non-tree edges.
There have been a number of proposals that fall under this class \cite{10.5555/1083592.1083651,1617443,10.1145/1247480.1247573,10.1145/1929934.1929941,10.1145/1376616.1376677,10.14778/1920841.1920879,6544893,Yildirim2013DAGGERAS}. 
These sometimes differ in the computation of the vertex intervals.

\subsubsection{\textbf{The approach}}\label{sec:tree_cover_pipline}
The tree cover index is computed in three steps: (i) Transforming a general graph (potentially) with cycles to a DAG; (ii) Computing spanning trees and the interval labeling over this DAG; (iii) Hooking up roots of spanning trees and indexing non-tree edges.

\begin{figure}
    \begin{minipage}{0.48\linewidth}
        \centering
        \vspace{+0.21cm}
        \resizebox{.9\textwidth}{!}{
        \includegraphics{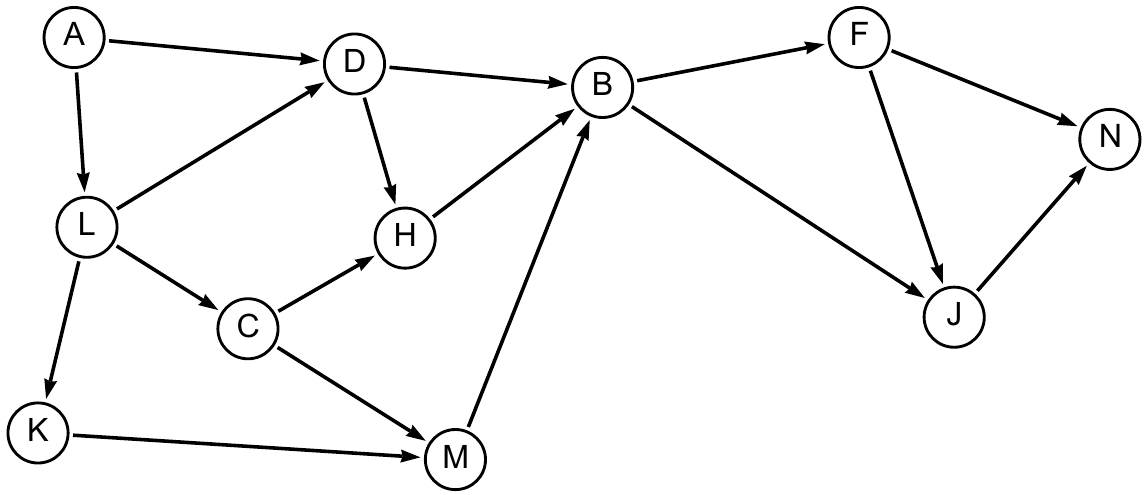}}
        \caption{The DAG obtained by transforming the plain graph in Fig. \ref{fig:graphs}(a)}
        \label{fig:DAG-of-the-plain-graph}
    \end{minipage}
    \hfill
    \begin{minipage}{0.48\linewidth}
        \centering
        \resizebox{\textwidth}{!}{
        \includegraphics{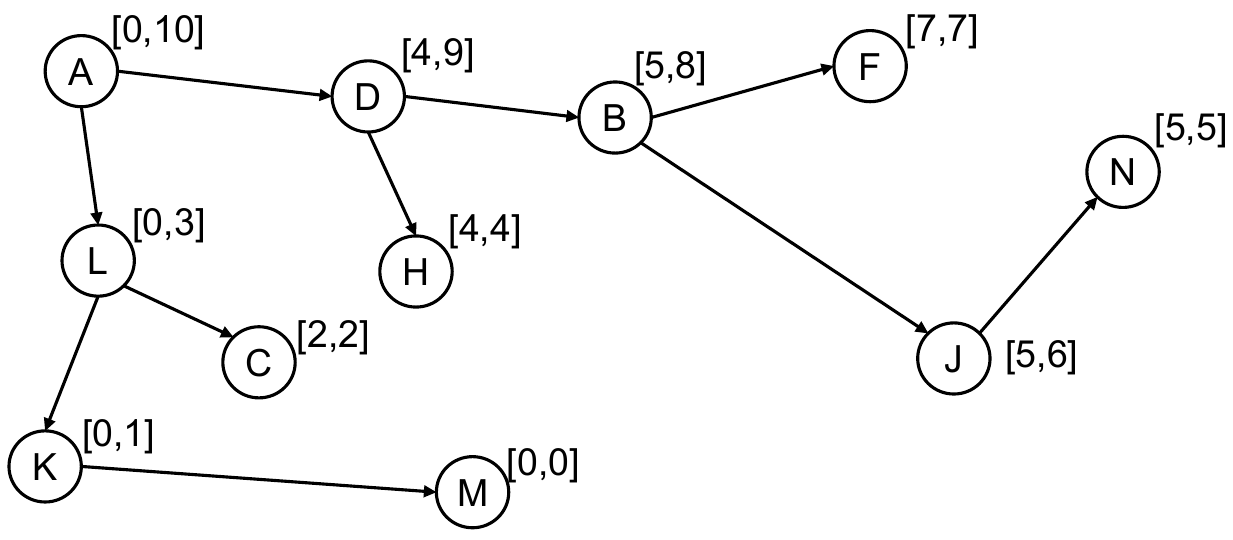}}
        \caption{\revision{The spanning tree of the DAG in Fig. \ref{fig:DAG-of-the-plain-graph} and the interval labeling of the spanning tree.}}
        \label{fig:interval-labeling-of-the-spanning-tree}
    \end{minipage}
\end{figure}

\textit{\textbf{Transformation}}.
A general graph is transformed into a DAG by identifying all the strongly connected components (SCCs) and selecting a representative vertex in each SCC. 
We use $Rep(v)$ to denote the representative of $v$ in the SCC containing $v$.
After the transformation, every edge $(u,v)$ in the general graph is transformed into an edge $(Rep(u),Rep(v))$ in the resulting DAG.
\revision{
For instance, the general graph in Fig. \ref{fig:graphs}(a) is transformed into the DAG in Fig. \ref{fig:DAG-of-the-plain-graph}, where $Rep(G)=Rep(I)=Rep(B)=B$. 
Then, the edge $(H, G)$ in Fig. \ref{fig:graphs}(a) is kept as the edge $(H, B)$ in Fig. \ref{fig:DAG-of-the-plain-graph}.}
The transformation can be done in $O(|V|+|E|)$ time using Tarjan's strongly connected components algorithm \cite{tarjan1972depth}.
Thus, most plain reachability indexes in the literature assume DAGs as input since generalization is trivial. 

\textit{\textbf{Query reduction with the transformation}}.
A reachability query on the general graph can be reduced to a corresponding query on the DAG, \textit{i.e.}, a query $Q_r(s,t)$ in the general graph is reduced to a query $Q_r(Rep(s),Rep(t))$
The case of $Rep(s)=Rep(t)$ indicates that $s$ and $t$ belong to the same SCC. Thus, $t$ is reachable from $s$. 
Otherwise, the query needs to be processed on the DAG. 

\textit{\textbf{Spanning trees and the interval labeling}}.
The tree cover approach first computes the spanning trees in the DAG and then computes the interval labeling according to the spanning trees.
For the DAG in Fig. \ref{fig:DAG-of-the-plain-graph}, a corresponding spanning tree is presented in Fig. \ref{fig:interval-labeling-of-the-spanning-tree}.
The interval labeling can be efficiently computed and is able to include all the reachability information in the spanning tree.
Specifically, the interval labeling assigns an interval to each vertex $v$ in the tree, which is computed based on a postorder traversal from the root in the tree.  
Let $\mathcal{L}_v=[a_v,b_v]$ be the interval assigned to vertex $v$. 
Then, the second endpoint $b_v$ is the postorder number of $v$, and the first endpoint $a_v$ is the lowest postorder number of all the descendants of $v$ in the tree. 
In Fig. \ref{fig:interval-labeling-of-the-spanning-tree}, we present the intervals that are computed based on the postorder traversal from root $A$.

\textit{\textbf{Query processing with the interval labeling}}.
Reachability query $Q_r(s,t)$ can be processed by checking whether $b_t\in [a_s,b_s]$. 
The intuition is processing $Q_r(s,t)$ by checking whether the subtree rooted at $s$ in the spanning tree contains $t$, and the intervals assigned to $s$ and $t$ encode sufficient information for such a checking.
Consider the query $Q_r(A, B)$. 
The postorder number of $B$ is contained in the interval of $A$ as $8\in [0,10] $. Thus, $Q_r(A, B)=True$.
We note $Q_r(s,t)$ can be also processed by checking whether the interval of $s$ subsumes the interval of $t$, \textit{e.g.}, $\mathcal{L}_B\subseteq\mathcal{L}_A$ as $[5,8] \subseteq [0,10]$.
These two approaches are essentially equivalent.

\textit{\textbf{Limitations of interval labeling}}. Two major problems arise with the interval labeling approach: (i) To assign an interval to every vertex in the graph, it may be necessary to construct a spanning forest, \textit{i.e.}, a set of spanning trees that together cover all the vertices; (ii) Interval labeling can only cover the reachability information through paths in a spanning tree.

\begin{figure}
    \centering
    \resizebox{\textwidth}{!}{
    \includegraphics{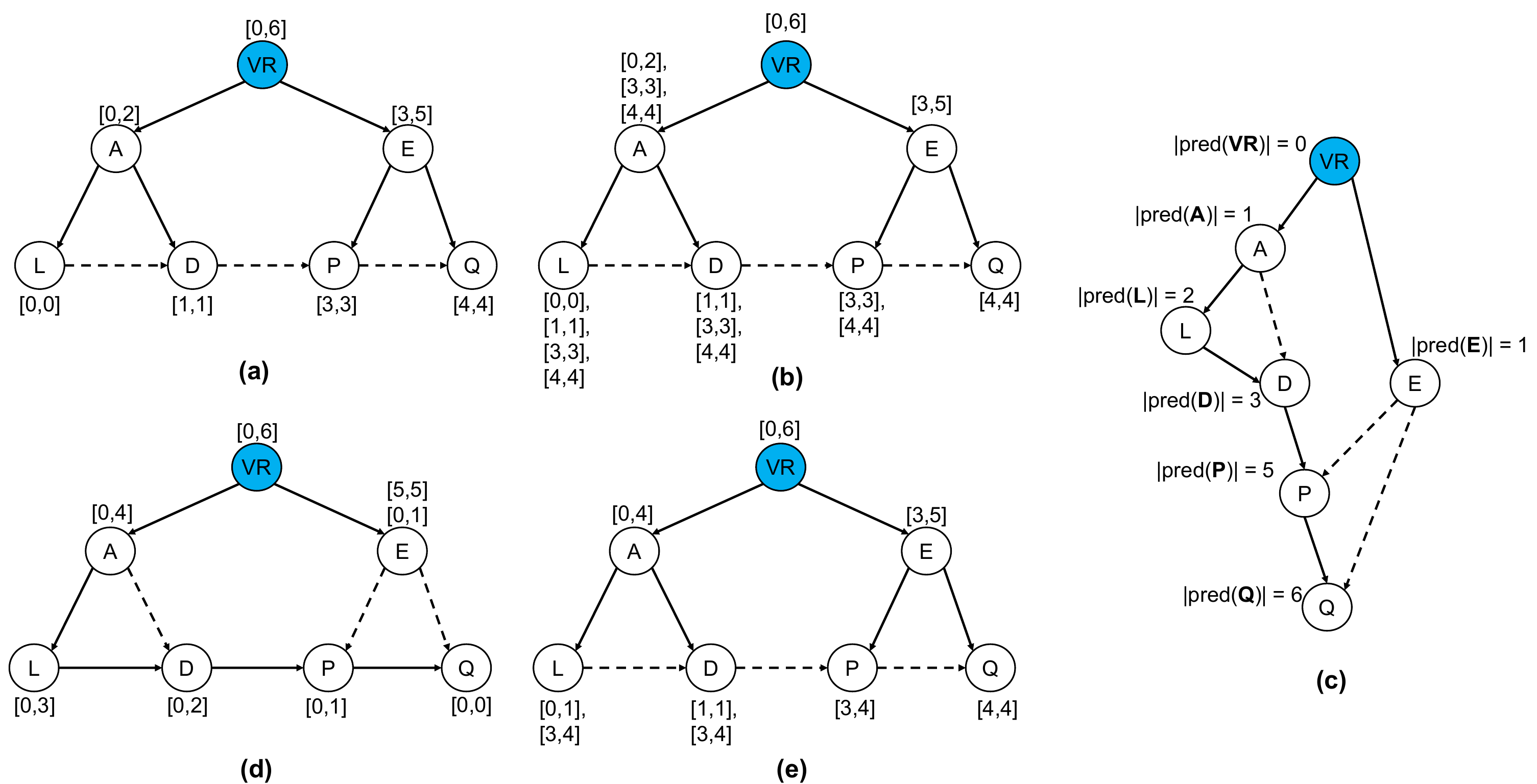}}
    \caption{\revision{Illustration of the tree cover index, where solid edges represent tree edges and dashed edges denote non-tree edges. (a) shows the initial interval labeling based on a spanning tree; (b) displays the vertex intervals after interval inheritance; (c) depicts the optimal tree cover (spanning tree); (d) presents the corresponding interval labeling based on the optimal tree cover; and (e) shows interval optimization on (b) by merging adjacent intervals to reduce index size.}}
    \label{fig:tree-cover-example}
\end{figure}

\textit{\textbf{Hooking up roots of spanning trees}}.
The implication of the first problem is the following. If  interval labeling is computed independently for each  spanning tree, then the intervals cannot be used to compute reachability queries between two vertices from different spanning trees.
To deal with this problem, a virtual root (VR) can be created, which is linked to the root of each spanning tree. 
This results in a unified spanning tree and the technique discussed above would then apply.
Consider the DAG in Fig. \ref{fig:tree-cover-example}(a). 
A VR is created to link $A$ and $E$ that are the roots of spanning trees. 

\textit{\textbf{Indexing non-tree edges}}.
The second problem implies that if a target vertex is only reachable from a source vertex via paths including non-tree edges, that reachability information is not handled in the interval labeling approach. 
Consider the example in Fig. \ref{fig:tree-cover-example}(a), vertex $Q$ is reachable from vertex $L$, but the interval $\mathcal{L}_L$ does not contain the postorder number $4$ of $Q$.
This issue  is addressed by first computing intervals according to a spanning tree and then inheriting intervals.
Specifically, vertices are examined in a reverse topological order (from  bottom to  top). For each edge $(u,v)$ in the DAG, $u$ inherits all the intervals of $v$.
There might be cases where, after performing the interval inheritance for all the outgoing edges of $u$, some intervals (of $u$) are subsumed by the other intervals. 
In this case, the subsumed intervals can  simply be deleted.
Fig. \ref{fig:tree-cover-example}(b) shows the intervals of all the vertices after performing the interval inheritance and removing subsumed intervals in Fig. \ref{fig:tree-cover-example}(a).

\textit{\textbf{The optimum tree cover}}.
The spanning tree, based on which the interval labeling is computed, has an important impact on the total number of intervals of vertices in the DAG. 
To understand this, consider the two sets of intervals computed for the same DAG but using different spanning trees in Fig. \ref{fig:tree-cover-example}(b) and Fig. \ref{fig:tree-cover-example}(d).
Both sets of intervals are able to cover  the reachability information in the DAG, but Fig. \ref{fig:tree-cover-example}(d) has less intervals than Fig. \ref{fig:tree-cover-example}(b).
So, an important question is how to compute a spanning tree that can lead to the minimum number of intervals.
\textit{The optimum tree cover} \cite{10.1145/67544.66950} is the spanning tree such that all the reachability information in the DAG can be recorded by using the minimum number of intervals. 
The spanning tree  in Fig. \ref{fig:tree-cover-example}(c)  is the optimum tree cover of the DAG. Fig. \ref{fig:tree-cover-example}(d) shows the set of intervals computed based on the optimum tree cover, which has less intervals than the one in Fig. \ref{fig:tree-cover-example}(b). 
The complexity of computing the minimum index based on the optimum tree cover is the same as the complexity of computing the transitive closure \cite{10.1145/67544.66950}.

\textit{\textbf{Merging adjacent intervals}}.
An optimization technique to reduce the number of intervals is to merge adjacent intervals.
Specifically, if vertex $v$ has two intervals $[a_v,b_v]$ and $[a'_v,b'_v]$ with $a'_v = b_v+1$ , then the two intervals can be merged into the interval $[a_v,b'_v]$.
This may result in overlapping intervals, \textit{i.e.}, $a_v\leq b_v\leq a'_v\leq b'_v$, and these can also be merged into $[a_v,b'_v]$.
For vertex $A$ in Fig. \ref{fig:tree-cover-example}(b), the intervals $[0,2], [3,3],$ and $[4,4]$ can be merged into one interval $[0,4]$, resulting in Fig. \ref{fig:tree-cover-example}(e). 
Merging reduces the total number of intervals in Fig. \ref{fig:tree-cover-example}(b) from $15$ to $9$ as shown in Fig. \ref{fig:tree-cover-example}(e). 
We note that the effect of merging intervals is not considered in the optimum tree cover. 

In essence, the tree cover index is an interval labeling approach that incorporates interval inheritance and merging. The primary limitation of the tree cover approach lies in the possibility of a substantial number of intervals.

\subsubsection{\textbf{Reducing the number of intervals}}
Several follow-up works \cite{10.5555/1083592.1083651,10.14778/1920841.1920879,6544893} aim at reducing the number of intervals for each vertex. 
\revision{The latest works adopt two different kinds of designs for reducing the number of intervals:} Recording exactly $k$ intervals for each vertex, \textit{e.g.}, GRAIL \cite{10.14778/1920841.1920879}; Recording at most $k$ intervals for each vertex, \textit{e.g.}, Ferrari \cite{6544893}.
In both approaches, $k$ is an input parameter.
Neither approach computes a complete index, and the query results using index lookups may contain false positives but no false negative. 
These partial indexes can be used to guide online traversal to compute correct query results.

\begin{figure}
    \centering
    \resizebox{0.9\textwidth}{!}{
    \includegraphics{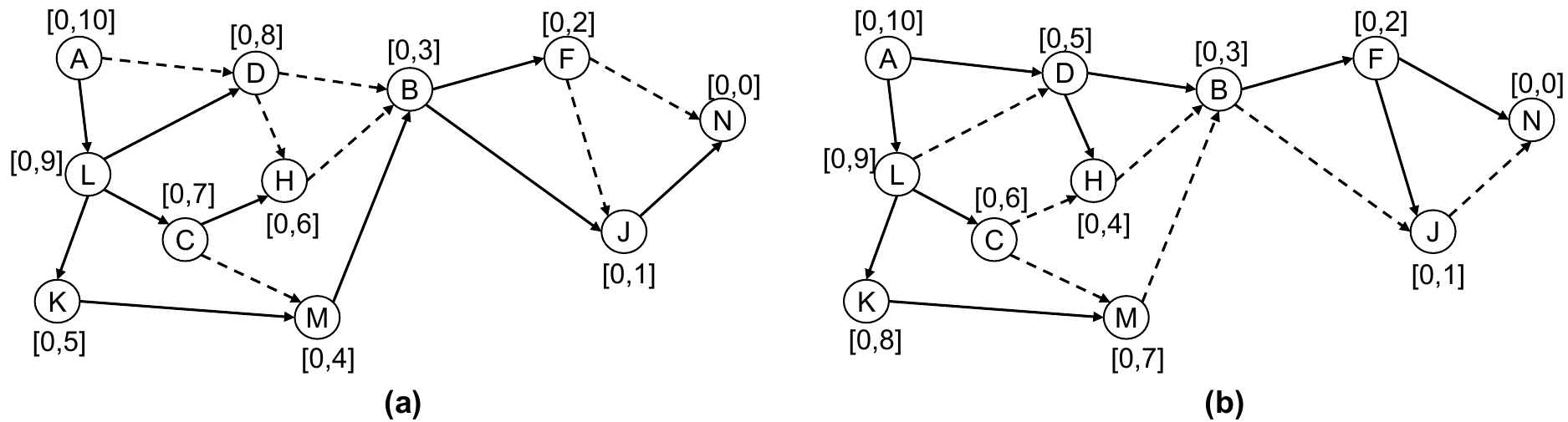}}
    \caption{\revision{Running examples of GRAIL. (a) and (b) show two intervals for each vertex in the DAG, which are computed based on two different spanning trees.}}
    \label{fig:example_grail}
\end{figure}

\textit{\textbf{Interval labeling in GRAIL}}.
Each vertex has $k$ intervals that are computed by using $k$ random spanning trees, referred to as $\mathcal{L}_{v}=(\mathcal{L}^1_{v},..,\mathcal{L}^k_{v})$, where $\mathcal{L}^i_{v}=[a^i_v, b^i_v]$ is the interval of $v$ computed based on the $i$-th spanning tree.
Similar to the interval labeling in the tree cover approach,  $\mathcal{L}^i_{v}$ is  computed based on a postorder traversal of the spanning tree, and $b^i_v$  records the postorder number of $v$ in the tree.
However, the computation of $a^i_v$ is different. 
For each vertex $v$, GRAIL gets the minimum first endpoint of all the outgoing neighbors of $v$ in the DAG, denoted as $c$, and then records the minimum of $c$ and $b^i_v$ as $a^i_v$, \textit{i.e.}, $a^i_v = min (c,b^i_v), c = min\{a^i_u: u\in outNei(v)\}$, where $outNei(v)$ is the set of outgoing neighbors of $v$ in the DAG. 
Fig. \ref{fig:example_grail}(a) shows an example of the interval labeling in GRAIL.
We note that GRAIL does not require interval inheritance.

\textit{\textbf{Query processing in GRAIL}}.
Query processing with the intervals in GRAIL does not have false negatives but may have false positives.
\revision{Consider the  query $Q_r(K,C)$ on the intervals computed in Fig. \ref{fig:example_grail}(a).} 
The interval of $C$ is not subsumed by the interval of $K$.
Therefore, the answer provided by the intervals is \textit{False}. 
Indeed, $C$ is not reachable from $K$ in the DAG. 
However, for  query  $Q_r(D,M)$, the interval of $M$ is subsumed by the interval of $D$. 
Thus, the answer provided by the intervals is \textit{True}. This does not correspond to the reachability information in the DAG as $M$ is not reachable from $D$. This is an example of false positive answer. 

\textit{\textbf{Guided DFS in GRAIL}}.
To have exact query results, a graph traversal is needed when the interval of $s$ subsumes the interval of $t$. 
The graph traversal, which is usually a DFS from $s$, can be guided by leveraging the partial index. 
Specifically, when the DFS visits vertex $v$, the intervals of $v$ and $t$ can be used to check whether $t$ is not reachable from $v$, which can be correctly processed by using the intervals. 
If so, $v$ can be pruned in the DFS.
\revision{Consider the query $Q_r(D,M)$ in the intervals computed in Fig. \ref{fig:example_grail}(a).}
The DFS is performed from vertex $D$, and then for each outgoing neighbor $v$ of $D$, we check whether $M$ is reachable from $v$ using the intervals. 
The outgoing neighbor $B$ cannot reach the target $M$ as the interval of $B$ does not subsume the interval of $M$. 
Then, $B$ is pruned in this DFS. 
For outgoing neighbor $H$, there might be false positives with the intervals. 
Then, $H$ is further explored by the DFS.
Since $B$, the the outgoing neighbor of $H$, cannot reach $M$. 
Therefore, the DFS terminates and returns \textit{False} as the query result. 

\textit{\textbf{Reducing the possibility of having false positives in GRAIL}}.
To reduce the possibility of false positives,  GRAIL computes $k$ intervals using $k$ random spanning trees. 
Consequently, each vertex has $k$ intervals, \textit{i.e.}, $\mathcal{L}_{v}=(\mathcal{L}^1_{v},..,\mathcal{L}^k_{v})$. 
With the $k$ intervals, the interval of $s$ subsumes the interval of $t$ if and only if  $\mathcal{L}^i_{s}$ subsumes $\mathcal{L}^i_{t}$ for each $i\in (1,...,k)$.
Note that the interval subsumption is checked on the intervals of $s$ and $t$ that are computed on the same spanning tree.
\revision{False positives may still exist in the query processing with $k$ intervals.}
Thus, the guided DFS is still necessary. 
\revision{Consider the running example in Fig. \ref{fig:example_grail}(a) and (b), where two spanning trees are computed on the same DAG, which give two intervals for each vertex.}
For query  $Q_r(D,M)$, with intervals  of $D$ and $M$ computed based on the second spanning tree shown in Fig. \ref{fig:example_grail}(b), the interval of $D$ does not subsume the interval of $M$.
Therefore, query result is \textit{False}. 
\revision{We note that a separately computed topological ordering of the DAG can be leveraged to further reduce the possibility of false positives. This ordering must be distinct from the postorder values used for interval labeling, even though the reverse of a postorder can also serve as a valid topological order.}

\begin{figure}
    \centering
    \resizebox{0.9\textwidth}{!}{
    \includegraphics{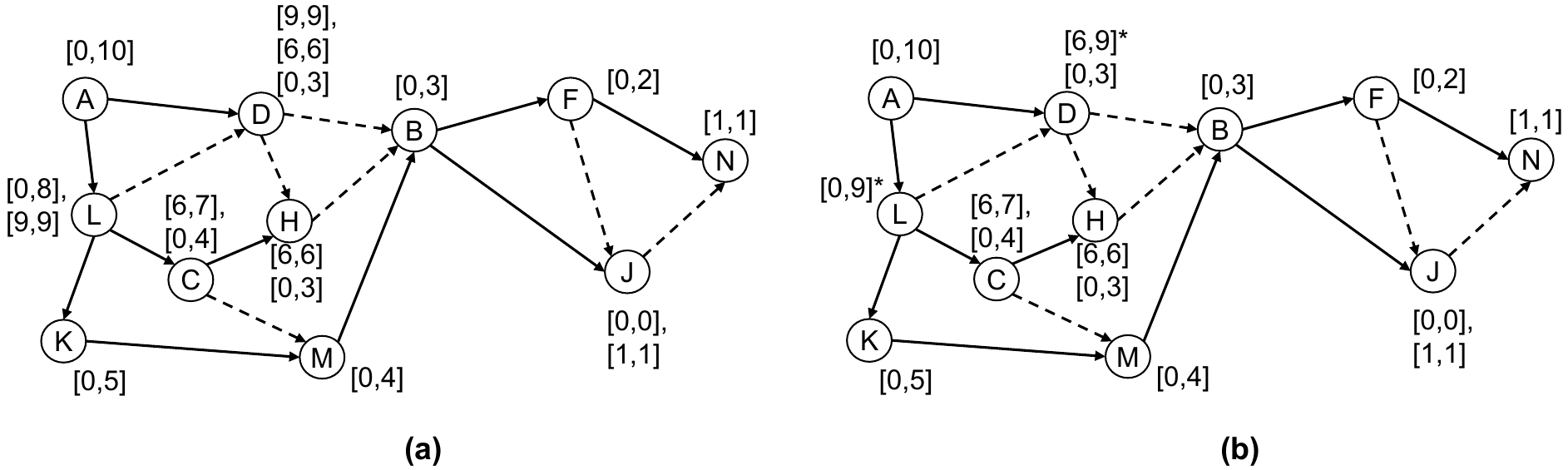}}
    \caption{\revision{Running examples of Ferrari. 
    (a) shows the Ferrari index before merging intervals and (b) shows the index after merging intervals, where each vertex has at most $2$ intervals, \textit{e.g.}, vertices D and L, }}
    \label{fig:example_ferrari}
\end{figure}

\textit{\textbf{Interval labeling in Ferrari}}.
Ferrari adopts a different design to reduce the number of intervals. 
The overall computations in Ferrari follow the steps in the tree cover approach, \textit{i.e.}, computing intervals based on spanning trees, followed by inheriting and merging intervals. 
The adjacent intervals, \textit{e.g.}, the intervals $[0,0]$ and $[1,2]$,  can be merged into $[0,2]$. 
This is termed \textit{merging without gaps}. 
Ferrari also merges non-adjacent intervals to get approximate intervals. This is termed \textit{merging with gaps}, and does not exist in the original tree cover proposal. 
Merging with gaps can be applied recursively such that each vertex has $k$ intervals.
Consider the example in Fig. \ref{fig:example_ferrari}, where intervals of each vertex in Fig. \ref{fig:example_ferrari}(a) are merged so that each vertex has at most $2$ intervals (Fig. \ref{fig:example_ferrari}(b)). 
Vertex $D$ in Fig. \ref{fig:example_ferrari}(a) has $3$ intervals, which is merged into $2$. 
Two options are available: 
(i) merging the first two intervals $[0,3]$ and $[6,6]$ into $[0,6]^*$ (starred intervals denote approximate ones);
(ii) merging the last two intervals $[6,6]$ and $[9,9]$ into $[6,9]^*$. 
In this example, option (ii) is adopted as the size of the corresponding approximated interval is smaller. The intuition is to add less error when merging non-adjacent intervals. 
When non-adjacent intervals are merged, reachability information that does not exist in the graphs is added into the intervals. 
Since the primary goal of Ferrari is to reduce the number of intervals,  the merging options that  introduce less error are preferable.

\textit{\textbf{Query processing in Ferrari}}.
Query processing in Ferrari first uses exact intervals to find reachability. 
If that is not possible, approximate intervals are used. As in GRAIL, using exact intervals may result in false negatives (but no false positives).
Using approximate intervals may result in false positives but no false negatives, because approximated intervals are obtained only by merging non-adjacent intervals. 
For the query $Q_r(D, N)$ in Fig. \ref{fig:example_ferrari}(b), the postorder number of $N$ is contained in the exact interval of $D$, thus the answer is \textit{True}. 
For the query $Q_r(D, K)$ in Fig. \ref{fig:example_ferrari}(b), we can not determine the query result with the exact interval of $D$ as it does not contain the postorder number of $K$.
In this case, the approximate interval $[6,9]^*$ of $D$ is used, resulting in the answer \textit{False}, since the postorder number is not contained in $[6,9]^*$. 
As false positives are still a possibility, graph traversal is still needed as in Ferrari.

Although partial indexes require additional graph traversals for query processing, their index building time and index size scale linearly with the input graph size, making them one of the first methods feasible for large graphs with millions of vertices.
\revision{
In addition, reachability processing using these indexes can be an order of magnitude faster than graph traversals \cite{10.14778/1920841.1920879}.}

\textit{\textbf{Other approaches for non-tree edges}}.
There also exist a few early approaches that are designed to deal with the problem of non-tree edges in the interval labeling technique. These include Tree-SSPI \cite{10.5555/1083592.1083651},  Dual-labeling \cite{1617443}, and  GRIPP \cite{10.1145/1247480.1247573}. 
These early indexes have specific designs that are different from the ones in GRAIL and Ferrari. We briefly discuss them below.  \\
\textbf{Tree-SSPI} \cite{10.5555/1083592.1083651} distinguishes two kinds of paths in a DAG: \textit{tree paths} of only tree edges, and \textit{remaining paths} of non-tree edges. A target vertex $t$ is reachable from a source vertex $s$ via one of these. Tree paths can be simply addressed by  interval labeling. For remaining paths, additional information is kept about the predecessors of the target vertex.\\
\textbf{Dual-labeling} \cite{1617443} also leverages the interval labeling technique to deal with reachability caused by tree paths. The interval $[v_a,v_b)$ is kept for each vertex $v$, where $v_a$ and $v_b-1$ are the preorder and postorder numbers of $v$ respectively (this is equivalent to the original one as $Q_r(s,t)$ can be simply processed by checking whether $t_a$ is in the range of $[s_a,s_b)$). Dual-labeling records a \textit{link table} to process reachability caused by paths of non-tree edges. Specifically, if there is a non-tree edge $(u,v)$ such that $v$ is not reachable from $u$ via tree paths, the \textit{link} $u_a\rightarrow [v_a,v_b)$ is recorded in the link table. Then, in the link table, if two links $u_a\rightarrow [v_a,v_b)$ and $x_a\rightarrow [y_a,y_b)$  satisfy the condition that $x_a$ is in the range of $[v_a,v_b)$, an additional link $u_a\rightarrow [y_a,y_b)$ is recorded in the link table as well. This essentially computes and stores the transitive closure of the link table. Given a query $Q_r(s,t)$ that cannot be processed by the intervals of $s$ and $t$, dual-labeling looks for a link $u_a\rightarrow [v_a,v_b)$ such that $u_a$ is in the range of $[s_a,s_b)$ and $t_a$ is in the range of $[v_a,v_b)$. \\
\textbf{GRIPP} \cite{10.1145/1247480.1247573} performs a DFS and directly computes intervals of vertices on a general graph that can have cycles. As the incoming degree of a vertex $v$ in the graph can be larger than $1$, $v$ will be considered having an instance for each incoming edge. One of these instances is treated as the \textit{tree instance} of $v$, denoted as $v'$. During the DFS, the outgoing neighbors of tree instances will be visited while the search will not be expanded for the case of non-tree instances.  
For processing $Q_r(s,t)$, GRIPP uses the interval of the tree instance $s'$ to retrieve a set of vertex instances that are reachable from $s'$, denoted as reachable instance set (RIS) of $s$ or $RIS(s)$. 
If any instance of $t$ is in $RIS(s)$, query result $True$ is immediately returned. 
Otherwise, the non-tree instances in $RIS(s)$ are retrieved to check reachability. \revision{This procedure is performed recursively until no instance of $t$ can be reached, in which case the query result is $False$.}

\subsubsection{\textbf{Advanced covers}}
The tree-cover-based approaches use spanning trees to cover an input DAG and computes the interval labeling on the spanning trees.
\revision{
Path-tree labeling \cite{10.1145/1376616.1376677,10.1145/1929934.1929941} uses a structure that is more advanced than a tree in the input DAG.
The approach first computes a disjoint set of paths in the graph such that each vertex belongs to a specific path. Each of these sets of paths is known as a \textit{path partition} \cite{10.1145/1376616.1376677}.  
Then, the advanced structure is computed by partitioning the DAG by the path partition to obtain a \textit{path-tree graph} \cite{10.1145/1376616.1376677}, where each vertex represents a path in the DAG.}
The interval labeling with additional information is computed on the spanning trees of the path-tree graph. 
\revision{
The main reason for adopting the advanced cover is that it results in fewer non-covered edges compared to a tree cover.}
The number of non-covered edges has an impact on the index size. 
Thus, the advanced cover can lead to an index of smaller size.
The advanced structure provides more comprehensive coverage of the DAG compared to  the spanning tree because: (i) an edge in a path of the path partition is considered covered; (ii) a pair of crossed edges from one path to another path in the path partition contains redundancy and keeping one of two edges is sufficient.
The reachability in the path-tree graph can be processed by computing the interval labeling augmented with additional information of vertices in paths. 
\revision{
Reachability caused by non-covered edges is handled by computing the corresponding transitive closure, which is essentially the approach used in dual-labeling \cite{1617443}.}

\begin{figure}
    \centering
    \resizebox{0.9\textwidth}{!}{
    \includegraphics{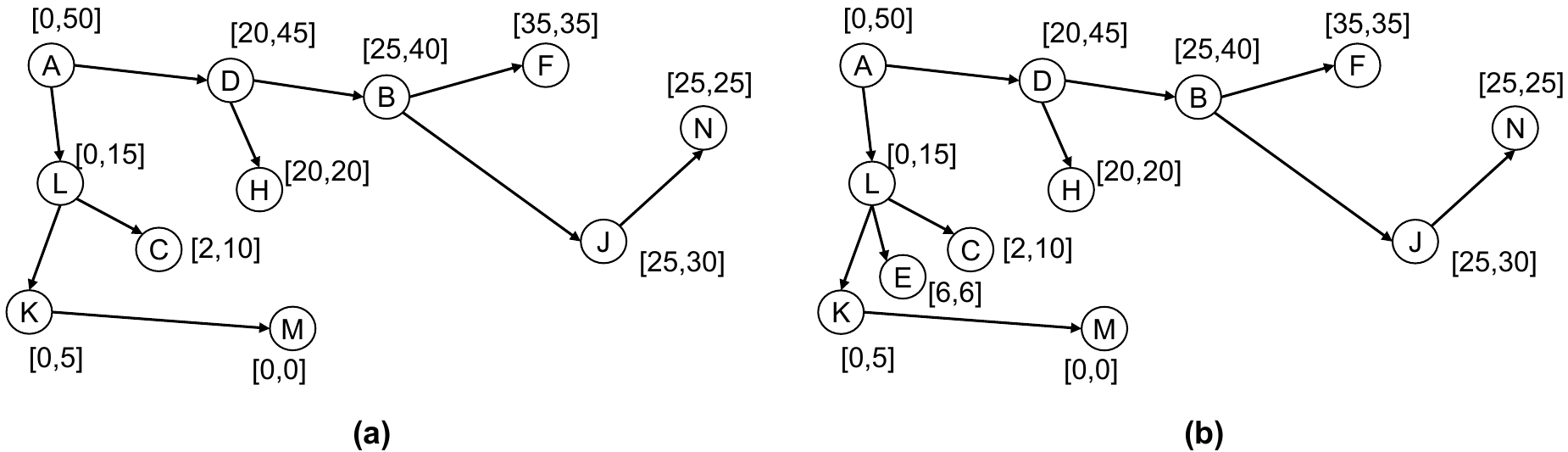}}
    \caption{\revision{Interval labeling based on non-contiguous postorder numbers. (a) shows the labeling before inserting edge (L,E); (b) shows the updated labeling after edge (L,E) is inserted, utilizing gaps between postorder numbers to avoid recomputation.}}
    \label{fig:non-contiguous_postorder}
\end{figure}

\subsubsection{\textbf{Techniques for maintaining intervals on dynamic graphs}}
\revision{
A few tree-cover-based approaches \cite{10.1145/67544.66950,Yildirim2013DAGGERAS} also address index maintenance as the graph is updated.} 
Due to their reliance on a DAG, tree-cover-based approaches need to maintain two kinds of reachability information: reachability within SCCs and reachability within DAGs. 
Maintaining SCCs over dynamic graphs is expensive, as edge insertions or deletions might lead to merging or splitting SCCs. 
For maintaining the second kind of reachability information, a few optimization techniques have been proposed.

\textit{\textbf{Non-contiguous postorder numbers}}.
In the original tree cover index, the postorder numbers computed for each vertex do not need to be contiguous. 
The intervals in Fig. \ref{fig:interval-labeling-of-the-spanning-tree} are computed based on the contiguous postorder numbers $(0,1,2,3,4,5,6,7,8,9,10)$.
If the contiguous sequence is replaced with the following non-contiguous sequence $(0,5,10,15,20,25,30,35,40,45,50)$, the corresponding intervals are still correct (Fig. \ref{fig:non-contiguous_postorder}(a)). 
The benefit of using a non-contiguous sequence is that the gaps between the non-contiguous postorder numbers can be used to deal with tree-edge insertions.
In Fig. \ref{fig:non-contiguous_postorder}, where tree-edge $(L, E)$ is inserted into the tree in Fig. \ref{fig:non-contiguous_postorder}(a), the gap between the postorder numbers of $K$ and $C$ can be leveraged to get the postorder number of $E$ ($6$ in this example), based on which the interval $[6,6]$ of $E$ can be computed. 
The intervals for the other vertices remain  unchanged.
If there is no gap between non-contiguous postorder numbers, then recomputing the postorder numbers is  necessary.
Determining a reasonable gap between non-contiguous postorder numbers for real-world applications has not been studied.

\textit{\textbf{Edge deletions in partial indexes allowing false positives}}.
Dagger \cite{Yildirim2013DAGGERAS} is an extension of GRAIL to dynamic graphs. Thus, it is a partial index without false negatives in the query result but may have false positives.
It also leverages the non-contiguous postorder numbers to deal with tree edge insertions.
The main issue addressed in Dagger is that when an edge in the DAG is deleted, the index does not require any updates. 
The reason is that deleting edges only introduces false positives in the index that are allowed in Dagger. 

\subsubsection{\textbf{Chain-cover indexes}}\label{sec:chain_cover_index}
Chain cover \cite{10.1145/99935.99944} index class is similar to the tree cover index class as both of them use specialized structures to cover the input graph, \textit{i.e.},  chains\footnote{A chain \cite{10.1145/99935.99944} is a sequence of vertices and for any adjacent pair of vertices $(u,v)$ in the sequence there exists a path or an edge from $u$ to $v$ in the graph.} in the former and trees in the latter.
Tree cover can be thought of as a variant of chain cover \cite{10.1145/99935.99944}.
\revision{
Tree cover can offer a better compression rate than chain cover \cite{10.1145/1929934.1929941}. This is because, unlike chains where each vertex has at most one immediate successor, trees allow multiple immediate successors per vertex, resulting in more efficient coverage and compression.}

\subsection{2-Hop-based indexes}\label{sec:plain_2-hop_index}
Given two paths $(s,u)$ and $(u,t)$, the existence of the path $(s,t)$ is obvious. Obtaining new paths by concatenating those with common vertices can save storage space in the transitive closure. 
The intuition behind the 2-hop index is to maintain paths that can be achieved via two hops\footnote{In the 2-hop index, a hop can be an edge or a path of an arbitrary length.}, and to use path concatenation to compress transitive closures.

\subsubsection{\textbf{The approach}}
In the 2-hop index (or the 2-hop labeling), each vertex $v\in V$ is labeled with two sets of vertices $L_{in}(v)$ and $L_{out}(v)$, such that  the following two conditions are satisfied: 
(i) $L_{in}(v),L_{out}(v)\subseteq V$, and there is a path in $G$ from every $x\in L_{in}(v)$ to $v$ and there is a path in $G$ from $v$ to every $x\in L_{out}(v)$; 
(ii) for any two vertices $s,t\in V$, $t$ is reachable from $s$ if and only if $L_{out}(s)\cap L_{in}(t)\neq \emptyset$.
Notice that the 2-hop index assumes\footnote{Without the assumption, condition (ii) is equivalent to: $t$ is reachable from $s$ if and only if  $t\in L_{out}(s)$, $s\in L_{in}(t)$ or $L_{out}(s)\cap L_{in}(t)\neq \emptyset$.} that $v$ is included in $L_{in}(v)$ and $L_{out}(v)$ for each $v$. 
Consider the index in Fig. \ref{fig:example-2-hop}(a) and the query $Q(A,N)$. Since $L_{out}(A)\cap L_{in}(N)\neq \emptyset$, $Q(A,N)=True$.

In constructing the 2-hop index, the transformation from a general graph to a DAG is optional. 

\subsubsection{\textbf{The minimum 2-hop index}}
The size of the 2-hop index is $ \sum_{i=1}^n |L_{out}(v_i)| + |L_{in}(v_i)|$ for a graph with $n$ vertices. 
Given an input graph, many 2-hop indexes can be built, such that each of them can satisfy the two conditions discussed above. The minimum 2-hop index is the one that has the smallest index size. Obviously, the minimum 2-hop index can maximumly compress the transitive closure of the input graph. 
\revision{
We use the following toy example to explain the intuition.
Consider a graph with two edges $(s,u)$ and $(u,t)$. 
One possible 2-hop index is to have $u$ in both $L_{out}(s)$ and $L_{in}(t)$.
Another option is to record $s$ in $L_{in}(u)$ and $L_{in}(t)$ and to record $t$ in $L_{out}(u)$. 
The first option is preferable since it has fewer number of index entries.}
The general goal of building the minimum 2-hop index is to maximumly compress the transitive closure. 
However, the problem of computing the minimum 2-hop index is NP-hard \cite{10.5555/545381.545503}. 
Efficient heuristics have been proposed to reduce the index size, \textit{e.g.}, TFL \cite{10.1145/2463676.2465286}, DL \cite{10.14778/2556549.2556578}, PLL \cite{10.1145/2505515.2505724}, and TOL \cite{10.1145/2588555.2612181}.
We take PLL as a representative for discussion. Then, we will mention the differences between different approaches.

\textbf{\textit{The PLL indexing algorithm}}.
The indexing algorithm of PLL \cite{10.1145/2505515.2505724} consists of three major steps: 
(i) computing a vertex accessing order; 
(ii) processing vertices iteratively according to the vertex accessing order; 
(iii) applying pruning rules during the processing of each vertex.
For computing a vertex accessing order, a simple strategy is to sort vertices according to vertex degree, \textit{i.e.}, $(d_{in}(v)+1)\times (d_{out}(v)+1)$, where $d_{in}(v)$ and $d_{out}(v)$ are in-degree and out-degree of each vertex $v$.
Vertices are then iteratively processed according to the vertex accessing order, and processing each vertex $v$ consists of performing backward and forward BFSs from $v$ to compute $L_{out}$-entries and $L_{in}$-entries. 
During each BFS, PLL prunes the search space guided by a set of rules. 
Consider the case of forward BFS (backward BFS works similarly).
When the forward BFS from $v$ visits $u$ and $u$ has been processed, \textit{i.e.}, $u$ ranks in front of $v$ in the vertex accessing order,  $u$ is pruned in the BFS. The reason is that the reachability information due to the paths from $v$ to any vertex via $u$ has been recorded in the backward and forward BFSs performed from $u$.

\begin{figure}
    \centering
    \resizebox{!}{0.5\textwidth}{
    \includegraphics{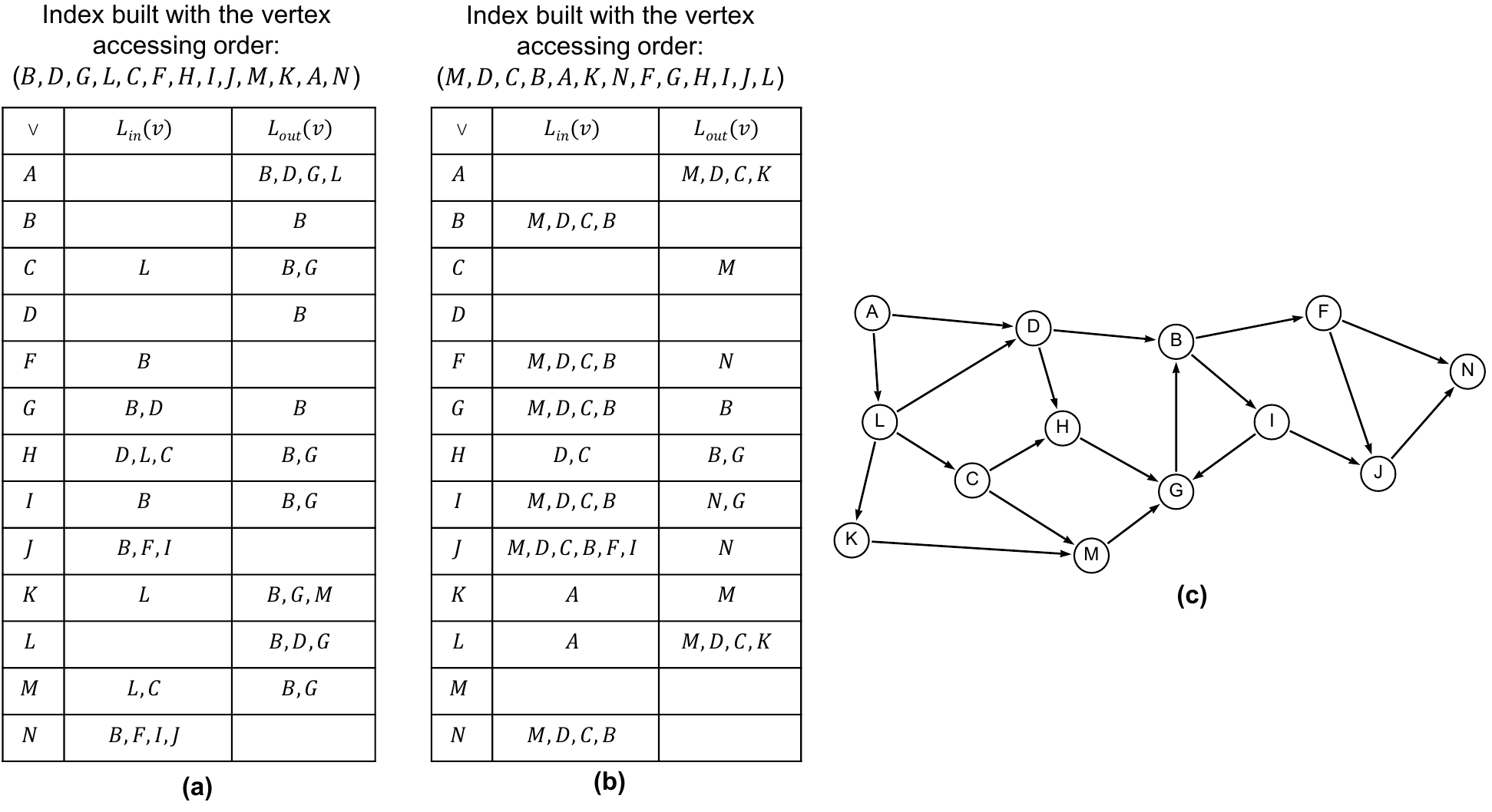}}
    \caption{\revision{The 2-hop indexes (a) and (b) built for the graph (c). Although both indexes can fully record the reachability information, the index in (a) has a smaller size than the index in (b).}}
    \label{fig:example-2-hop}
\end{figure}

\revision{Consider the 2-hop indexes shown in Fig.~\ref{fig:example-2-hop}(a) and Fig.~\ref{fig:example-2-hop}(b), both constructed for the graph in Fig.~\ref{fig:example-2-hop}(c). These two indexes are built using different vertex accessing orders but are functionally equivalent, as both fully encode the reachability information. The index in Fig.~\ref{fig:example-2-hop}(a) contains 39 entries, which is fewer than the 47 entries in Fig.~\ref{fig:example-2-hop}(b). In general, determining an optimal vertex access order that minimizes index size is NP-hard~\cite{weller2014optimal}. The order based on vertex degree, as used in Fig.~\ref{fig:example-2-hop}(a), serves as an example of an efficient heuristic~\cite{10.1145/2505515.2505724}.}

\textbf{\textit{Query optimization based on vertex accessing order}}.
\revision{Query processing using the 2-hop index needs to check whether one of the following conditions can be satisfied: $t\in L_{out}(s)$, $s\in L_{in}(t)$, or $L_{out}(s)\cap L_{in}(t)\neq \emptyset$.}
In order to efficiently perform the search in $L_{out}(s)$ and $L_{in}(t)$, the access order of vertices in $L_{out}(s)$ and $L_{in}(t)$ can be leveraged.
This is feasible because vertices are added into $L_{out}$-entries and $L_{in}$-entries according to the vertex accessing order.
Let $rank(v)$ be the vertex accessing order of $v$, and 
$rank(L_{out}(v)) = \{rank(u)|u\in L_{out}(v)\}$ and 
$rank(L_{in}(v)) = \{rank(w)|w\in L_{in}(v)\}$.
Then, we can store $rank(L_{out}(v))$ and $rank(L_{in}(v))$ instead of $L_{out}(v)$ and $L_{in}(v)$, and the elements in $rank(L_{out}(v))$ and $rank(L_{in}(v))$ are already sorted.
Given a query $Q_r(s,t)$, the search in $rank(L_{out}(s))$ and $rank(L_{in}(t))$ is performed. Checking whether $t\in L_{out}(s)$ or $s\in L_{in}(t)$ can be processed by the binary search of $rank(t)\in rank(L_{out}(s))$ or $rank(s)\in rank(L_{in}(t))$.
In order to check whether $L_{out}(s)\cap L_{in}(t)\neq \emptyset$,  merge-join can be performed without the need for sorting.
 

\textbf{\textit{The TOL indexing framework}}.
\revision{TOL  \cite{10.1145/2588555.2612181} is a general index class that generalizes the indexing algorithm of PLL.} 
TOL computes a total order of vertices and then iteratively processes vertices to compute $L_{out}$-entries and $L_{in}$-entries according to the total order, performing pruning during processing. 
The vertex accessing order computed based on vertex degree is an instantiation of the total order, which is the strategy used in DL and PLL\footnote{It has been proven that DL and PLL are equivalent \cite{10.14778/2556549.2556578}.}. 
An alternative instantiation is the use of topological folding numbers, which are derived from a topological ordering of the transformed DAG and used in the TFL index \cite{10.1145/2463676.2465286}.
An advanced instantiation of the total order introduced in TOL is based on the contribution score of vertices, which is defined according to the number of the vertices that can reach $v$ and the number of the vertices that are reachable from $v$. The contribution score can be approximated by a linear scan of the DAG. 
\revision{
The TOL index built according to approximate contribution scores exhibits superior performance in terms of index construction time over those built based on previous heuristics  \cite{10.1145/2588555.2612181}.} 
In addition, it can support graph updates.
We note that among existing heuristics, only the simple strategy based on vertex degree can be applied to the case of general graphs, while other strategies require DAG as input.

\subsubsection{\textbf{Partial 2-hop indexes}}
The 2-hop indexes presented previously are complete indexes in that they record complete reachability information. Partial 2-hop indexes are also possible (\textit{e.g.},  DBL \cite{10.1007/978-3-030-73197-7_52} and O'Reach \cite{10.1145/3556540}) where the index is built only for a selected subset of vertices in the graph.
In DBL, the subset of vertices are selected as those in the top-$k$ degree ranking, aka \textit{landmark vertices}. 
In O'Reach, the subset of vertices are those residing in the middle of a topological ordering, referred as \textit{supportive vertices}.
Query processing in DBL and O'Reach might require graph traversals. 
DBL also adds  BFL that is a partial index without false negatives (presented later in Section \ref{sec:approximate_tc}) to prune the search space when traversing the graph.

\subsubsection{\textbf{Extended hop-based indexes}}
\revision{A few indexes in the literature extend the 2-hop index.}
With the 2-hop index, a query $Q_r(s,t)$ is processed by checking whether there are common vertices in $L_{out}(s)$ and $L_{in}(t)$. Some algorithms extend the definition of ``common vertices'' by replacing them with common chains (the 3-hop index \cite{10.1145/1559845.1559930}) or common trees (the path-hop index \cite{10.1145/1871437.1871457}). 
Although these indexes are able to further compress transitive closure,  the corresponding indexing cost is higher than  that of the 2-hop index, which makes them less applicable.

\subsubsection{\textbf{2-Hop indexes on dynamic graphs}}\label{sec:2-hop_dynamic}
A few 2-hop indexes can support dynamic graphs, including TOL and DBL.
TOL assumes a DAG as input, such that the input general graph is first transformed into a DAG. Thus, maintaining SCCs is necessary for the index updates.
TOL relies on existing approaches \cite{Yildirim2013DAGGERAS,10.14778/3364324.3364329,10.1137/1.9781611976007.9} to deal with the problem of maintaining SCCs and designs an incremental algorithm to perform index updates on the DAG.
The algorithm can support both edge insertions and deletions.  
We use the case of edge insertion to discuss the underlying intuition.
Consider inserting an edge $(u,v)$ into a graph. 
Let $in(u)$ be the set of vertices that can reach $u$ and $out(v)$ be the set of vertices that are reachable from $v$. 
After inserting $(u,v)$, every vertex  in $out(v)$ is reachable from every vertex in $in(u)$, and this reachability information needs to be recorded in the index. 
TOL computes $in(u)$ and $out(v)$ by using the snapshot of the index that exists before inserting $(u,v)$.
In general, $out(v)$ is computed by using $L_{out}(v)$ and $W=\{w| \exists x\in L_{out}(v), w \text{ is reachable from }x\}$.
$W$ can be efficiently computed by recording \textit{inverted index entries} for each vertex, \textit{i.e.}, if $x\in L_{in}(w)$, then TOL records $w\in L^{-1}_{in}(x)$.
The inserted index entries can make some of the index entries that exist prior to the edge insertion redundant. Such redundancy can be addressed by the TOL indexing algorithm efficiently.
\revision{The  edge deletion case works in a similar way in TOL.}
DBL can only support edge insertions, but does not assume a DAG as input, so there is no need to maintain SCCs. 
In DBL, $in(u)$ and $out(v)$ are computed by performing backward and forward BFSs, respectively.
Early works also discussed how to maintain the 2-hop index on dynamic graphs, including the U2-hop index \cite{4912201} and the index proposed by Ralf et al. \cite{1410143}. However, it has been shown that they cannot scale to large graphs \cite{10.1145/2588555.2612181}.

\begin{figure}
    \centering
    \resizebox{0.85\textwidth}{!}{
    \includegraphics{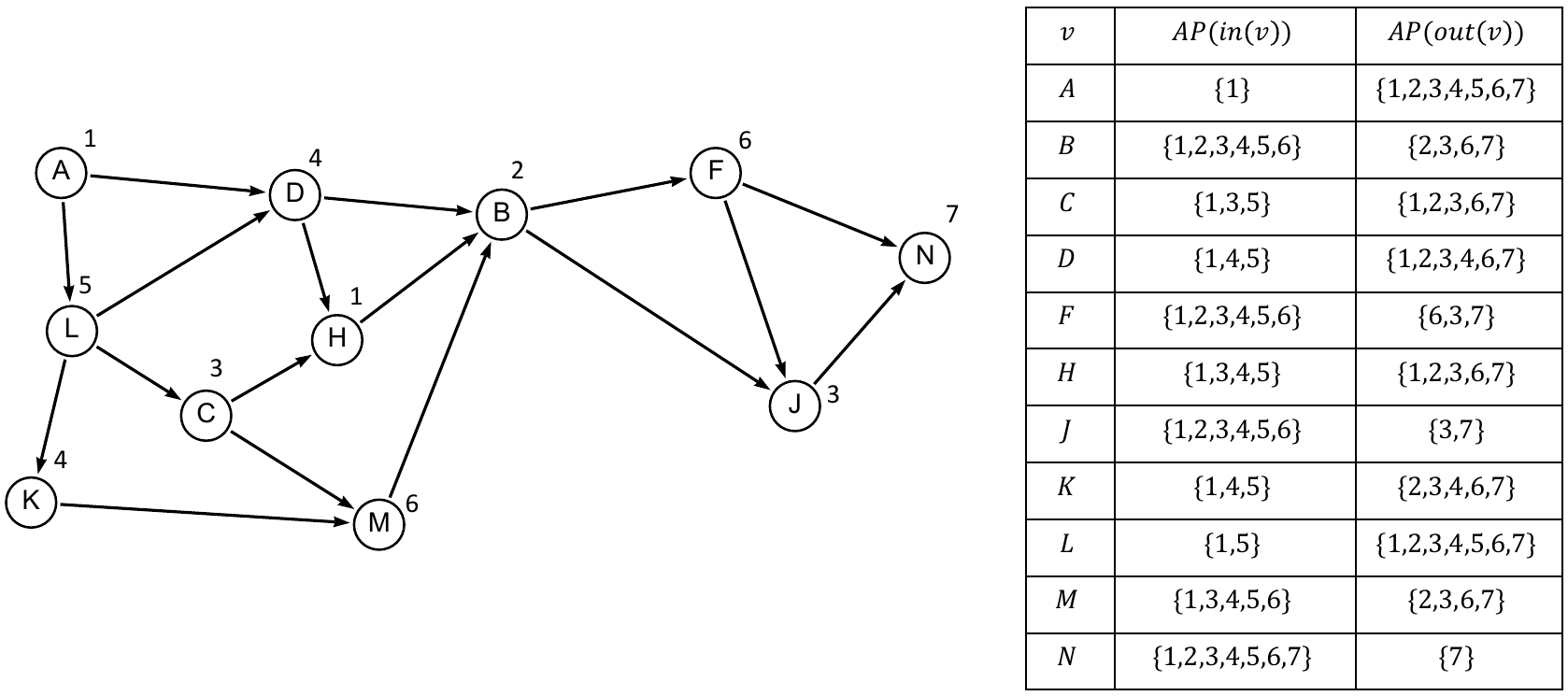}}
    \caption{\revision{Running example of the BFL index.}}
    \label{fig:example-BFL}
\end{figure}

\subsection{Approximate transitive closures}\label{sec:approximate_tc}
\subsubsection{\textbf{The approach}}
Note the following:  
If vertex $u$ can reach vertex $v$, then $out(v)\subseteq out(u)$, because there might exist a vertex $w$ such that $w$ is reachable from $u$ but $w$ is not reachable from $v$.
For example, in Fig. \ref{fig:DAG-of-the-plain-graph}, $C$ can reach $B$, such that $out(B)\subseteq out(C)$.
Based on the observation, we can have the following condition: if $out(v)\nsubseteq out(u)$, then $u$ cannot reach $v$.
Similarly, if $in(u)\nsubseteq in(v)$, then $u$ cannot reach $v$.
The two conditions are not usable for designing a reachability index as computing $out(v)$ or $in(v)$ for each vertex $v$ is equivalent to computing the transitive closure, which is unrealistic in practice. 
This is the basis of approximate transitive closure algorithms.
The primary goal is to compute an approximate version of $out(v)$ and $in(v)$, such that they can be efficiently computed and also have a much smaller size.
We use $AP()$ to denote the approximating function.
The main problem is how to design $AP()$. 
It should satisfy the following condition: if $AP(out(v))\nsubseteq AP(out(u))$ then $out(v)\nsubseteq out(u)$. Therefore, we can derive that $u$ cannot reach $v$. 
Consequently, the approximate transitive closure will not have false negatives in the query results.

\subsubsection{\textbf{The design of the AP() function}}
Two approximate functions have been proposed. The first one is based on the min-wise independent permutation (IP index \cite{10.14778/2732977.2732992,10.1007/s00778-017-0468-3}). 
\revision{
The approximate function computes and records in $out(v)$ and $in(v)$ a fixed number of smallest integers obtained by a permutation of vertices.}
The second  is based on Bloom filter (BFL index  \cite{7750623}). The approximate function computes and records in $out(v)$ and $in(v)$ the hash codes of vertices.
Consequently, both indexes are partial indexes, and the query processing will not have false negatives but may have false positives. This is because  the indexes are designed based on the condition discussed previously. 
Thus, if the index lookup returns \textit{True}, graph traversal is necessary.
In this case, the guided DFS designed for GRAIL can also be applied, \textit{i.e.}, pruning the search space by leveraging the fact that index lookups do not provide false negatives.

\revision{
We use a toy example to illustrate BFL. 
We compute a hash code for each vertex in the DAG in Fig. \ref{fig:DAG-of-the-plain-graph}, and the resulting DAG and the corresponding BFL index are shown in Fig. \ref{fig:example-BFL}. $ AP(out(v))$ and $ AP(in(v))$ record the hash codes of vertices, instead of vertex identifiers. 
Consider the query $Q_r(B,C)$. 
Since $out(C)\nsubseteq out(B)$, we can determine that $C$ is not reachable from $B$.
For the query $Q_r(D,M)$, index scan shows $out(M)\subseteq out(D)$ and $in(D)\subseteq in(M)$. Therefore, the query result cannot be determined by using only index lookups as there might be false positives.
In this case, the guided DFS from $D$ needs to be performed. The DFS can be accelerated by leveraging the partial index to prune search space, \textit{i.e.}, none of the outgoing neighbors of $D$ can reach $M$, thus the query result $False$ can be immediately returned.}

\subsection{Other techniques}
A few other indexing techniques have also been proposed that do not fall into the three main classes discussed above.
We briefly discuss these techniques below.

\subsubsection{\textbf{Dominance drawing}}
Feline \cite{Veloso2014ReachabilityQI} is designed based on dominance drawing, \textit{i.e.}, assigning coordinates $(v_x,v_y)$ to every vertex $v$ in the graph. In Feline, the assigned coordinates satisfy the following attributes: if $v$ is reachable from $u$, then $u_x\leq v_x$ and $u_y\leq v_y$.
Consequently, Feline is a partial index. The query results with index lookups computed based on the coordinates do not have false negatives but may have false positives, requiring graph traversal. 

\subsubsection{\textbf{Reachability contraction hierarchies}}
Preach \cite{merz2014preach}  is designed based on reachability contraction hierarchies. Vertices with out-degree of $0$ (\textit{sink} vertices) and those with in-degree of $0$ (\textit{source} vertices) are recursively removed from the graph. 
This produces a vertex ordering that will be used in answering query $Q_r(s,t)$.
The idea is to use the vertex ordering to prune the search space during the graph traversal for answering $Q_r(s,t)$.
Specifically, queries are processed by bidirectional search from $s$ and $t$, where only paths of increasing vertex orderings are visited. Preach is essentially a partial index without false positives, and the bidirectional search needs to be exhausted to return \textit{False}. Pruning rules are designed to accelerate the searching.

\subsubsection{\textbf{Reachability-specific graph reduction}}
A few graph reduction techniques are proposed to accelerate reachability indexing, including SCARAB \cite{10.1145/2213836.2213856}, ER \cite{zhou2017dag}, and RCN \cite{zhou2021fast}. 
The main idea is to reduce the size of the input graph by removing vertices/edges while preserving reachability information.
The reduced graph will be taken as input for building reachability indexes.
These reduction techniques are orthogonal to the indexing techniques.

\subsubsection{\textbf{Index-free approaches}}
Recent works propose advanced traversal approaches for plain reachability query processing without using indexes (arrow \cite{8731546}, IFCA \cite{ICDE55515.2023.00172}). These approaches provide optimized traversal algorithms that are slightly faster than the naive traversal. 
However, query performance of these approaches is still not comparable with indexing methods as index-free methods do not leverage any offline computation.

\subsection{\revision{Hop-constrained reachability}}
\revision{
In addition to plain Boolean reachability, many real-world scenarios require path existence checks under length constraints. These \textit{hop-constrained reachability queries} \cite{10.14778/2350229.2350247} ask (i) whether a path exists from a source vertex $s$ to a target vertex $t$, and (ii) the path contains at most $k$ edges.
Such constraints arise in a variety of application domains, including social networks (\textit{e.g.}, friends-of-friends recommendations) \cite{10.14778/2350229.2350247}, and financial systems for fraud detection and tracing \cite{10.1007/978-3-031-00129-1_37}.}

\revision{
Traditional reachability indexes are designed to answer unbounded paths and therefore cannot directly support hop-constrained semantics. To address this gap, K-Reach \cite{cheng2014efficient,10.14778/2350229.2350247} introduces an efficient indexing scheme based on vertex covers. It constructs a reduced subgraph by selecting high-impact vertices and precomputing $k$-hop reachability among them. During query evaluation, it classifies the source and target vertices relative to the vertex cover and applies either local search or index-based lookup accordingly. This hybrid approach achieves a favorable balance between indexing cost and query performance.
SQLG+ \cite{10.1007/978-3-031-00129-1_37} takes a system-oriented perspective by enabling $k$-hop query processing directly within relational database systems. It translates queries into recursive SQL with pruning optimizations, leveraging standard indexing and execution engines. }


\subsection{\revision{Practical trade-offs and guidance}}\label{sec:practical_discussion_pr}
\revision{
While Tables \ref{table:pr_indexes} summarizes various indexing techniques, practitioners often need additional guidance when choosing an appropriate index for a specific application. We discuss practical trade-offs by distinguishing between static and dynamic graph settings and considering structural characteristics, performance needs, and query types.}

\revision{
For static graphs, one key factor is graph sparsity.
If the graph is sparse or structurally close to a tree, tree-cover–based methods are generally more suitable. Interval labeling schemes can compactly encode reachability information  with low space overhead and fast query performance.
For graphs that are moderately sparse, and when indexing time is a major concern, approximate transitive closure methods such as BFL are preferable, as they offer a good balance between construction cost and query performance.
In contrast, for denser graphs, 2-hop labeling approaches tend to perform better. Although their construction time can be higher, they provide excellent query efficiency for such graphs.}

\revision{
In dynamic settings, maintaining certain index structures becomes more challenging.
Indexes that depend heavily on precomputing SCCs become less suitable, as SCC maintenance under frequent updates is computationally expensive.
In these cases, 2-hop indexes are often more effective as most of them do not require such DAG transformation.
If the graph is guaranteed to remain a DAG even after updates, and the update frequency is low, tree-cover–based methods, especially those using non-contiguous postorder optimizations, remain a promising option.}

\revision{
If the indexing method is expected to support multiple types of queries \textit{(e.g.}, reachability and shortest-path queries), 2-hop labeling methods offer the most versatility. They can be extended to encode distance or path information without requiring major structural changes to the index.}

%% file: sections/path-constraint-reachability-index.tex
\section{Path-Constrained Reachability Indexes}\label{sec:pcr_indexes}
Compared to  plain reachability queries, the main difference in path-constrained reachability queries is that these  have a path constraint that is defined by a regular expression $\alpha$. 
In this survey, we consider the basic regular expressions, where we have edge labels as literal characters and concatenation, alternation, and the Kleene operators (star and plus) as the meta characters (see Section \ref{sec:pcr-queries} for the grammar of $\alpha$).
As discussed in Section \ref{sec:pcr-queries}, a path-constrained reachability query $Q_r(s,t,\alpha)$ checks two conditions: (i) whether $t$ is reachable from $s$; (ii) whether edge labels in the path from $s$ to $t$ can satisfy the constraint specified by $\alpha$.

The existing path-constrained reachability indexes are categorized in Table \ref{table:pcr_indexes} according to $5$ metrics, four of which are those used for plain reachability indexes \revision{(see the beginning of Section \ref{sec:pr_indexes})}. The additional one is the  path constraint column that indicates whether the indexing approach is designed for \textit{alternation-based} or \textit{concatenation-based} reachability queries. In general, existing approaches are specifically designed for a subclass of path-constrained reachability queries, and there is currently no reachability index that supports general path constraints.

\revision{In Table \ref{table:pcr_indexes}, the construction time, index size, and query time columns reflect the complexity of each approach. Due to space constraints, a detailed discussion is provided in the online appendix~\cite{zhang2023indexingtechniquesgraphreachability}.}

A naive index for path-constrained reachability queries is a generalized transitive closure (GTC), which extends the transitive closure by adding edge labels. 
GTCs are more expensive to compute than TCs.
The main reason is that paths from $s$ to $t$ have to be distinguished in GTCs as they can satisfy different path constraints.
Existing indexes are specifically designed for different types of queries according to the path constraints:  \\
    \textit{\textbf{Alternation-based queries}} \cite{10.1145/1807167.1807183, ZOU201447, 10.1145/3035918.3035955, 10.14778/3380750.3380753,10.1145/3451159}: \revision{The path constraint is alternation of edge labels under  Kleene star, \textit{i.e.}, $\alpha=(l_1 \cup l_2\cup...)^*$, $\forall l_i\in \mathcal{L}$.} 
For example, $Q_r(A,N,(\texttt{worksFor}\cup\texttt{friendOf})^*)=True$  in Fig. \ref{fig:graphs}(b) as the following path only contains labels $friendOf$ and $worksFor$: $(A,\texttt{friendOf},D,\texttt{worksFor},B,\texttt{friendOf},F,\texttt{friendOf},J,\texttt{worksFor},N)$.\\
    \textit{\textbf{Concatenation-based queries}} \cite{10.1109/ICDE55515.2023.00013}: \revision{The path constraint is concatenation of edge labels under Kleene star, \textit{i.e.}, $\alpha=(l_1 \cdot l_2\cdot...)^*$, $\forall l_i\in \mathcal{L}$.}
    For instance, $Q_r(L,B,(\texttt{worksFor}\cdot\texttt{friendOf})^*)=True$ in Fig. \ref{fig:graphs}(b) as the label sequence in the path $(L,\texttt{worksFor},D,\texttt{friendOf},$ $H,\texttt{worksFor},G,\texttt{friendOf},B)$ are repeats of $(\texttt{worksFor},\texttt{friendOf})$.    
\\
The different path constraints will lead to different edge label information recorded in the indexes. In the remainder of this section, we discuss each of these index classes.

\subsection{Indexes for alternation-based queries}\label{sec:alternation-based_indexes}
\textit{Alternation-based queries} (also known as  \textit{label-constrained reachability} (LCR) queries) are widely supported by practical query languages, including SPARQL \cite{sparql}, SQL/PGQ \cite{10.1145/3514221.3526057}, GQL \cite{10.1145/3514221.3526057}, and openCypher \cite{openCypher}. 
For instance, $Q_r(A,N,(\texttt{worksFor}\cup\texttt{friendOf})^*)$ can be expressed in SPARQL by the following statement:
\texttt{ASK WHERE {:A (:friendOf | :worksFor)* :N}}.

\textit{\textbf{Foundations for indexing alternation-based path constraints}}. 
To index alternation-based reachability queries, path-label sets have to be recorded in the index. They are basically the set of edge labels in a path. 
The first alternation-based reachability index designed by Jin et al. \cite{10.1145/1807167.1807183} lays two foundations on storing and computing path-label sets: (i) \textit{sufficient path-label sets} (SPLSs) for removing redundancy, and (ii) \textit{transitivity of SPLSs} for efficient computation. 

    \textbf{\textit{SPLSs}}. Only recording the minimal sets of all the path-label sets from a source to a target is sufficient. For example, in Fig. \ref{fig:graphs}(b), there exist two path-label sets from $L$ to $M$: $\{\texttt{worksFor}, \texttt{follows}\}$ and $\{\texttt{worksFor}\}$, and recording $\{\texttt{worksFor}\}$ in the index is sufficient as   given any $\mathcal{L}'$ defined by $\alpha$,  $\{\texttt{worksFor}\}\subset \{\texttt{worksFor}, \texttt{friendOf}\} \subseteq \mathcal{L}'$, \textit{i.e.}, $\{\texttt{worksFor}\}$ can always satisfy the constraint that $\{\texttt{worksFor}, \texttt{friendOf}\}$ satisfies. Thus, $\{\texttt{worksFor}\}$ is belongs to the SPLS from $L$ to $M$, referred to as $S_{L,M}$. 
    Note that $S_{u,v}$ is a set of path-label sets as there can be multiple minimal sets of path-labels sets from $u$ to $v$.
    
    \textit{\textbf{The transitivity of SPLSs}}. Given $S_{u,v}$ from $u$ to $v$ and  $S_{v,w}$ from $v$ to $w$, the SPLS $S_{u,w}$ of paths via $v$ can be derived by first computing the cross product of $S_{u,v}$ and $S_{v,w}$, followed by computing the minimal sets of $S_{u,v}\times S_{v,w}$.
    For example, in the graph in Fig. \ref{fig:graphs}(b), with $S_{A,L}=\{\{\texttt{follows}\}\}$ and $S_{L,M}=\{\{\texttt{worksFor}\}\}$, $S_{A,M}=\{\{\texttt{follows},\texttt{worksFor}\}\}$.  This property is referred to as the transitivity of SPLSs. It can be leveraged to efficiently compute $S_{s,t}$ without traversing the input graph \cite{10.1145/1807167.1807183}. 

With the SPLSs, the GTC can be designed with the following schema \texttt{(Source, Target, SPLSs)}, where for each pair of reachable vertices \textit{(Source, Target)}, the corresponding $SPLSs$ are recorded.
Due to the computation of SPLSs, GTCs are more expensive to compute than TCs, which is already impractical to compute. 
The primary goal of designing alternation-based reachability indexes is to efficiently compute and effectively compress GTC.
The existing indexes presented in Table \ref{table:pcr_indexes} can be categorized into three classes according to the underlying index framework: the tree cover, GTC, and the 2-hop index. 
Each class  uses a plain reachability index as the underlying class and then adds additional information to handle the evaluation of path constraints.  
There are problems in designing the indexes based on each specific index class. 
For the tree cover indexes,  how to combine interval labeling with SPLSs and how to deal with the reachability caused by non-tree edges need to be resolved. 
When using a GTC as the framework, efficient computation of the GTC is the major challenge. 
For the 2-hop indexes, the problems are the combination with SPLSs and index updates for dynamic graphs. 
These questions will be discussed in the remainder of this section.

\begin{table}
    \centering
    \caption{\revision{Overview of path-constrained reachability indexes.
Notations: NT: partial transitive closure \cite{10.1145/1807167.1807183};
$|NET|$: \#non-empty entries in NT;
$|NEP|$: \#non-empty pairs in NT;
$T_i$: extended spanning trees;
$b_i$: \#back edges attached to $T_i$;
$\chi_i$:\#forward edges in $T_i$;
$|m_i|$: \#edges in   $T_i$ plus cross edges attached to $T_i$; 
$h_i$: max\#forward edges in a path in $T_i$;
$|LV|$:\#landmark vertices;
$|NL|$:\#non-landmark index entries;
$d$: diameter of the graph;
$MOD$: maximum outgoing degree;
$|\mathcal{L}_0|$:  \#edge labels in a query;
$M$: maximum length of minimum repeats;
    }} \label{table:pcr_indexes}
    \resizebox{\linewidth}{!}{
    \begin{tabular}{|p{2cm}|c|p{2cm}|c|c|c|p{3cm}|p{3cm}|p{3cm}|} \hline
        \textbf{Indexing Technique} & \textbf{Index Class}   & \textbf{Path Constraint} & \textbf{Index Type} & \textbf{Input} & \textbf{Dynamic} & \textbf{Construction Time}  & \textbf{Index size}  & \textbf{Query Time}\\\hline \hline
        Jin et al. \cite{10.1145/1807167.1807183}       & Tree cover  & Alternation   & Complete  & General & No & $O(|V|^2|E|\binom{|\mathcal{L}|}{|\mathcal{L}|/2})$ &$O(|V|^2\binom{|\mathcal{L}|}{|\mathcal{L}|/2})$& $O(\log |NET| + \sum_{i=1}^{|NEP|}|NT(s_i,t_i)|)$\\ \hline
        Chen et al. \cite{10.1145/3451159}              & Tree cover & Alternation   & Complete  & General & No & $O(\sum_{i}(|m_i|+b_i+\chi_i|\mathcal{L}|+\chi_ih_i))$ &$O(\sum_{i=1}^{d}(T_i + b_i +\chi_i|\mathcal{L}| + \chi_ih_i))$& $O(\sum_{i=1}^{k}b_i(h^2_i + h_i|\mathcal{L}|)$\\ \hline
        Zou et al. \cite{10.1145/2063576.2063807,ZOU201447} &     GTC                & Alternation   & Complete  & General   & \begin{tabular}{@{}c@{}}Yes \\ (I\&D)\end{tabular} &$O(max(|V|^3,|V|^2D^d))$&$O(|V|^2\binom{|\mathcal{L}|}{|\mathcal{L}|/2})$& $O(\binom{|\mathcal{L}|}{|\mathcal{L}|/2})$\\ \hline
        \revision{Landmark index \cite{10.1145/3035918.3035955}}   
            & GTC & Alternation   &\begin{tabular}{@{}c@{}}Partial \\ (no FP)\end{tabular}   & General & No & $O((|V|\log|V|+|E|+2^{|\mathcal{L}|}|LV|+|NL|(|V|-k))|V|2^{|\mathcal{L}|})$ & $O(|V|(|LV|2^{|\mathcal{L}|}+|NL|))$& $O(|E|+|LV|(2^{|\mathcal{L}|}+|V|))$\\ \hline        
        \revision{P2H+  \cite{10.14778/3380750.3380753}} 
            & 2-Hop & Alternation   & Complete  & General & No & $O(|E||V|^22^{2|\mathcal{L}|})$&$O(|V|^22^{|\mathcal{L}|})$&$O((|L_{out}(s)| + |L_{in}(t)|)|\mathcal{L}_0|)$\\ \hline 
        \revision{DLCR \cite{10.14778/3529337.3529348}} 
            & 2-Hop & Alternation   & Complete  & General & \begin{tabular}{@{}c@{}}Yes \\ (I\&D)\end{tabular} &$O(|E||V|^22^{2|\mathcal{L}|})$&$O(|V|^22^{|\mathcal{L}|})$&$O((|L_{out}(s)| + |L_{in}(t)|)|\mathcal{L}_0|)$ \\ \hline \hline
        \revision{RLC index \cite{10.1109/ICDE55515.2023.00013}} 
            & 2-Hop & Concatenation & Complete  & General & No &$O(|V|^{M+1}|\mathcal{L}|+|\mathcal{L}|^M|V||E|M)$&$O(|V|^2|\mathcal{L}|^M)$&$O(|L_{out}(s)| + |L_{in}(t)|)$\\ \hline 
    \end{tabular}}
\end{table}

\subsubsection{\textbf{Tree-based indexes: Jin et al.} \cite{10.1145/1807167.1807183}} \label{sec:jin-et-al}
\begin{figure}
    \centering
    \resizebox{0.9\textwidth}{!}{
    \includegraphics{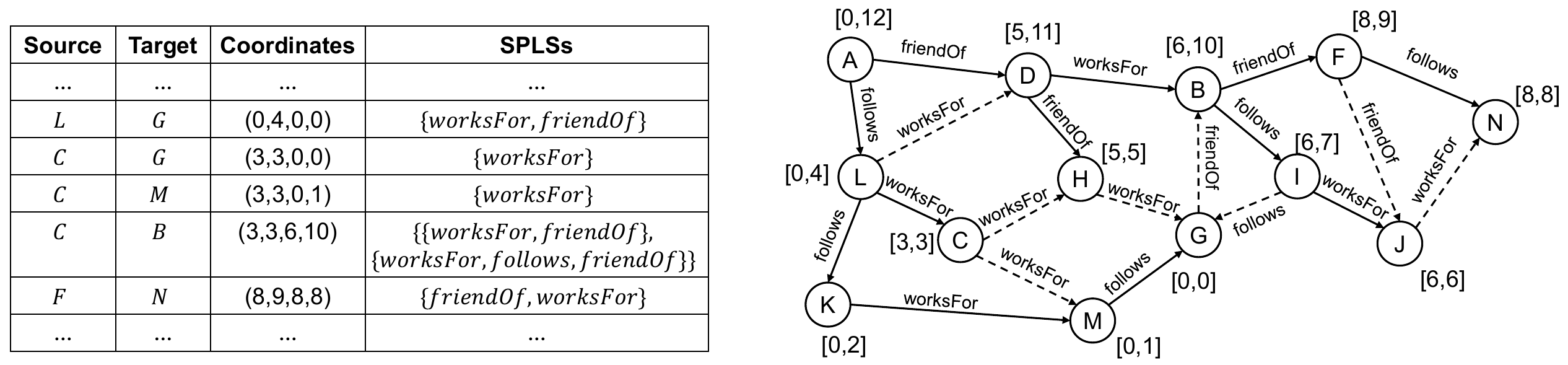}}
    \caption{\revision{Running example of the index proposed by Jin et al. \cite{10.1145/1807167.1807183}.}}
    \label{fig:example-jin-et-al}
\end{figure}

The first index for alternation-based queries compresses GTC using spanning trees. 
The general idea is to distinguish two different classes of paths and compute a partial GTC  for only one  of them, potentially reducing the index size. 
The path categorization is based on spanning trees. 
After computing a spanning tree of the input graph, each edge is either a tree edge or a non-tree edge. 
\revision{
A path is classified as belonging to the first class if either its first or last edge is a tree edge; otherwise, it is categorized as belonging to the second class.
\revision{Consider the example in Fig. \ref{fig:example-jin-et-al}, where tree edges and non-tree edges are highlighted in solid and dashed edges, respectively.}
 From $L$ to $G$, there are four possible paths, and only the path $(L,\texttt{worksFor},D,\texttt{friendOf},H,\texttt{worksFor},G)$ belongs to the second case.
}
Thus, the partial GTC only needs to record the SPLS of the path, as shown in Fig  \ref{fig:example-jin-et-al}. 
\revision{
The partial GTC shown in Fig. \ref{fig:example-jin-et-al} has an additional column \texttt{Coordinates}, which is needed for query optimization   discussed  below.}

\textit{\textbf{Query processing}}.
Query processing starts by checking whether the reachability is recorded in the partial 
GTC. 
If it is not, then online traversal is performed, where the partial transitive closure is leveraged to answer the reachability query. Online traversal only needs to visit tree edges as the reachability information caused by non-tree edges is already recorded in the partial GTC.
In general, a query $Q_r(s,t,\alpha)$ is processed by looking for a path from $s$ to $t$ by three steps: online traversal on the spanning tree from $s$ to $u$, partial GTC lookups for the path from $u$ to $v$, and online traversal on the spanning tree from $u$ to $t$. 
%

\textit{\textbf{Query optimization}}.
Given a query $Q_r(s, t, \alpha)$, the query processing approach outlined above necessitates the execution of online traversals within the spanning tree. These traversals are aimed at identifying a pair of vertices $(u, v)$, where $u$ and $t$ are reachable from $s$ and $v$, respectively, as per $\alpha$, while also confirming the presence of the entry $(u, v)$ within the partial GTC.
For example, an instance of $(u,v)$ are $(C,M)$ in the query $Q_r(L,G,(\texttt{worksFor},\texttt{follows})^*)$.
\revision{
To avoid online graph traversals, two optimization techniques are proposed \cite{10.1145/1807167.1807183}, based on the interval labeling and the histograms of path-label sets, respectively.
These optimizations techniques are designed to address tow key problems: (1) searching $(u,v)$ in the partial GTC such that  $u$ and $t$ are reachable from $s$ and $v$, respectively, in the spanning tree; (2) ensuring that the path-label sets from $s$ to $u$ and the ones from $v$ to $t$  satisfy the given constraint. 
As a result, queries can be processed with only index lookups without any online traversal. We discuss these optimizations in detail below.
}

\textit{\textbf{Query optimization: the interval labeling}}.
This addresses the issue of checking whether there exists $(u,v)$ in the partial GTC such that $u$ and $t$ are reachable from $s$ and $v$ in the spanning tree.
The idea is to achieve this by a multi-dimensional search based on  interval labeling.
Specifically, after computing the interval labeling on the spanning tree, each vertex obtains an interval  (see Section \ref{sec:tree-cover}).
Then, given $s$ and $t$, the search looks for $(u,v)$ such that the corresponding intervals can satisfy the following constraint: $[u_a,u_b]$ is subsumed by $[s_a,s_b]$, and $[t_a,t_b]$ is subsumed by $[v_a,v_b]$, which represents that $u$ and $t$ are reachable from $s$ and $v$, respectively, on the spanning tree.
Therefore, $(u_a,u_b,v_a,v_b)$ should satisfy the following $4$ ranges:   \revision{   
$s_a\leq u_a\leq s_b$; 
$s_a\leq  u_b\leq s_b$;
$v_a\leq t_a$; 
$t_b\leq  v_b$.}
With these four ranges, the search can be reduced to a four-dimensional search problem.
Fig.~\ref{fig:example-jin-et-al} presents a running example of the partial GTC, where the \texttt{Coordinates} column contains four-dimensional coordinate vectors. To accelerate search operations, a k-d tree \cite{10.1145/361002.361007} can be constructed over this column.

\textit{\textbf{Query optimization: histograms of path-label sets}}.
Interval labeling can deal with the plain reachability information from $s$ to $u$ and from $v$ to $t$.
However,  the path-label sets from $s$ to $u$ and from $v$ to $t$ on the spanning tree need to be checked for the evaluation of the path constraint.
To avoid  online traversal for computing such path-label sets, a histogram-based optimization can be adopted.
The idea is to record for each vertex $x$ in the tree the occurrence of each edge label in the path from the root $r$ of the tree to $x$, \textit{e.g.}, if the sequence of the edge labels in the path from $r$ to $x$ is $(\texttt{friendOf}, \texttt{worksFor}, \texttt{friendOf})$, the histogram of $x$, referred to as $H(x)$, is $\{\texttt{friendOf}:2, \texttt{worksFor}:1\}$. 
With the histograms, the path-label set of the path from $s$ to $u$ (or from $v$ to $t$) in the spanning tree can be computed by $H(u)-H(s)$ (or $H(t)-H(v)$).


The index construction process begins by computing a spanning tree, followed by the iterative computation of the partial GTC starting from each vertex.
The impact of the spanning trees on the index size of partial GTC is extensively studied \cite{10.1145/1807167.1807183}. 

\subsubsection{\textbf{Tree-based indexes: Chen et al.} \cite{10.1145/3451159}}
Another tree-based index for alternation-based queries uses graph decomposition to deal with the problem of reachability caused by non-tree edges.
An input graph $G$ is decomposed into $\mathcal{T}$ and $G_c$. 
$\mathcal{T}$ is an extended spanning tree, and  interval labeling can be used to process reachability within $\mathcal{T}$. 
$G_c$ is a graph summary, including reachability that cannot be covered by  interval labeling. 
The decomposition is done  recursively generating $(\mathcal{T}^i,G^i_c)$ where $i$ denotes the iteration, and continues until the decomposition result of a graph summary only contains an extended tree. 
The objective is to maximally leverage  interval labeling to compress the transitive closure. 
Specifically, for each decomposition result $(\mathcal{T}^i,G^i_c)$, the interval labeling with histograms (see Section \ref{sec:jin-et-al}) is computed on $\mathcal{T}^i$, which is used as the index for $\mathcal{T}^i$. 
The series of the indexes on $(\mathcal{T}^1, ..., \mathcal{T}^i,...,\mathcal{T}^n)$ collectively constitutes the total index.
Then, a query $Q_r(s,t,\alpha)$ is decomposed into subqueries that are processed using the series of the indexes.

\begin{figure}
    \centering
    \resizebox{\textwidth}{!}{
    \includegraphics{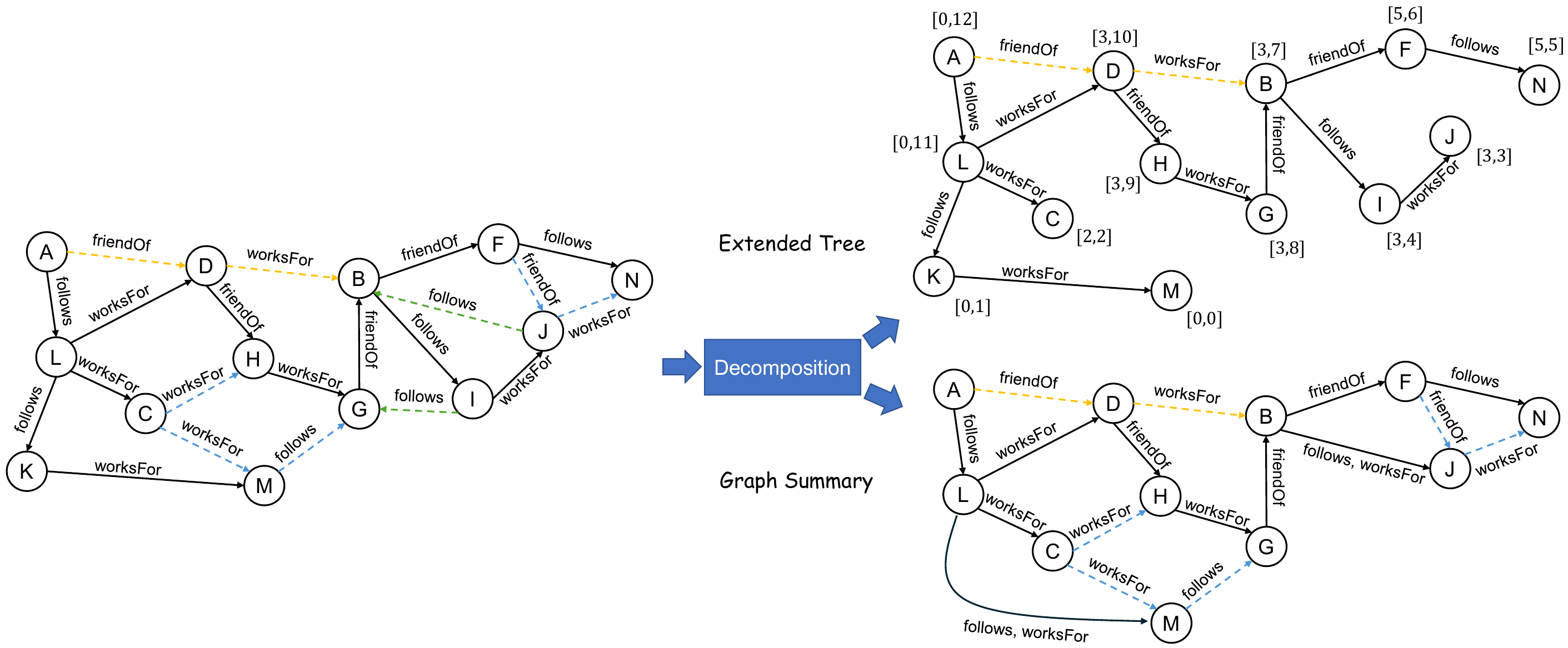}}
    \caption{\revision{Running example of the index proposed by Chen et al. \cite{10.1145/3451159}.}}
    \label{fig:example-chen-et-al}
\end{figure}

\textit{\textbf{Edge classification and extended trees}}.
\revision{
The definition of extended trees is based on classifying each edge as either a \textit{tree edge} (part of the spanning tree) or a \textit{non-tree edge}. In Fig.~\ref{fig:example-chen-et-al}, tree edges are shown as solid black lines. Each non-tree edge $(u, v)$, illustrated with dashed lines, falls into one of the following categories: a \textit{forward edge}, if there is a path from $u$ to $v$ in the tree (\textit{e.g.}, $(A, D)$ in dashed orange); a \textit{cross edge}, if $u$ and $v$ are not on the same path in the tree (\textit{e.g.}, $(C, H)$ in dashed blue); and a \textit{back edge}, if there is a path from $v$ to $u$ in the tree (\textit{e.g.}, $(I, G)$ in dashed green).}
Under the edge classification, an \textit{extended tree} is a spanning tree with the corresponding forward edges.

\textit{\textbf{Querying processing in extended trees}}.
\revision{
Interval labeling using histograms is sufficient to capture all reachability information in an extended tree. For each forward edge $(u, v)$, a path from $u$ to $v$ already exists in the tree. The forward edge only introduces additional path-label sets from $u$ to $v$. These additional sets can be fully recorded using the histograms $H(u)$ and $H(v)$.}
Since there can be multiple paths from the root to each vertex in the tree, $H(u)$ (or $H(v)$) in this approach are sets of path-label sets augmented with occurrences of edge labels.

\textit{\textbf{Query decomposition and processing}}.
A query is decomposed into several subqueries, and each of them is evaluated using either an extended tree or a graph summary.
As the graph summary can be further decomposed, the subquery on the graph summary will be decomposed correspondingly.
\revision{
Consider Fig. \ref{fig:example-chen-et-al} and the query $Q_r(L,F,\alpha)$ where $ \alpha  = \{\texttt{worksFor}\cup \texttt{friendOf}\}^*$ is decomposed into  $Q_1:Q_r(L,C,\alpha)$,  $Q_2:Q_r(C,G,\alpha)$, and  $Q_3:Q_r(G,F,\alpha)$.} $Q_1$ and $Q_3$ are processed on the extended tree while $Q_2$ on the graph summary.
The decomposition is performed by computing two specific sets of vertices for each vertex in the extended tree, denoted as \textit{dominant} and \textit{transferring} vertices (see  \cite{10.1145/3451159} for  formal definitions).
$C$ is a dominant vertex for vertex $L$, which means that outgoing reachability with cross edges can be found via $C$. 
$G$ is a transferring vertex for vertex $F$, which means incoming reachability with cross edges can be found via $G$.
The dominant and transferring vertices are computed for vertices in each extended tree.

\subsubsection{\textbf{GTC-based indexes: Zou et al.} \cite{10.1145/2063576.2063807,ZOU201447}} 
The first index in this class is designed based on a transformation from a general graph into an augmented DAG (ADAG) with path-label set information. 
The transformation also identifies SCCs. 
However, this transformation is more complicated than the one for plain graphs as paths with different sets of labels need to be distinguished. 
The benefit of performing this transformation is that the GTC on the augmented DAG can be computed incrementally in a bottom-up manner. 
In general, the GTC of the general graph can be computed by combining three kinds of reachability information: reachability inside each SCC, reachability in the ADAG, and reachability between vertices in the ADAG and vertices in the SCCs.

\begin{figure}
    \centering
    \resizebox{0.85\textwidth}{!}{
    \includegraphics{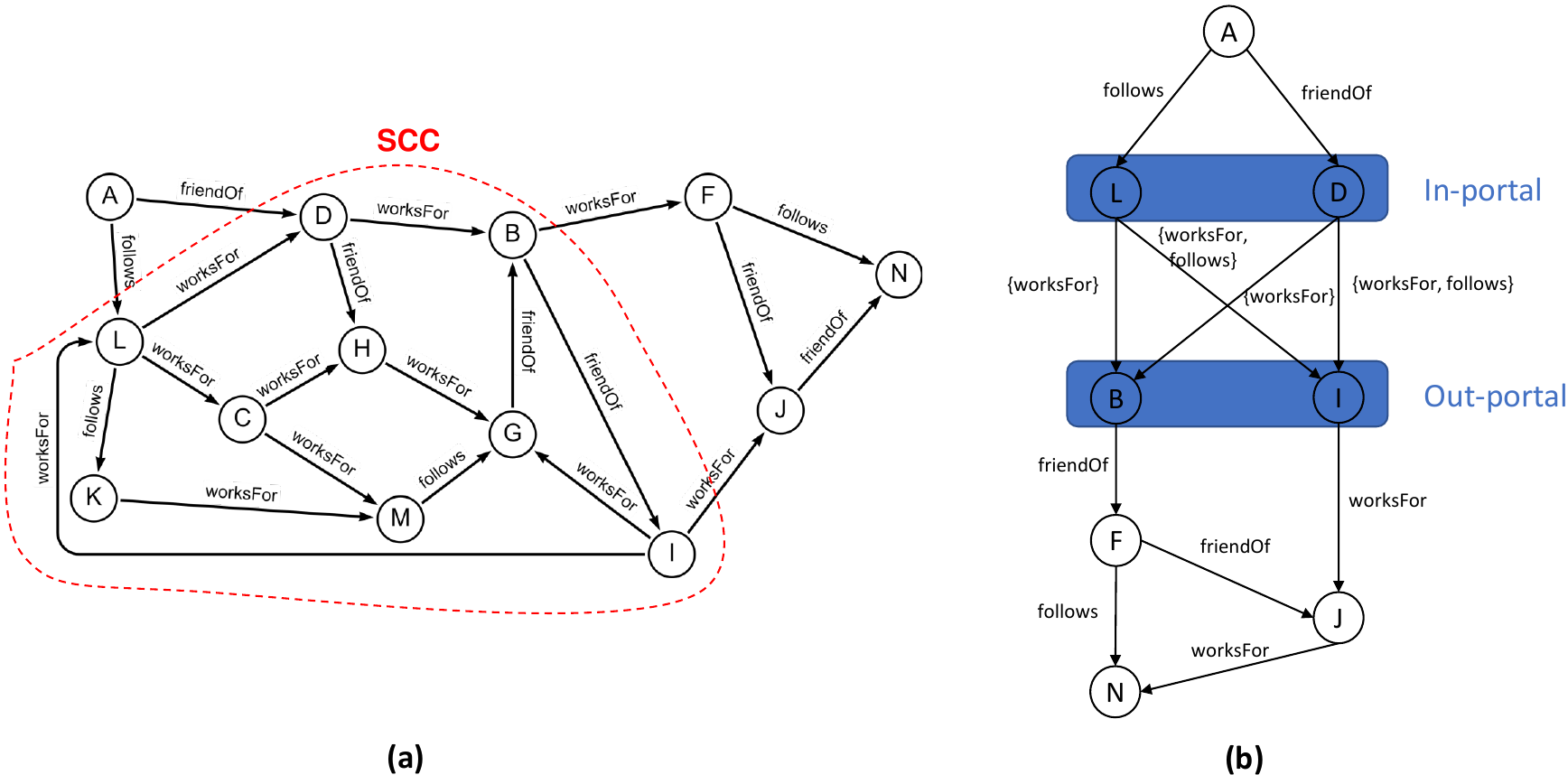}}
    \caption{\revision{A running example of the index proposed by Zou et al. \cite{10.1145/2063576.2063807,ZOU201447}. (a) highlights the strongly connected component on the graph. (b) presents the augmented DAG.}}
    \label{fig:example-zou-et-al}
\end{figure}

\textit{\textbf{Transformation into the ADAG}}.
The transformation first identifies SCCs in the input graph and then computes two kinds of vertices in each SCC: \textit{in-portal} vertices and \textit{out-portal} vertices. 
In-portal vertices of an SCC have incoming edges from vertices out of the SCC while out-portal vertices of an SCC have outgoing edges to vertices out of the SCC.
Then, each SCC is replaced with a bipartite graph with in-portal and out-portal vertices, where edges are labeled with sufficient path-label sets from in-portal to out-portal vertices. 
Consider the example in Fig. \ref{fig:example-zou-et-al}.
The only SCC in the example is highlighted, where in-portal vertices are $\{L,D\}$ and out-portal vertices are $\{B,I\}$.
\revision{
The ADAG is shown in Fig. \ref{fig:example-zou-et-al}(b) consisting of a bipartite subgraph with $\{L,D\}$ and $\{B,I\}$. 
The edge $(L,I)$ in the bipartite subgraph is labeled  $\{\texttt{worksFor,follows}\}$ because it is the only sufficient path-label set from $L$ to $I$ in the SCC.}

\textit{\textbf{Incremental computation of reachability in the ADAG}}.
Teachability in the ADAG can be computed in a bottom-up manner according to a topological ordering.
Specifically, vertices are examined in the reverse topological order and the single-source GTCs are shared in a bottom-up manner. Note that a single-source GTC is a GTC with a fixed source vertex, \textit{i.e.}, a single-source GTC contains all the vertices that are reachable from a source vertex and the corresponding SPLSs. 

\textit{\textbf{Computing the reachability within each SCC: a Dijkstra-like algorithm}}.
An interesting algorithm for computing reachability within each SCC is proposed in this approach. The key idea is to use the number of distinct edge labels along a path as a measure of distance between vertices.
Consider the SCC in Fig. \ref{fig:example-zou-et-al}(a),  two paths exist from $L$ to $H$: $P1=\{L,\texttt{worksFor},D,\texttt{friendOf},H\}$ and $P2=\{L,\texttt{worksFor}, C,\texttt{worksFor}, H\}$.
In this case, $P2$ is considered as the shorter one as it only has one distinct label. 
By doing this,  Dijkstra's algorithm can be extended to compute the single-source GTC for vertices in SCCs.
In general, `shorter' paths represent paths with sufficient path label sets, which only need to be traversed and recorded in the index.

\textit{\textbf{The reachability from vertices in a SCC to vertices in the ADAG}}.
To compute reachability information, the single-source GTC of in-portal or out-portal vertices are leveraged. Consider the example in Fig. \ref{fig:example-zou-et-al}.
\revision{
To compute the single-source GTC of $G$ over the entire graph,  the sufficient path-label sets from $G$ to the out-portal vertices $B$ and $I$ are first computed using the Dijkstra-like algorithm. These sets are then combined with the single-source GTCs of $B$ and $I$, which are computed incrementally in the ADAG.}

\revision{
The approach enhances scalability by partitioning the graph, building local GTCs, and using a skeleton graph for inter-partition query processing. It also supports dynamic updates, including both edge insertions and deletions.}


\subsubsection{\textbf{GTC-based indexes: \revision{Landmark index}} \cite{10.1145/3035918.3035955}}
\begin{figure}
    \centering
    \resizebox{0.75\textwidth}{!}{
    \includegraphics{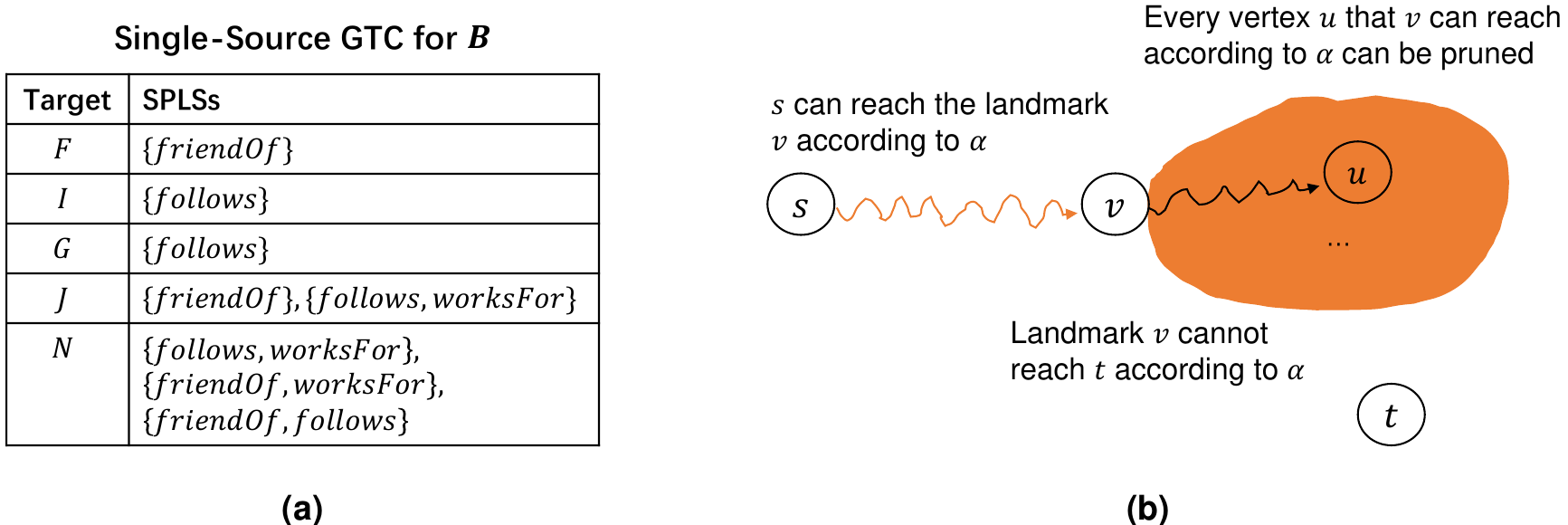}}
    \caption{\revision{Illustration of the landmark index. (a) shows the index constructed for landmark B using the edge-labeled graph from Fig. \ref{fig:graphs}(b). (b) demonstrates the pruning technique used during query processing with the landmark index.}}
    \label{fig:example-landmark-index}
\end{figure}

Landmark index is a partial index, which computes single-source GTCs for a subset of vertices in the graph. 
The subset of vertices are selected as the top-k vertices in degree ranking, denoted as \textit{landmarks}.
Query $Q_r(s,t,\alpha)$ is processed by performing BFS from $s$, which can be accelerated if a landmark vertex is visited by the BFS because the single-source GTC of the landmark records all reachability information from the landmark.
Fig. \ref{fig:example-landmark-index}(a) shows the single-source GTC for  the graph of Fig. \ref{fig:graphs}(b), where $B$ is the landmark vertex.

\textit{\textbf{Indexing non-landmark vertices}}.
For each non-landmark vertex, the approach precomputes a fixed number of SPLSs of paths from the non-landmark vertex to each landmark. 
Consider the example in Fig. \ref{fig:example-landmark-index}(a), $A$ is a non-landmark vertex, and the approach precomputes $2$ SPLSs from $A$ to $B$, \textit{i.e.}, $\{\texttt{follows,worksFor}\}$ and $\{\texttt{friendOf, worksFor}\}$.
These SPLSs can be leveraged to speed up graph traversal.
In case the index entries of a non-landmark vertex cannot satisfy the path constraint in a query, a BFS traversal is still necessary.

\textit{\textbf{Query optimization: pruning techniques for false queries}}.
The major issue in  query processing is that the index is not fully leveraged if the query result is actually \textit{False}. 
To deal with this issue, a pruning technique is proposed in the approach. 
\revision{Consider a toy example in Fig. \ref{fig:example-landmark-index}(b) and the  query $Q_r(s,t,\alpha)$.
Assume that $s$ can reach the landmark $v$ according to the path constraint specified by $\alpha$. 
Then, the single-source GTC of $v$ can be used to check whether $t$ is reachable from $v$ according to the path constraint.
In case this is $False$, every vertex $u$ that is reachable from $v$ according to the path constraint can be pruned in the subsequent search of the BFS.}

\subsubsection{\textbf{2-Hop-based indexes: \revision{P2H+}} \cite{10.14778/3380750.3380753}}
P2H+ is a 2-hop index for alternation-based queries,  designed based on the transitivity of plain reachability and SPLSs.
Given the reachability from $s$ to $u$ and from $u$ to $t$, it follows that $t$ is reachable from $s$.
In addition, if both paths from $s$ to $u$ and  from $u$ to $t$ satisfy the path constraint specified by $\alpha$, the path from $s$ to $t$ can also satisfy the constraint, due to the transitivity of SPLSs (see  Section \ref{sec:alternation-based_indexes}).
Therefore, SPLSs can be added to the index entries of the 2-hop index for processing alternation-based queries.

P2H+ indexing extends the PLL indexing algorithm \cite{10.1145/2505515.2505724} to deal with alternation-based reachability. 
The P2H+ indexing algorithm intervenes in the second step of the PLL indexing algorithm by computing the SPLSs while performing backward and forward BFS.

P2H+ applies pruning rules to accelerate index building as well as remove redundancy in the index.
\revision{The first pruning rule states that if a vertex 
$v$ encountered during the current BFS has already been processed (\textit{i.e.}, a BFS from $v$ was performed before the current BFS), $v$ can be safely pruned from further processing in the current BFS. This is because its reachability information has already been recorded.}    
The second pruning rule is that an index entry can be skipped if it can be derived from the current snapshot of index. 
The third pruning rule is to remove redundant index entries while computing SPLSs and inserting index entries. 
To avoid the overhead of insertion followed by deletion, an optimization technique based on edge visiting priority is developed.
The idea is to prioritize the visiting of the edges with labels that have been visited in the BFS. 
Consequently, the index entries with minimal label sets will be inserted first, which can avoid deleting redundant index entries.

\subsubsection{\textbf{2-Hop-based indexes: \revision{DLCR}} \cite{10.14778/3529337.3529348}}
\begin{figure}[!]
    \centering
    \resizebox{0.35\textwidth}{!}{
    \includegraphics{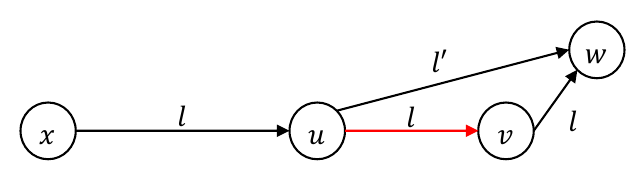}}
    \caption{A toy example for illustrating index updates in DLCR \cite{10.14778/3529337.3529348}.}
    \label{fig:illustration-dlcr}
\end{figure}

DLCR is the latest and the most advanced indexing technique for alternation-based queries. 
It is an extension of P2H+ to dynamic graphs, in that DLCR is also designed based on the 2-hop index, but allows index updates. 
DLCR ensures that the index is free of any redundancy following index updates. 
Consider the toy example in Fig. \ref{fig:illustration-dlcr}.
After inserting edge $(u,v)$ with label $l$ in the graph, $w$ is reachable from $u$ with SPLS $\{l\}$, which needs to be recorded in the index. 
In addition, $x$ can reach $w$ with $\{l\}$ after the insertion, such that the path from $x$ to $w$ with the label set $\{l,l'\}$ becomes redundant, which needs to be deleted from the index. 
For the case of deleting edge $(u,v)$ with label $l$ in the graph, the reachability information from $x$ to $w$ with $\{l\}$ and from $u$ to $w$ with $\{l\}$ need to be deleted accordingly. 
In addition, the path from $x$ to $w$ with $\{l,l'\}$, which was redundant because of the reachability from $x$ to $w$ with $\{l\}$ before the deletion, needs to be inserted into the index after the edge deletion. 

DLCR efficiently updates the index by leveraging the current snapshot of the index to reflect the changes. When inserting an edge $(u,v)$ with label ${l}$, DLCR performs forward BFS from vertex $u$ and backward BFS from vertex $v$ to record additional reachability information. It is important to note that these BFS operations do not start from scratch; instead, they leverage the index entries in $L_{in}(u)$ and $L_{out}(v)$. During the BFS operations, if the corresponding index entries change for any visited vertex, that vertex is marked as an \textit{affected vertex}, which is used to identify and remove redundant index entries. The process for edge deletion is similar to that of edge insertion.
It is worth emphasizing that to completely remove redundancies after edge insertion or to add all necessary index entries after edge deletion, the maintenance of \textit{inverted index entries} (see Section \ref{sec:2-hop_dynamic}) is crucial. For more in-depth information, we direct the readers to the paper~\cite{10.14778/3529337.3529348}.

\subsection{Indexes for concatenation-based queries}
In concatenation-based reachability queries, the path constraint defines a label sequence, and the edge labels in the path from the source to the target are repeats of the defined label sequence. 
In Fig. \ref{fig:graphs}(b), $Q_r(L,B,(\texttt{worksFor}\cdot\texttt{friendOf})^*)=True$ because the sequence of edge labels in the path from $L$ to $B$ via $D$, $H$, and $G$ are repeats of $(\texttt{worksFor},\texttt{friendOf})$.
Concatenation-based reachability are supported by SPARQL \cite{sparql}, SQL/PGQ \cite{10.1145/3514221.3526057}, and GQL \cite{10.1145/3514221.3526057}.

\textit{\textbf{Minimum repeats}}.
To index concatenation-based reachability, path-label sequences from a source to a target are necessary to record \cite{10.1109/ICDE55515.2023.00013}. Recording only the \textit{minimum repeats}, instead of the raw label sequences, is sufficient for concatenation-based reachability query processing \cite{10.1109/ICDE55515.2023.00013}, 
For instance, given the path-label sequence $(\texttt{worksFor},\texttt{friendOf},\texttt{worksFor},\texttt{friendOf})$, the minimum repeat  is $(\texttt{worksFor},\texttt{friendOf})$.

\textit{\textbf{Kernel-based search}}.
With  minimum repeats, the generalized transitive closure that extends the transitive closure with the minimum repeats can be used to process concatenation-based queries.
The problem is how to compute the minimum repeats from a source to a target as there can be an infinite number of minimum repeats due to the existence of cycles in the path. 
Consider the graph in Fig. \ref{fig:graphs}(b) and the problem of computing the minimum repeats from $D$ to $J$. 
Due to the existence of the cycle including vertices $B$, $I$, and $G$, the potential number of minimum repeats from $D$ to $J$ is infinite (assuming arbitrary path semantics -- see Section \ref{sec:background}).
The solution to the problem is based on a practical observation, that the number of concatenated labels under the Kleene star is usually bounded. 
Based on the observation, the first reachability index for concatenation-based queries \cite{10.1109/ICDE55515.2023.00013} designs a \textit{kernel-based search} to compute minimum repeats. 
The idea is to first compute kernels that are essentially label sequences on the fly and then use the kernels to guide the subsequent search. 
%
\revision{
Consider the above example with minimum repeats of length up to $2$.
The kernel-based search starts the BFS from $D$. When the BFS  visits $I$ with the label sequence $(\texttt{worksFor}, \texttt{friendOf})$, the label sequence is treated as the kernel of this search. 
Subsequent search will be guided by the kernel and can visit $J$, which means $(\texttt{worksFor}, \texttt{friendOf})$ is a minimum repeat from $D$ to $J$.}

\textit{\textbf{Transitive minimum repeats}}.
\revision{
To utilize an efficient index framework, such as 2-hop labeling, minimum repeats are required to be transitive, since advanced indexes rely on local path information to infer reachability along global paths. However, this transitivity property does not generally hold.
For example, in Fig.~\ref{fig:graphs}(b), it is not possible to derive the minimum repeat from $L$ to $I$ along paths passing through $G$, even though the minimum repeat from $L$ to $G$ is \{\texttt{worksFor}\} and the minimum repeat from $G$ to $I$ is \{\texttt{friendOf}\}. To address this issue, the key idea is to use only equivalent minimum repeats when attempting derivation~\cite{10.1109/ICDE55515.2023.00013}. 
Given that the minimum repeat from $s$ to $v$ is $MR1$ and from $v$ to $t$ is $MR2$, we can derive the minimum repeat from $s$ to $t$ only when $MR1 = MR2$.
}

\subsubsection{\textbf{2-Hop-based indexes: \revision{RLC index}} \cite{10.1109/ICDE55515.2023.00013}}
The RLC index is designed based on the 2-hop index for evaluating  plain reachability, and the kernel-based search and transitive minimum repeats for computing minimum repeats so as to evaluate concatenation-based path constraint.
For instance, in the graph in Fig. \ref{fig:graphs}(a), $B$ is reachable from $L$ with minimum repeat $(\texttt{worksFor},\texttt{friendOf})$, which can be recorded as $(B, (\texttt{worksFor},\texttt{friendOf}))\in L_{out}(L)$. Similarly, $B$ can reach $J$ with the same minimum repeat, which can be recorded as $(B, (\texttt{worksFor},\texttt{friendOf}))\in L_{in}(J)$.
The RLC indexing algorithm extends the PLL indexing algorithm by using kernel-based search to compute minimum repeats. 
Each kernel-based search consists of two phases: kernel search and kernel BFS, which are used to compute kernels and guide the subsequent search using kernels, respectively.
The RLC indexing algorithm is equipped with pruning rules to accelerate the index building process, \textit{e.g.}, skipping index entries and vertices during the search.
In general, the pruning rules are similar in spirit to the pruning rules in the PLL indexing class for plain reachability, but adapted to the computation of minimum repeats, \textit{e.g.}, pruning rules for skipping vertices can only be applied in the phase of kernel BFS in the RLC indexing algorithm.
For processing $Q_r(s,t,\alpha)$ with the RLC index, 
the index entries of $L_(out)(s)$ and $L_{in}(t)$ are examined to see whether 
$(t,MR)\in L_{out}(s)$, $(s,MR)\in L_{in}(t)$, or there exist index entries $(v,MR)$ in both $L_{out}(s)$ and $L_{in}(t)$.

\begin{figure}
    \centering\resizebox{0.75\textwidth}{!}{
    \includegraphics{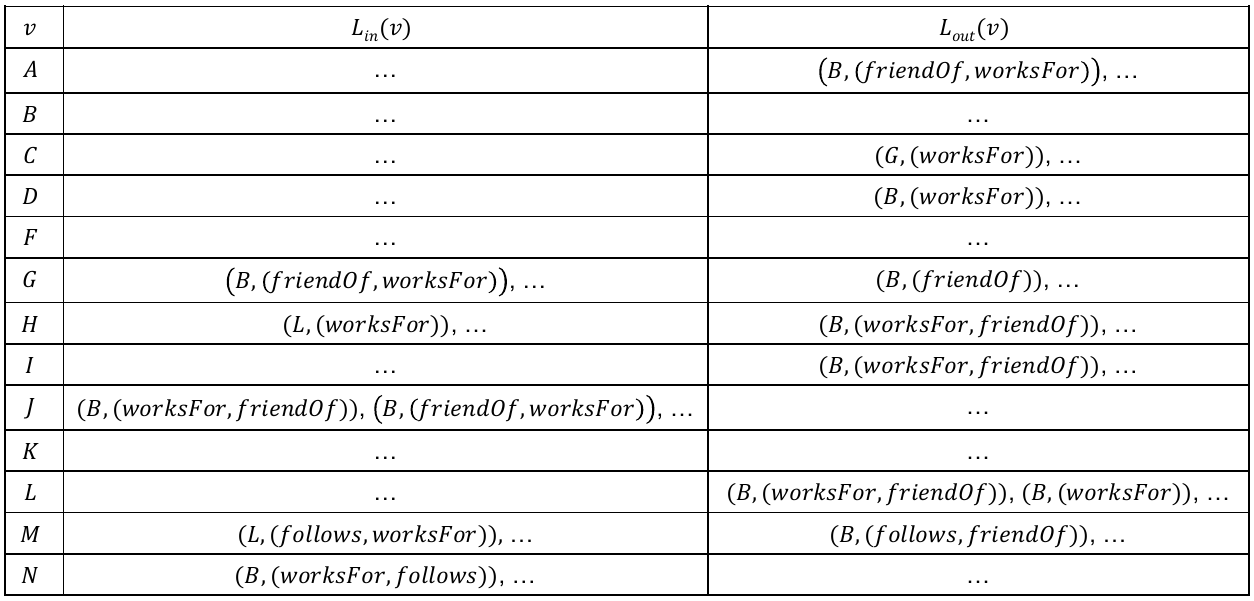}}
    \caption{\revision{A running example of the RLC index built for the graph in Fig. \ref{fig:graphs}(b).}}
    \label{fig:example_rlc_index}
\end{figure}

We present a running example of the RLC index in Fig. \ref{fig:example_rlc_index}, designed to encompass all minimum repeats involving at most $2$ edge labels. This illustrative example exclusively includes index entries relevant to vertices $B$ and $L$.
\revision{
For $Q_r(L,J,(\texttt{worksFor}\cdot\texttt{friendOf})^*)$, as there exists $(B,(\texttt{worksFor},\texttt{friendOf}))$ in both $L_{out}(L)$ and $L_{in}(J)$, such that the query result is $True$.}

\subsection{\revision{Practical trade-offs and guidance}}
\revision{
Following the discussion of practical trade-offs in the context of plain reachability indexes (see Section \ref{sec:practical_discussion_pr}), tree-cover–based approaches are generally suitable for sparse and static graphs, where interval labeling can be highly effective. In contrast, 2-hop approaches tend to perform well across a broader range of scenarios, including relatively dense graphs and dynamic updates. In particular, the DLCR index has demonstrated competitive performance in alternation-based reachability.
2-hop indexes are practically appealing due to their broad support for path constraints. They can be extended to handle both alternation and concatenation, making them a solid foundation for path-constrained reachability.
}

%% file: sections/challenge.tex
\section{Open Challenges}\label{sec:our_vision}
%
\revision{
Despite many proposed methods, significant challenges remain in integrating full-fledged indexes into modern graph database management systems (GDBMSs). These include handling dynamic updates, supporting path constraints, and achieving scalability and parallelism. This section discusses these challenges and future research opportunities.}

\subsection{Large dynamic graphs}
\revision{\textit{\textbf{Design trade-offs in practical reachability indexes}}}.
State-of-the-art plain reachability indexes can be built efficiently even on large graphs. 
For example, given a graph with 3 million vertices and 16 million edges, BFL~\cite{7750623} can be constructed in just 1.3 seconds, with an index size of 132 megabytes.
Query processing using BFL takes only 0.26 microseconds per query on average. 
This level of efficiency is achieved by using partial indexes instead of complete indexes.
%
%
We analyzed existing techniques and observed a clear evolution in the design of plain reachability indexes. 
\textit{Early plain reachability indexes were complete, such as tree cover~\cite{10.1145/67544.66950} and 2-hop labeling~\cite{10.5555/545381.545503}. 
These were followed by partial indexes without false positives, such as GRIPP~\cite{10.1145/1247480.1247573}, and more recently by partial indexes without false negatives, such as BFL~\cite{7750623}.}
\revision{
Although both types of partial indexes may require graph traversal, the design of partial indexes without false negatives is generally preferable to those without false positives.}
This is because the reachability ratio in real-world graphs, which is defined as the number of reachable vertex pairs divided by the total number of vertex pairs, is typically very low~\cite{7750623}. 
As a result, most queries are likely to return \texttt{False}. Thus, partial indexes without false negatives can return the correct result without traversal in most cases.

\textit{\textbf{Challenges for plain reachability indexes}}.
The main challenge for plain reachability indexes is that partial indexes without false negatives are mainly designed for static graphs, and a recent work, \textit{i.e.}, DBL \cite{10.1007/978-3-030-73197-7_52}, can only support insertion-only graphs. 
Designing partial indexes without false negatives for dynamic graphs remains an open challenge.

\textit{\textbf{Challenges for path-constrained reachability indexes}}.
The reachability ratio for path-constrained reachability queries can be even lower because the queries will further apply a path constraint to all retrieved paths from source to target. 
This calls for the design of a partial index without false negatives. 
However, the only partial index for path-constrained reachability in the literature is landmark index \cite{10.1145/3035918.3035955}, which is a partial index without false positives and can only support static graphs. 
Although DLCR \cite{10.14778/3529337.3529348} can support dynamic graphs, it is a complete index and the indexing cost can be high on large graphs. 
For path-constrained reachability queries, designing partial indexes without false negatives for dynamic  graphs remains an open challenge.

\revision{
\textit{\textbf{Streaming graph processing}}.
Streaming graphs \cite{10.1145/3318464.3389733}, characterized by a continuous stream of vertices and edges, are crucial for real-time analytical applications such as e-commerce \cite{10.1145/3318464.3389733}, social network \cite{9835463}, fraud detection \cite{10.14778/3675034.3675040}.
Due to their unbounded nature, computations over streaming graphs are typically performed using a sliding window model, where only the most recent changes within a defined time window are considered. This requires continuously adding and removing edges and vertices, placing considerable pressure on the efficiency of index maintenance for query processing.
Most indexing methods designed for static graphs or graphs with rare updates cannot meet the performance demands of real-time streaming scenarios due to the frequent updates.
To address this challenge, incremental computation has emerged as an effective strategy for sliding window graph analytics. It enables computation reuse across overlapping windows, proving efficiency for various queries, including regular path queries \cite{10.1145/3318464.3389733,10.14778/3641204.3641214,10.1145/3639260} and connectivity queries \cite{10.14778/3675034.3675040,zhang2025lowlatencyslidingwindowconnectivity}.
In particular, the Bidirectional Incremental Computation (BIC) model \cite{10.14778/3675034.3675040} presents a general framework that avoids the physical deletion of edges. By focusing solely on incremental edge insertions, it simplifies index maintenance and enhances computational efficiency. Investigating the use of the BIC model for maintaining reachability indexes and supporting general graph queries in sliding window contexts is a promising research direction.
}

\subsection{Various formalisation in path-constrained reachability queries}
\textit{\textbf{General path constraints}}.
Existing path-constrained reachability indexes can only support a specific type of path constraints: either alternation-based or concatenation-based.
However, path constraints occurring in real-world query logs belong to several classes \cite{10.1145/3308558.3313472}, \textit{e.g.}, $a^*\cdot b^*$, $(a_1 \cup ... \cup a_k)\cdot b^*$, etc., where $a$ and $b$ represent edge labels.
Designing reachability indexes for queries with more general path constraints is an open research direction.

\textit{\textbf{Different path semantics}}.
Evaluation of path-constrained queries can be based on various path semantics.
However, current indexes can only support specific path semantics.
Concatenation-based indexes \cite{10.1109/ICDE55515.2023.00013}  are designed for arbitrary path semantics, where vertices can be traversed an arbitrary number of times.
Alternation-based indexes \cite{10.1145/1807167.1807183, ZOU201447, 10.1145/3035918.3035955, 10.14778/3380750.3380753,10.1145/3451159} can be used to evaluate queries based on both simple (non-repeated edges) and arbitrary path semantics. This is because the foundation for building these indexes, \textit{i.e.}, sufficient path label sets, can naturally remove the repeated visiting of edges. 
In the new standard graph query languages, such as GQL and SQL/PGQ \cite{10.1145/3514221.3526057}, queries can combine \textit{restrictors} and \textit{selectors}, which leads to query processing using different path semantics. 
Restrictors are path predicates so that the number of matches is not infinite. 
For instance, \texttt{ACYCLIC} imposes non-repeated vertices including origin and destination. 
Selectors are the algorithms that compute the partitions of paths. For instance,  \texttt{ALL SHORTEST} will partition the solution space into groups of equal-length paths and eventually select the shortest ones. 
Therefore, choosing any combinations of restrictors and selectors controls the finite sets of matches and also the results.
A plethora of indexes can be designed to cover various combinations of selectors and restrictors, out of which those studied in current standard graph query languages.

\subsection{Reachability indexes on property graphs}\label{sec:index_property_graphs}
Reachability queries on property graphs include predicates on the edge and vertex properties, referred to as \textit{property predicates}. 
Property predicates impose additional constraints on the selection of paths from the source to the target. 
Existing indexes are not sufficient to process reachability queries on property graphs since they only index graph topology.
A theoretical study \cite{10.1145/2274576.2274585} characterizes property predicates that are tractable based on register automata. However, indexing property graphs for reachability has not been studied and remains an important challenge.

\subsection{Parallel computing}
\revision{
Graph processing has significantly advanced with parallel frameworks tailored for both CPU and GPU architectures. CPU-based frameworks such as GRAPE \cite{10.1145/3282488} facilitate parallelization by allowing sequential graph algorithms to be integrated through incremental computation and localized evaluation. Ligra \cite{10.1145/2442516.2442530}  provides a lightweight shared-memory solution optimized for graph traversal. GraphMat \cite{10.14778/2809974.2809983} integrates vertex-centric programming convenience with the efficiency of sparse matrix operations. On the GPU side, inherent advantages like massive parallelism and high memory bandwidth have driven adoption despite persistent challenges, particularly GPU memory constraints and CPU-GPU data transfer bottlenecks. Recent solutions like HyTGraph \cite{10184762} and CGgraph \cite{10.14778/3648160.3648179}  employ hybrid strategies, dynamically optimizing data transfers based on workload demands to maximize performance. For a deeper insight into GPU graph processing techniques, readers are referred to the comprehensive survey \cite{10.1145/3128571}.
}

\revision{
In the context of reachability indexing, parallelism remains relatively underexplored but shows growing potential. Traditional tree-cover methods, such as GRAIL, rely on DFS-based interval labeling, which poses challenges for parallel execution. A recent GPU-friendly approach~\cite{computation8040103} addresses this limitation by replacing DFS with multiple rounds of BFS, enabling concurrent labeling across the graph.
Recent advancements in the parallel construction of 2-hop indexes~\cite{10.1145/3299869.3319877,10.1145/3392717.3392745} adopt the vertex-centric paradigm to compute the label sets $ L_{out}(v)$  and $L_{in}(v)$  efficiently. These developments in parallel graph processing provide a foundation for designing scalable indexing schemes for path-constrained reachability queries, particularly as modern indexes increasingly build upon tree-cover and 2-hop index frameworks. Furthermore, it would be valuable to explore recent system-level advances in parallel graph processing to design scalable reachability indexes and investigate how such indexes can accelerate broader graph analytics workloads. For example, recent work has leveraged parallel reachability techniques to speed up the computation of strongly connected components \cite{10.1145/3589259}.
}

\subsection{\revision{Integrating reachability indexes in modern GDBMS architectures}}
\revision{
GDBMS research has advanced significantly across both centralized and distributed systems. Centralized GDBMSs have introduced techniques such as predefined joins \cite{10.14778/3510397.3510400} to accelerate traversals by treating edges as built-in relational joins, columnar and list-based storage layouts \cite{10.14778/3476249.3476297} for efficient memory access, and worst-case optimal join algorithms \cite{10.14778/3342263.3342643} for complex subgraph queries. Systems like Kùzu \cite{feng2023kuzu} integrate many of these ideas into a graph-native engine. 
Concurrently, distributed graph data systems have evolved to handle scale and concurrency with innovations like G-TRAN \cite{10.14778/3551793.3551813}, which uses RDMA and decentralized coordination to support high-throughput graph transactions, and GDI \cite{10.1145/3581784.3607068}, a portable RDMA-based interface for building scalable graph engines. 
These system-level innovations present new opportunities to design scalable reachability indexes that align with modern GDBMS execution and storage models. Integrating such indexes into these architectures can further enhance their analytical capabilities, providing a foundation for complex graph queries with minimal overhead.}

\revision{
Despite these advances in system design, the integration of reachability indexes into general query processing within modern GDBMSs remains underexplored.
Most existing work considers reachability indexes in isolation, without considering their role in the overall query execution.
A few approaches~\cite{Kuijpers2021PathII,10.1007/978-3-319-58943-5_43,fletcher2016efficient} focus on materializing paths of fixed length and indexing them using B+-trees. 
However, incorporating reachability indexes into modern GDBMSs and exploiting them to accelerate the evaluation of recursive graph queries remain largely unexplored.
An early effort~\cite{10.1145/2484425.2484443} integrates the Ferrari index~\cite{6544893} into RDF-3X~\cite{10.14778/1453856.1453927}, showing promising results for processing SPARQL queries with Kleene star semantics. 
A promising research direction is to investigate how recent indexing techniques for path-constrained reachability can be integrated into GDBMS architectures to support complex recursive graph query processing.}

%% file: sections/appendix.tex
\newpage
\appendix

\section{Property Graphs}
\begin{figure}
    \centering
    \resizebox{0.8\linewidth}{!}{\includegraphics{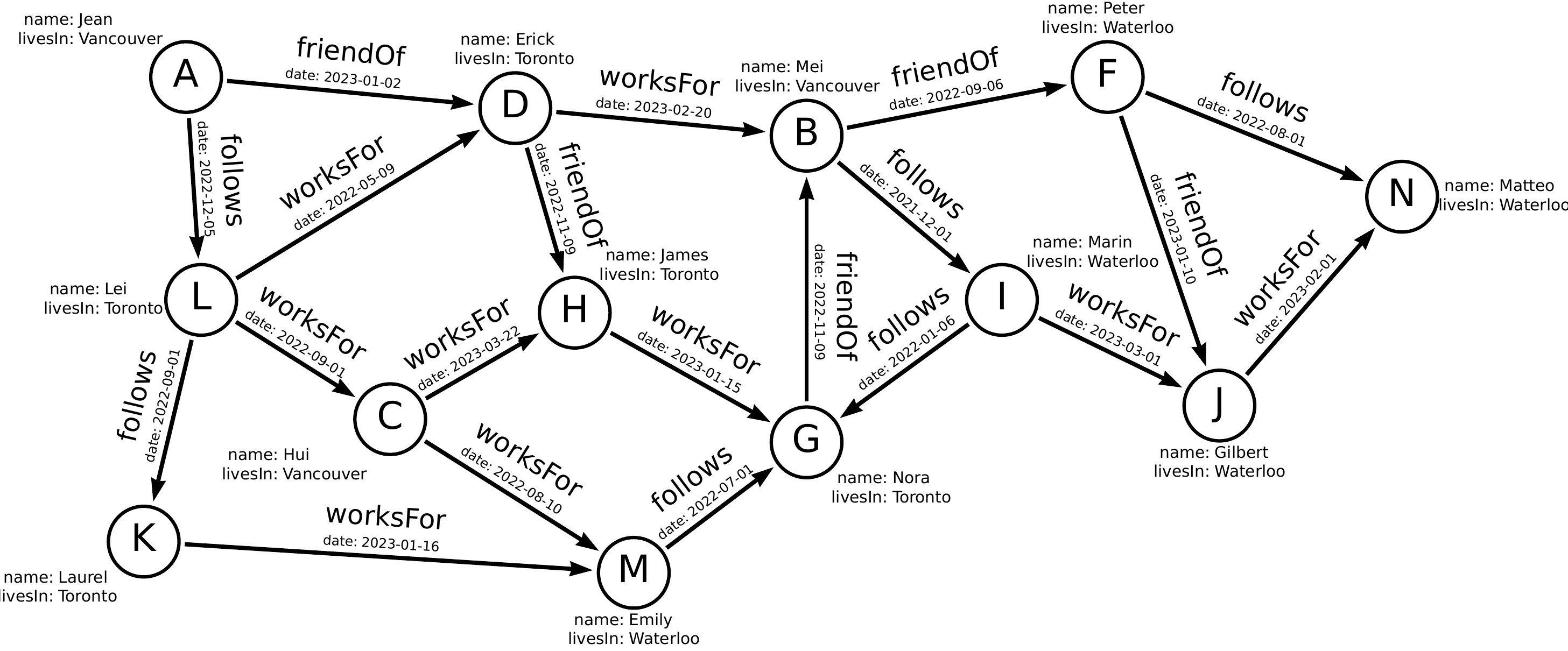}} 
    \caption{\revision{Example of a property graph.}}
    \label{fig:property graph}
\end{figure}

Fig. \ref{fig:property graph} presents an example of a property graph, which extends the edge-labeled graph in Fig. \ref{fig:graphs}(b) by adding attributes \texttt{name} and \texttt{livesIn} to  each vertex and attribute \texttt{date} to each edge.
Queries in property graphs might exhibit additional predicates on the properties.

\section{Complexity of plain reachability indexes}\label{sec:pr_complexity}
The complexities of plain reachability indexes are presented in Table \ref{table:pr_indexes}. 
We briefly present the complexities and refer readers to each approach for detailed analysis.

\revision{
Tree cover \cite{10.1145/67544.66950} has high complexity in index construction time due to the computation of minimum index. 
Dual labeling \cite{1617443} is influenced by the number of non-tree edges $|NTE|$ , while path-tree \cite{10.1145/1376616.1376677,10.1145/1929934.1929941} depends on the number of disjoint paths $|DP|$.
The other approaches based on tree cover are partial indexes, which have the linear complexity with the number of spanning trees $k$, such as GRAIL \cite{10.14778/1920841.1920879} and DAGGER \cite{Yildirim2013DAGGERAS}, Ferrari \cite{6544893} respectively. }

\revision{
2-Hop labeling \cite{10.5555/545381.545503}  has high construction time. 
The indexing techniques  are proposed to reduce the index sizes of 2-hop labeling. 
The bounds of index size of TFL \cite{10.1145/2463676.2465286}, DL \cite{10.14778/2556549.2556578}, HL \cite{10.14778/2556549.2556578}, and TOL \cite{10.1145/2588555.2612181} are unknown as they are heuristics to address the minimum 2-hop labeling problem.
They have the same query time $O(|L_{out}(s)|+|L_{out}(t)|)$ because of applying merge-join. 
Both DBL \cite{10.1007/978-3-030-73197-7_52} and O'Reach \cite{10.1145/3556540} are partial indexes.
The number of chains $|C|$ has an impact on the complexity of 3-hop labeling \cite{10.1145/1559845.1559930}. 
}

\revision{
Both IP \cite{10.14778/2732977.2732992,10.1007/s00778-017-0468-3} and BFL \cite{7750623} are partial indexing approaches.
The performance of IP primarily depends on the number of minimum elements ($ME$) in the MinHash scheme, the constrained index size for high-degree vertices ($|HLV|$), and the reachability ratio ($r$).
In contrast, BFL’s performance is influenced by the size of the Bloom filter in bits ($|BF|$).}

\revision{
The number of chains has a significant impact on approaches based on chain covers.
Both Feline \cite{Veloso2014ReachabilityQI} and Preach \cite{merz2014preach} are partial indexes, and they have linear construction time, index size, and $O(1)$ or $O(|V|+|E|)$ query time.}

\section{DLCR Example}

\begin{figure}
    \centering
    \resizebox{\textwidth}{!}{
    \includegraphics{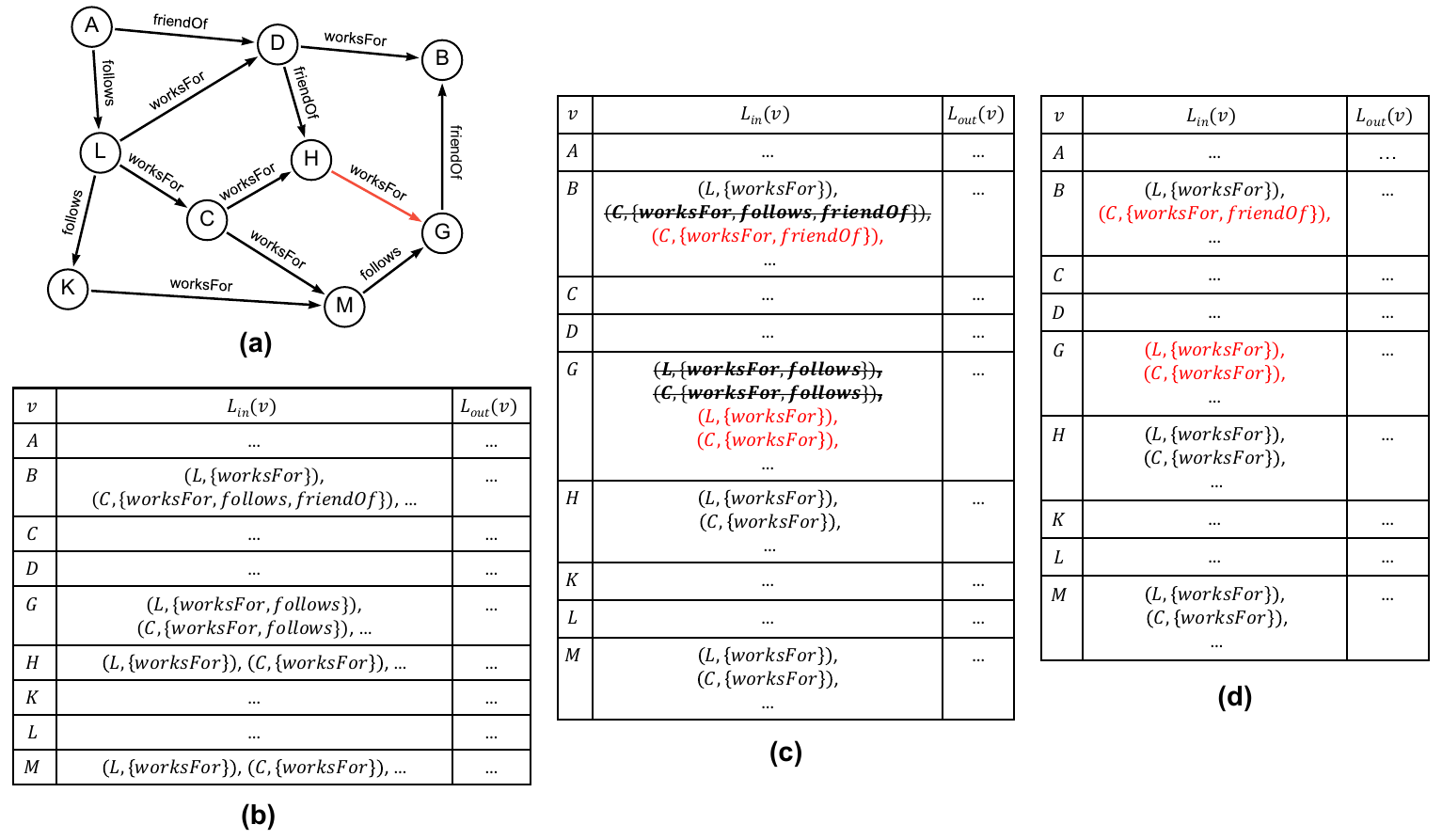}}
    \caption{
\revision{Running example of index updates in DLCR \cite{10.14778/3529337.3529348}. (a) shows the graph after inserting the edge $(H,\texttt{worksFor},G)$. (b) and (d) present the index snapshots before and after the edge insertion, respectively. (c) shows an intermediate snapshot during the update process, prior to removing any redundant entries.}
    }
    \label{fig:example-dlcr}
\end{figure}

We present a running example of index updates in DLCR in Fig. \ref{fig:example-dlcr}, which only includes the information for updating $L_{in}$-entries.
For inserting edge $(H,\texttt{worksFor},G)$ in (a), the snapshots of the index are changed from (b) to (d), and (c) is an intermediate snapshot during the updates, which exists after inserting index entries but before deleting any redundancies. For deleting edge $(H,\texttt{worksFor},G)$ in (a), the index is changed from (d) to (b).

\section{Complexity of path-constrained reachability indexes}\label{sec:pcr_complexity}
We discuss the complexity of path-constrained reachability indexes shown in Table \ref{table:pcr_indexes}.

For the index proposed by Jin et al. \cite{10.1145/1807167.1807183}, the index construction time is dominated by a procedure for computing all the vertices that a specific vertex can reach, \textit{i.e.}, the \textit{SingleSourcePathLabel} algorithm \cite{10.1145/1807167.1807183}, which takes $O(|V||E|\binom{|\mathcal{L}|}{|\mathcal{L}|/2})$ time and is repeated for each vertex.
The index size is bounded by $O(|V|^2\binom{|\mathcal{L}|}{|\mathcal{L}|/2})$.
The query time is $O(log(k) + \sum_{i=1}^{k'}|NT(u_i,v_i)|)$ (see \cite{10.1145/1807167.1807183} for the specific factors in the bound.)
As the approach by Chen et al. \cite{10.1145/3451159}  is based on recursive graph decomposition, the complexity included factors related to such decomposition (see \cite{10.1145/3451159} for detail).

The approach by Zou et al. \cite{10.1145/2063576.2063807,ZOU201447} computes and stores the generalized transitive closure of input graph. The index size is $O(|V|^2\binom{|\mathcal{L}|}{\mathcal{L}/2})$, and the query time is $O(\binom{|\mathcal{L}|}{\mathcal{L}/2})$, \textit{i.e.}, looking for sufficient path-label sets recorded under $(s,t)$.
The landmark index \cite{10.1145/3035918.3035955} computes the single-source GTC for landmark vertices, which takes $O((|V|(\log|V|+2^{|\mathcal{L}|})+|E|)k2^{|\mathcal{L}|})$ time for $k$ landmarks. 
Indexing $b$ entries for each of $|V|-k$ non-landmark vertices takes $O((|V|(\log|V|+b)+|E|)(|V|-k)2^{|\mathcal{L}|})$ time. The space for landmark vertices is $O(kn2^{|\mathcal{L}|})$ and the space for non-landmark vertices is $O((n-k)b)$. 
Query processing can visit $k$ landmarks in the worst case, which takes $O(|E|+k(2^{|\mathcal{L}|}+|V|))$ time.

In the P2H+ index \cite{10.14778/3380750.3380753}, each BFS for computing index entries takes $O(|E||V|2^{2|\mathcal{L}|})$ time, which is repeated for each vertex. Each $L_{in}(v)$ or $L_{out}(v)$ can take $O(n2^{|\mathcal{L}|})$ space. For the query processing, a merge join is performed between $L_{out}(s)$ and $L_{in}(t)$.
As DLCR \cite{10.14778/3529337.3529348} is essentially equivalent to P2H+ on static graphs, it has the same complexity. The main advantage of DLCR over P2H+ is index updates on dynamic graphs. 
Edge insertion and deletion take $O(2^{2|\mathcal{L}|}|V|(|\mathcal{L}_{add}||V|+|B|(|V|+|E|)))$ and  $O((|B|(|\mathcal{L}|+\log|V|)+|C|2^{2|\mathcal{L}|}|V|)(|V|+|E|))$ time, respectively (see \cite{10.14778/3529337.3529348} for the specific factors).

The RLC index \cite{10.1109/ICDE55515.2023.00013}  is built by performing kernel search and kernel BFS from each vertex.
The kernel search from each vertex takes $O(|\mathcal{L}|^k|V|^k)$ time, which can generate $O(|\mathcal{L}|^k)$ kernels, and the kernel BFS for each kernel takes $O(|E|k)$ time.
Each $L_{out}(v)$ or $L_{in}(v)$ can have $O(|V||\mathcal{L}|^k)$ entries.
Query processing is based on a merge join, which takes $O(|L_{out}(s)| + |L_{in}(t)|)$ time. 

%% file: main.bbl

\begin{thebibliography}{118}


\ifx \showCODEN    \undefined \def \showCODEN     #1{\unskip}     \fi
\ifx \showDOI      \undefined \def \showDOI       #1{#1}\fi
\ifx \showISBNx    \undefined \def \showISBNx     #1{\unskip}     \fi
\ifx \showISBNxiii \undefined \def \showISBNxiii  #1{\unskip}     \fi
\ifx \showISSN     \undefined \def \showISSN      #1{\unskip}     \fi
\ifx \showLCCN     \undefined \def \showLCCN      #1{\unskip}     \fi
\ifx \shownote     \undefined \def \shownote      #1{#1}          \fi
\ifx \showarticletitle \undefined \def \showarticletitle #1{#1}   \fi
\ifx \showURL      \undefined \def \showURL       {\relax}        \fi
\providecommand\bibfield[2]{#2}
\providecommand\bibinfo[2]{#2}
\providecommand\natexlab[1]{#1}
\providecommand\showeprint[2][]{arXiv:#2}

\bibitem[Abiteboul and Vianu(1997)]%
        {abiteboul1997regular}
\bibfield{author}{\bibinfo{person}{Serge Abiteboul} {and}
  \bibinfo{person}{Victor Vianu}.} \bibinfo{year}{1997}\natexlab{}.
\newblock \showarticletitle{Regular Path Queries with Constraints}. In
  \bibinfo{booktitle}{\emph{Proc. 16th ACM SIGACT-SIGMOD-SIGART Symp. on
  Principles of Database Systems}}. \bibinfo{pages}{122–133}.
\newblock
\showISBNx{0897919106}
\urldef\tempurl%
\url{https://doi.org/10.1145/263661.263676}
\showDOI{\tempurl}


\bibitem[Agrawal et~al\mbox{.}(1989)]%
        {10.1145/67544.66950}
\bibfield{author}{\bibinfo{person}{R. Agrawal}, \bibinfo{person}{A. Borgida},
  {and} \bibinfo{person}{H.~V. Jagadish}.} \bibinfo{year}{1989}\natexlab{}.
\newblock \showarticletitle{Efficient Management of Transitive Relationships in
  Large Data and Knowledge Bases}. In \bibinfo{booktitle}{\emph{Proc. ACM
  SIGMOD Int. Conf. on Management of Data}}. \bibinfo{pages}{253–262}.
\newblock
\showISBNx{0897913175}
\urldef\tempurl%
\url{https://doi.org/10.1145/67544.66950}
\showDOI{\tempurl}


\bibitem[Angles(2012)]%
        {6313676}
\bibfield{author}{\bibinfo{person}{Renzo Angles}.}
  \bibinfo{year}{2012}\natexlab{}.
\newblock \showarticletitle{A Comparison of Current Graph Database Models}. In
  \bibinfo{booktitle}{\emph{Proc. Workshops of 28th Int. Conf. on Data
  Engineering}}. \bibinfo{pages}{171--177}.
\newblock
\urldef\tempurl%
\url{https://doi.org/10.1109/ICDEW.2012.31}
\showDOI{\tempurl}


\bibitem[Angles et~al\mbox{.}(2017)]%
        {10.1145/3104031}
\bibfield{author}{\bibinfo{person}{Renzo Angles}, \bibinfo{person}{Marcelo
  Arenas}, \bibinfo{person}{Pablo Barcel\'{o}}, \bibinfo{person}{Aidan Hogan},
  \bibinfo{person}{Juan Reutter}, {and} \bibinfo{person}{Domagoj Vrgo\v{c}}.}
  \bibinfo{year}{2017}\natexlab{}.
\newblock \showarticletitle{Foundations of Modern Query Languages for Graph
  Databases}.
\newblock \bibinfo{journal}{\emph{ACM Comput. Surv.}} \bibinfo{volume}{50},
  \bibinfo{number}{5}, Article \bibinfo{articleno}{68} (\bibinfo{year}{2017}),
  \bibinfo{numpages}{40}~pages.
\newblock
\urldef\tempurl%
\url{https://doi.org/10.1145/3104031}
\showDOI{\tempurl}


\bibitem[Angles and Gutierrez(2008)]%
        {10.1145/1322432.1322433}
\bibfield{author}{\bibinfo{person}{Renzo Angles} {and} \bibinfo{person}{Claudio
  Gutierrez}.} \bibinfo{year}{2008}\natexlab{}.
\newblock \showarticletitle{Survey of Graph Database Models}.
\newblock \bibinfo{journal}{\emph{ACM Comput. Surv.}} \bibinfo{volume}{40},
  \bibinfo{number}{1}, Article \bibinfo{articleno}{1} (\bibinfo{date}{feb}
  \bibinfo{year}{2008}), \bibinfo{numpages}{39}~pages.
\newblock
\showISSN{0360-0300}
\urldef\tempurl%
\url{https://doi.org/10.1145/1322432.1322433}
\showDOI{\tempurl}


\bibitem[Angles and Gutierrez(2018)]%
        {Angles2018}
\bibfield{author}{\bibinfo{person}{Renzo Angles} {and} \bibinfo{person}{Claudio
  Gutierrez}.} \bibinfo{year}{2018}\natexlab{}.
\newblock \showarticletitle{An Introduction to Graph Data Management}.
\newblock In \bibinfo{booktitle}{\emph{Graph Data Management, Fundamental
  Issues and Recent Developments}}. \bibinfo{pages}{1--32}.
\newblock
\urldef\tempurl%
\url{https://doi.org/10.1007/978-3-319-96193-4\_1}
\showDOI{\tempurl}


\bibitem[Barcel\'{o} et~al\mbox{.}(2012)]%
        {10.1145/2389241.2389250}
\bibfield{author}{\bibinfo{person}{Pablo Barcel\'{o}}, \bibinfo{person}{Leonid
  Libkin}, \bibinfo{person}{Anthony~W. Lin}, {and} \bibinfo{person}{Peter~T.
  Wood}.} \bibinfo{year}{2012}\natexlab{}.
\newblock \showarticletitle{Expressive Languages for Path Queries over
  Graph-Structured Data}.
\newblock \bibinfo{journal}{\emph{ACM Trans. Database Syst.}}
  \bibinfo{volume}{37}, \bibinfo{number}{4}, Article \bibinfo{articleno}{31}
  (\bibinfo{year}{2012}), \bibinfo{numpages}{46}~pages.
\newblock
\showISSN{0362-5915}
\urldef\tempurl%
\url{https://doi.org/10.1145/2389241.2389250}
\showDOI{\tempurl}


\bibitem[Barcel\'{o}~Baeza(2013)]%
        {10.1145/2463664.2465216}
\bibfield{author}{\bibinfo{person}{Pablo Barcel\'{o}~Baeza}.}
  \bibinfo{year}{2013}\natexlab{}.
\newblock \showarticletitle{Querying Graph Databases}. In
  \bibinfo{booktitle}{\emph{Proc. 32nd ACM SIGACT-SIGMOD-SIGART Symp. on
  Principles of Database Systems}}. \bibinfo{pages}{175–188}.
\newblock
\showISBNx{9781450320665}
\urldef\tempurl%
\url{https://doi.org/10.1145/2463664.2465216}
\showDOI{\tempurl}


\bibitem[Barthélemy(2011)]%
        {barthelemy2011spatial}
\bibfield{author}{\bibinfo{person}{Marc Barthélemy}.}
  \bibinfo{year}{2011}\natexlab{}.
\newblock \showarticletitle{Spatial networks}.
\newblock \bibinfo{journal}{\emph{Phys. Rep.}} \bibinfo{volume}{499},
  \bibinfo{number}{1} (\bibinfo{year}{2011}), \bibinfo{pages}{1--101}.
\newblock
\showISSN{0370-1573}
\urldef\tempurl%
\url{https://doi.org/10.1016/j.physrep.2010.11.002}
\showDOI{\tempurl}


\bibitem[Bentley(1975)]%
        {10.1145/361002.361007}
\bibfield{author}{\bibinfo{person}{Jon~Louis Bentley}.}
  \bibinfo{year}{1975}\natexlab{}.
\newblock \showarticletitle{Multidimensional Binary Search Trees Used for
  Associative Searching}.
\newblock \bibinfo{journal}{\emph{Commun. ACM}} \bibinfo{volume}{18},
  \bibinfo{number}{9} (\bibinfo{year}{1975}), \bibinfo{pages}{509–517}.
\newblock
\showISSN{0001-0782}
\urldef\tempurl%
\url{https://doi.org/10.1145/361002.361007}
\showDOI{\tempurl}


\bibitem[Besta et~al\mbox{.}(2023a)]%
        {10.1145/3581784.3607068}
\bibfield{author}{\bibinfo{person}{Maciej Besta}, \bibinfo{person}{Robert
  Gerstenberger}, \bibinfo{person}{Marc Fischer}, \bibinfo{person}{Michal
  Podstawski}, \bibinfo{person}{Nils Blach}, \bibinfo{person}{Berke Egeli},
  \bibinfo{person}{Georgy Mitenkov}, \bibinfo{person}{Wojciech Chlapek},
  \bibinfo{person}{Marek Michalewicz}, \bibinfo{person}{Hubert Niewiadomski},
  \bibinfo{person}{Juergen Mueller}, {and} \bibinfo{person}{Torsten Hoefler}.}
  \bibinfo{year}{2023}\natexlab{a}.
\newblock \showarticletitle{The Graph Database Interface: Scaling Online
  Transactional and Analytical Graph Workloads to Hundreds of Thousands of
  Cores}. In \bibinfo{booktitle}{\emph{Proc. ACM/IEEE Conf. on High Performance
  Computing, Networking, Storage and Analysis}}. Article
  \bibinfo{articleno}{22}, \bibinfo{numpages}{18}~pages.
\newblock


\bibitem[Besta et~al\mbox{.}(2023b)]%
        {10.1145/3604932}
\bibfield{author}{\bibinfo{person}{Maciej Besta}, \bibinfo{person}{Robert
  Gerstenberger}, \bibinfo{person}{Emanuel Peter}, \bibinfo{person}{Marc
  Fischer}, \bibinfo{person}{Micha\l{} Podstawski}, \bibinfo{person}{Claude
  Barthels}, \bibinfo{person}{Gustavo Alonso}, {and} \bibinfo{person}{Torsten
  Hoefler}.} \bibinfo{year}{2023}\natexlab{b}.
\newblock \showarticletitle{Demystifying Graph Databases: Analysis and Taxonomy
  of Data Organization, System Designs, and Graph Queries}.
\newblock \bibinfo{journal}{\emph{ACM Comput. Surv.}} \bibinfo{volume}{56},
  \bibinfo{number}{2}, Article \bibinfo{articleno}{31} (\bibinfo{year}{2023}),
  \bibinfo{numpages}{40}~pages.
\newblock
\urldef\tempurl%
\url{https://doi.org/10.1145/3604932}
\showDOI{\tempurl}


\bibitem[Bonifati and Dumbrava(2019)]%
        {10.1145/3335409.3335411}
\bibfield{author}{\bibinfo{person}{Angela Bonifati} {and}
  \bibinfo{person}{Stefania Dumbrava}.} \bibinfo{year}{2019}\natexlab{}.
\newblock \showarticletitle{Graph Queries: From Theory to Practice}.
\newblock \bibinfo{journal}{\emph{ACM SIGMOD Rec.}} \bibinfo{volume}{47},
  \bibinfo{number}{4} (\bibinfo{date}{may} \bibinfo{year}{2019}),
  \bibinfo{pages}{5–16}.
\newblock
\showISSN{0163-5808}
\urldef\tempurl%
\url{https://doi.org/10.1145/3335409.3335411}
\showDOI{\tempurl}


\bibitem[Bonifati et~al\mbox{.}(2018)]%
        {10.5555/3307192}
\bibfield{author}{\bibinfo{person}{Angela Bonifati}, \bibinfo{person}{George
  Fletcher}, \bibinfo{person}{Hannes Voigt}, {and} \bibinfo{person}{Nikolay
  Yakovets}.} \bibinfo{year}{2018}\natexlab{}.
\newblock \bibinfo{booktitle}{\emph{Querying Graphs}}.
\newblock \bibinfo{publisher}{Morgan \& Claypool Publishers}.
\newblock
\showISBNx{168173432X}
\urldef\tempurl%
\url{https://doi.org/10.2200/S00873ED1V01Y201808DTM051}
\showURL{%
\tempurl}


\bibitem[Bonifati et~al\mbox{.}(2019)]%
        {10.1145/3308558.3313472}
\bibfield{author}{\bibinfo{person}{Angela Bonifati}, \bibinfo{person}{Wim
  Martens}, {and} \bibinfo{person}{Thomas Timm}.}
  \bibinfo{year}{2019}\natexlab{}.
\newblock \showarticletitle{Navigating the Maze of Wikidata Query Logs}. In
  \bibinfo{booktitle}{\emph{Proc. 28th Int. World Wide Web Conf.}}
  \bibinfo{pages}{127–138}.
\newblock
\showISBNx{9781450366748}
\urldef\tempurl%
\url{https://doi.org/10.1145/3308558.3313472}
\showDOI{\tempurl}


\bibitem[Bramandia et~al\mbox{.}(2010)]%
        {4912201}
\bibfield{author}{\bibinfo{person}{Ramadhana Bramandia}, \bibinfo{person}{Byron
  Choi}, {and} \bibinfo{person}{Wee~Keong Ng}.}
  \bibinfo{year}{2010}\natexlab{}.
\newblock \showarticletitle{Incremental Maintenance of 2-Hop Labeling of Large
  Graphs}.
\newblock \bibinfo{journal}{\emph{IEEE Trans. Knowl. and Data Eng.}}
  \bibinfo{volume}{22}, \bibinfo{number}{5} (\bibinfo{year}{2010}),
  \bibinfo{pages}{682--698}.
\newblock
\urldef\tempurl%
\url{https://doi.org/10.1109/TKDE.2009.117}
\showDOI{\tempurl}


\bibitem[Bunke(2000)]%
        {bunke00}
\bibfield{author}{\bibinfo{person}{Horst Bunke}.}
  \bibinfo{year}{2000}\natexlab{}.
\newblock \showarticletitle{Graph Matching: Theoretical Foundations,
  Algorithms, and Applications}. In \bibinfo{booktitle}{\emph{Vision
  Interface}}. \bibinfo{pages}{82--88}.
\newblock


\bibitem[Cai and Poon(2010)]%
        {10.1145/1871437.1871457}
\bibfield{author}{\bibinfo{person}{Jing Cai} {and} \bibinfo{person}{Chung~Keung
  Poon}.} \bibinfo{year}{2010}\natexlab{}.
\newblock \showarticletitle{{Path-Hop}: Efficiently Indexing Large Graphs for
  Reachability Queries}. In \bibinfo{booktitle}{\emph{Proc. 19th ACM Int. Conf.
  on Information and Knowledge Management}}. \bibinfo{pages}{119–128}.
\newblock
\showISBNx{9781450300995}
\urldef\tempurl%
\url{https://doi.org/10.1145/1871437.1871457}
\showDOI{\tempurl}


\bibitem[Calvanese et~al\mbox{.}(2002)]%
        {CALVANESE2002443}
\bibfield{author}{\bibinfo{person}{Diego Calvanese}, \bibinfo{person}{Giuseppe
  {De Giacomo}}, \bibinfo{person}{Maurizio Lenzerini}, {and}
  \bibinfo{person}{Moshe~Y. Vardi}.} \bibinfo{year}{2002}\natexlab{}.
\newblock \showarticletitle{Rewriting of Regular Expressions and Regular Path
  Queries}.
\newblock \bibinfo{journal}{\emph{J. Comput. Syst. Sci}} \bibinfo{volume}{64},
  \bibinfo{number}{3} (\bibinfo{year}{2002}), \bibinfo{pages}{443--465}.
\newblock
\showISSN{0022-0000}
\urldef\tempurl%
\url{https://doi.org/10.1006/jcss.2001.1805}
\showDOI{\tempurl}


\bibitem[Calvanese et~al\mbox{.}(2003)]%
        {10.1145/959060.959076}
\bibfield{author}{\bibinfo{person}{D. Calvanese}, \bibinfo{person}{G.
  De~Giacomo}, \bibinfo{person}{M. Lenzerini}, {and} \bibinfo{person}{M.~Y.
  Vardi}.} \bibinfo{year}{2003}\natexlab{}.
\newblock \showarticletitle{Reasoning on Regular Path Queries}.
\newblock \bibinfo{journal}{\emph{ACM SIGMOD Rec.}} \bibinfo{volume}{32},
  \bibinfo{number}{4} (\bibinfo{date}{dec} \bibinfo{year}{2003}),
  \bibinfo{pages}{83–92}.
\newblock
\showISSN{0163-5808}
\urldef\tempurl%
\url{https://doi.org/10.1145/959060.959076}
\showDOI{\tempurl}


\bibitem[Chen et~al\mbox{.}(2022a)]%
        {10.14778/3551793.3551813}
\bibfield{author}{\bibinfo{person}{Hongzhi Chen}, \bibinfo{person}{Changji Li},
  \bibinfo{person}{Chenguang Zheng}, \bibinfo{person}{Chenghuan Huang},
  \bibinfo{person}{Juncheng Fang}, \bibinfo{person}{James Cheng}, {and}
  \bibinfo{person}{Jian Zhang}.} \bibinfo{year}{2022}\natexlab{a}.
\newblock \showarticletitle{G-tran: a high performance distributed graph
  database with a decentralized architecture}.
\newblock \bibinfo{journal}{\emph{Proc. VLDB Endowment}} \bibinfo{volume}{15},
  \bibinfo{number}{11} (\bibinfo{date}{July} \bibinfo{year}{2022}),
  \bibinfo{pages}{2545–2558}.
\newblock
\showISSN{2150-8097}


\bibitem[Chen et~al\mbox{.}(2005)]%
        {10.5555/1083592.1083651}
\bibfield{author}{\bibinfo{person}{Li Chen}, \bibinfo{person}{Amarnath Gupta},
  {and} \bibinfo{person}{M.~Erdem Kurul}.} \bibinfo{year}{2005}\natexlab{}.
\newblock \showarticletitle{Stack-Based Algorithms for Pattern Matching on
  DAGs}. In \bibinfo{booktitle}{\emph{Proc. 31st Int. Conf. on Very Large Data
  Bases}}. \bibinfo{pages}{493–504}.
\newblock
\showISBNx{1595931546}
\urldef\tempurl%
\url{https://dl.acm.org/doi/10.5555/1083592.1083651}
\showURL{%
\tempurl}


\bibitem[Chen et~al\mbox{.}(2022b)]%
        {10.14778/3529337.3529348}
\bibfield{author}{\bibinfo{person}{Xin Chen}, \bibinfo{person}{You Peng},
  \bibinfo{person}{Sibo Wang}, {and} \bibinfo{person}{Jeffrey~Xu Yu}.}
  \bibinfo{year}{2022}\natexlab{b}.
\newblock \showarticletitle{DLCR: Efficient Indexing for Label-Constrained
  Reachability Queries on Large Dynamic Graphs}.
\newblock \bibinfo{journal}{\emph{Proc. VLDB Endowment}} \bibinfo{volume}{15},
  \bibinfo{number}{8} (\bibinfo{year}{2022}), \bibinfo{pages}{1645–1657}.
\newblock
\showISSN{2150-8097}
\urldef\tempurl%
\url{https://doi.org/10.14778/3529337.3529348}
\showDOI{\tempurl}


\bibitem[{Chen} and {Chen}(2008)]%
        {4497498}
\bibfield{author}{\bibinfo{person}{Y. {Chen}} {and} \bibinfo{person}{Y.
  {Chen}}.} \bibinfo{year}{2008}\natexlab{}.
\newblock \showarticletitle{An Efficient Algorithm for Answering Graph
  Reachability Queries}. In \bibinfo{booktitle}{\emph{Proc. 24th Int. Conf. on
  Data Engineering}}. \bibinfo{pages}{893--902}.
\newblock
\urldef\tempurl%
\url{https://doi.org/10.1109/ICDE.2008.4497498}
\showDOI{\tempurl}


\bibitem[Chen and Singh(2021)]%
        {10.1145/3451159}
\bibfield{author}{\bibinfo{person}{Yangjun Chen} {and}
  \bibinfo{person}{Gagandeep Singh}.} \bibinfo{year}{2021}\natexlab{}.
\newblock \showarticletitle{Graph Indexing for Efficient Evaluation of
  Label-constrained Reachability Queries}.
\newblock \bibinfo{journal}{\emph{ACM Trans. Database Syst.}}
  \bibinfo{volume}{46}, \bibinfo{number}{2}, Article \bibinfo{articleno}{8}
  (\bibinfo{year}{2021}), \bibinfo{numpages}{50}~pages.
\newblock
\showISSN{0362-5915}
\urldef\tempurl%
\url{https://doi.org/10.1145/3451159}
\showDOI{\tempurl}


\bibitem[Cheng et~al\mbox{.}(2013)]%
        {10.1145/2463676.2465286}
\bibfield{author}{\bibinfo{person}{James Cheng}, \bibinfo{person}{Silu Huang},
  \bibinfo{person}{Huanhuan Wu}, {and} \bibinfo{person}{Ada Wai-Chee Fu}.}
  \bibinfo{year}{2013}\natexlab{}.
\newblock \showarticletitle{{TF-Label}: A Topological-Folding Labeling Scheme
  for Reachability Querying in a Large Graph}. In
  \bibinfo{booktitle}{\emph{Proc. ACM SIGMOD Int. Conf. on Management of
  Data}}. \bibinfo{pages}{193–204}.
\newblock
\showISBNx{9781450320375}
\urldef\tempurl%
\url{https://doi.org/10.1145/2463676.2465286}
\showDOI{\tempurl}


\bibitem[Cheng et~al\mbox{.}(2012)]%
        {10.14778/2350229.2350247}
\bibfield{author}{\bibinfo{person}{James Cheng}, \bibinfo{person}{Zechao
  Shang}, \bibinfo{person}{Hong Cheng}, \bibinfo{person}{Haixun Wang}, {and}
  \bibinfo{person}{Jeffrey~Xu Yu}.} \bibinfo{year}{2012}\natexlab{}.
\newblock \showarticletitle{K-reach: who is in your small world}.
\newblock \bibinfo{journal}{\emph{Proc. VLDB Endowment}} \bibinfo{volume}{5},
  \bibinfo{number}{11} (\bibinfo{date}{July} \bibinfo{year}{2012}),
  \bibinfo{pages}{1292–1303}.
\newblock
\showISSN{2150-8097}
\urldef\tempurl%
\url{https://doi.org/10.14778/2350229.2350247}
\showDOI{\tempurl}


\bibitem[Cheng et~al\mbox{.}(2014)]%
        {cheng2014efficient}
\bibfield{author}{\bibinfo{person}{James Cheng}, \bibinfo{person}{Zechao
  Shang}, \bibinfo{person}{Hong Cheng}, \bibinfo{person}{Haixun Wang}, {and}
  \bibinfo{person}{Jeffrey~Xu Yu}.} \bibinfo{year}{2014}\natexlab{}.
\newblock \showarticletitle{Efficient processing of k-hop reachability
  queries}.
\newblock \bibinfo{journal}{\emph{VLDB J.}} \bibinfo{volume}{23},
  \bibinfo{number}{2} (\bibinfo{year}{2014}), \bibinfo{pages}{227--252}.
\newblock


\bibitem[Cohen et~al\mbox{.}(2003)]%
        {10.5555/545381.545503}
\bibfield{author}{\bibinfo{person}{Edith Cohen}, \bibinfo{person}{Eran
  Halperin}, \bibinfo{person}{Haim Kaplan}, {and} \bibinfo{person}{Uri Zwick}.}
  \bibinfo{year}{2003}\natexlab{}.
\newblock \showarticletitle{Reachability and Distance Queries via 2-Hop
  Labels}.
\newblock \bibinfo{journal}{\emph{SIAM J. on Comput.}} \bibinfo{volume}{32},
  \bibinfo{number}{5} (\bibinfo{year}{2003}), \bibinfo{pages}{1338--1355}.
\newblock
\urldef\tempurl%
\url{https://doi.org/10.1137/S0097539702403098}
\showDOI{\tempurl}


\bibitem[Cui et~al\mbox{.}(2024)]%
        {10.14778/3648160.3648179}
\bibfield{author}{\bibinfo{person}{Pengjie Cui}, \bibinfo{person}{Haotian Liu},
  \bibinfo{person}{Bo Tang}, {and} \bibinfo{person}{Ye Yuan}.}
  \bibinfo{year}{2024}\natexlab{}.
\newblock \showarticletitle{CGgraph: An Ultra-Fast Graph Processing System on
  Modern Commodity CPU-GPU Co-processor}.
\newblock \bibinfo{journal}{\emph{Proc. VLDB Endowment}} \bibinfo{volume}{17},
  \bibinfo{number}{6} (\bibinfo{date}{Feb.} \bibinfo{year}{2024}),
  \bibinfo{pages}{1405–1417}.
\newblock


\bibitem[Deutsch et~al\mbox{.}(2022)]%
        {10.1145/3514221.3526057}
\bibfield{author}{\bibinfo{person}{Alin Deutsch}, \bibinfo{person}{Nadime
  Francis}, \bibinfo{person}{Alastair Green}, \bibinfo{person}{Keith Hare},
  \bibinfo{person}{Bei Li}, \bibinfo{person}{Leonid Libkin},
  \bibinfo{person}{Tobias Lindaaker}, \bibinfo{person}{Victor Marsault},
  \bibinfo{person}{Wim Martens}, \bibinfo{person}{Jan Michels},
  \bibinfo{person}{Filip Murlak}, \bibinfo{person}{Stefan Plantikow},
  \bibinfo{person}{Petra Selmer}, \bibinfo{person}{Oskar van Rest},
  \bibinfo{person}{Hannes Voigt}, \bibinfo{person}{Domagoj Vrgo\v{c}},
  \bibinfo{person}{Mingxi Wu}, {and} \bibinfo{person}{Fred Zemke}.}
  \bibinfo{year}{2022}\natexlab{}.
\newblock \showarticletitle{Graph Pattern Matching in GQL and SQL/PGQ}. In
  \bibinfo{booktitle}{\emph{Proc. ACM SIGMOD Int. Conf. on Management of
  Data}}. \bibinfo{pages}{2246–2258}.
\newblock
\showISBNx{9781450392495}
\urldef\tempurl%
\url{https://doi.org/10.1145/3514221.3526057}
\showDOI{\tempurl}


\bibitem[Erling et~al\mbox{.}(2015)]%
        {10.1145/2723372.2742786}
\bibfield{author}{\bibinfo{person}{Orri Erling}, \bibinfo{person}{Alex
  Averbuch}, \bibinfo{person}{Josep Larriba-Pey}, \bibinfo{person}{Hassan
  Chafi}, \bibinfo{person}{Andrey Gubichev}, \bibinfo{person}{Arnau Prat},
  \bibinfo{person}{Minh-Duc Pham}, {and} \bibinfo{person}{Peter Boncz}.}
  \bibinfo{year}{2015}\natexlab{}.
\newblock \showarticletitle{The LDBC Social Network Benchmark: Interactive
  Workload}. In \bibinfo{booktitle}{\emph{Proc. ACM SIGMOD Int. Conf. on
  Management of Data}}. \bibinfo{pages}{619–630}.
\newblock
\showISBNx{9781450327589}
\urldef\tempurl%
\url{https://doi.org/10.1145/2723372.2742786}
\showDOI{\tempurl}


\bibitem[Fan et~al\mbox{.}(2011)]%
        {5767858}
\bibfield{author}{\bibinfo{person}{Wenfei Fan}, \bibinfo{person}{Jianzhong Li},
  \bibinfo{person}{Shuai Ma}, \bibinfo{person}{Nan Tang}, {and}
  \bibinfo{person}{Yinghui Wu}.} \bibinfo{year}{2011}\natexlab{}.
\newblock \showarticletitle{Adding regular expressions to graph reachability
  and pattern queries}. In \bibinfo{booktitle}{\emph{Proc. 27th Int. Conf. on
  Data Engineering}}. \bibinfo{pages}{39--50}.
\newblock
\urldef\tempurl%
\url{https://doi.org/10.1109/ICDE.2011.5767858}
\showDOI{\tempurl}


\bibitem[Fan et~al\mbox{.}(2018)]%
        {10.1145/3282488}
\bibfield{author}{\bibinfo{person}{Wenfei Fan}, \bibinfo{person}{Wenyuan Yu},
  \bibinfo{person}{Jingbo Xu}, \bibinfo{person}{Jingren Zhou},
  \bibinfo{person}{Xiaojian Luo}, \bibinfo{person}{Qiang Yin},
  \bibinfo{person}{Ping Lu}, \bibinfo{person}{Yang Cao}, {and}
  \bibinfo{person}{Ruiqi Xu}.} \bibinfo{year}{2018}\natexlab{}.
\newblock \showarticletitle{Parallelizing Sequential Graph Computations}.
\newblock \bibinfo{journal}{\emph{ACM Trans. Database Syst.}}
  \bibinfo{volume}{43}, \bibinfo{number}{4}, Article \bibinfo{articleno}{18}
  (\bibinfo{date}{Dec.} \bibinfo{year}{2018}), \bibinfo{numpages}{39}~pages.
\newblock


\bibitem[Feng et~al\mbox{.}(2023)]%
        {feng2023kuzu}
\bibfield{author}{\bibinfo{person}{Xiyang Feng}, \bibinfo{person}{Guodong Jin},
  \bibinfo{person}{Ziyi Chen}, \bibinfo{person}{Chang Liu}, {and}
  \bibinfo{person}{Semih Saliho{\u{g}}lu}.} \bibinfo{year}{2023}\natexlab{}.
\newblock \showarticletitle{K{\`u}zu graph database management system}. In
  \bibinfo{booktitle}{\emph{The Conf. on Innovative Data Systems Research}},
  Vol.~\bibinfo{volume}{7}. \bibinfo{pages}{25--35}.
\newblock


\bibitem[Fletcher and Theobald(2018)]%
        {Fletcher2018}
\bibfield{author}{\bibinfo{person}{George Fletcher} {and}
  \bibinfo{person}{Martin Theobald}.} \bibinfo{year}{2018}\natexlab{}.
\newblock \showarticletitle{Indexing for Graph Query Evaluation}.
\newblock In \bibinfo{booktitle}{\emph{Encyclopedia of Big Data Technologies}}.
  \bibinfo{pages}{1--9}.
\newblock
\showISBNx{978-3-319-63962-8}
\urldef\tempurl%
\url{https://doi.org/10.1007/978-3-319-63962-8_212-1}
\showDOI{\tempurl}


\bibitem[Fletcher et~al\mbox{.}(2016)]%
        {fletcher2016efficient}
\bibfield{author}{\bibinfo{person}{George~HL Fletcher}, \bibinfo{person}{Jeroen
  Peters}, {and} \bibinfo{person}{Alexandra Poulovassilis}.}
  \bibinfo{year}{2016}\natexlab{}.
\newblock \showarticletitle{Efficient regular path query evaluation using path
  indexes}. In \bibinfo{booktitle}{\emph{Proc. 19th Int. Conf. on Extending
  Database Technology}}. \bibinfo{pages}{636--639}.
\newblock


\bibitem[Foulds(1992)]%
        {Foulds1992}
\bibfield{author}{\bibinfo{person}{L.~R. Foulds}.}
  \bibinfo{year}{1992}\natexlab{}.
\newblock \showarticletitle{Digraphs}.
\newblock In \bibinfo{booktitle}{\emph{Graph Theory Applications}}.
  \bibinfo{pages}{93--122}.
\newblock
\showISBNx{978-1-4612-0933-1}
\urldef\tempurl%
\url{https://doi.org/10.1007/978-1-4612-0933-1_7}
\showDOI{\tempurl}


\bibitem[Gallagher(2006)]%
        {gallagher2006matching}
\bibfield{author}{\bibinfo{person}{Brian Gallagher}.}
  \bibinfo{year}{2006}\natexlab{}.
\newblock \showarticletitle{Matching Structure and Semantics: A Survey on
  Graph-Based Pattern Matching.}. In \bibinfo{booktitle}{\emph{Proceedings of
  the AAAI Fall Symposium}}. \bibinfo{pages}{43--53}.
\newblock


\bibitem[Gou et~al\mbox{.}(2024)]%
        {10.14778/3641204.3641214}
\bibfield{author}{\bibinfo{person}{Xiangyang Gou}, \bibinfo{person}{Xinyi Ye},
  \bibinfo{person}{Lei Zou}, {and} \bibinfo{person}{Jeffrey~Xu Yu}.}
  \bibinfo{year}{2024}\natexlab{}.
\newblock \showarticletitle{LM-SRPQ: Efficiently Answering Regular Path Query
  in Streaming Graphs}.
\newblock \bibinfo{journal}{\emph{Proc. VLDB Endowment}} \bibinfo{volume}{17},
  \bibinfo{number}{5} (\bibinfo{date}{Jan.} \bibinfo{year}{2024}),
  \bibinfo{pages}{1047–1059}.
\newblock


\bibitem[Gubichev et~al\mbox{.}(2013)]%
        {10.1145/2484425.2484443}
\bibfield{author}{\bibinfo{person}{Andrey Gubichev},
  \bibinfo{person}{Srikanta~J. Bedathur}, {and} \bibinfo{person}{Stephan
  Seufert}.} \bibinfo{year}{2013}\natexlab{}.
\newblock \showarticletitle{Sparqling Kleene: Fast Property Paths in RDF-3X}.
  In \bibinfo{booktitle}{\emph{Proc. 1st Int. Workshop on Graph Data Management
  Experiences \& Systems}}. \bibinfo{numpages}{7}~pages.
\newblock
\showISBNx{9781450321884}
\urldef\tempurl%
\url{https://doi.org/10.1145/2484425.2484443}
\showDOI{\tempurl}


\bibitem[Gupta et~al\mbox{.}(2021)]%
        {10.14778/3476249.3476297}
\bibfield{author}{\bibinfo{person}{Pranjal Gupta}, \bibinfo{person}{Amine
  Mhedhbi}, {and} \bibinfo{person}{Semih Salihoglu}.}
  \bibinfo{year}{2021}\natexlab{}.
\newblock \showarticletitle{Columnar storage and list-based processing for
  graph database management systems}.
\newblock \bibinfo{journal}{\emph{Proc. VLDB Endowment}} \bibinfo{volume}{14},
  \bibinfo{number}{11} (\bibinfo{date}{July} \bibinfo{year}{2021}),
  \bibinfo{pages}{2491–2504}.
\newblock
\showISSN{2150-8097}
\urldef\tempurl%
\url{https://doi.org/10.14778/3476249.3476297}
\showDOI{\tempurl}


\bibitem[Gutierrez and Sequeda(2021)]%
        {10.1145/3418294}
\bibfield{author}{\bibinfo{person}{Claudio Gutierrez} {and}
  \bibinfo{person}{Juan~F. Sequeda}.} \bibinfo{year}{2021}\natexlab{}.
\newblock \showarticletitle{Knowledge Graphs}.
\newblock \bibinfo{journal}{\emph{Commun. ACM}} \bibinfo{volume}{64},
  \bibinfo{number}{3} (\bibinfo{year}{2021}), \bibinfo{pages}{96–104}.
\newblock
\showISSN{0001-0782}
\urldef\tempurl%
\url{https://doi.org/10.1145/3418294}
\showDOI{\tempurl}


\bibitem[{Haixun Wang} et~al\mbox{.}(2006)]%
        {1617443}
\bibfield{author}{\bibinfo{person}{{Haixun Wang}}, \bibinfo{person}{{Hao He}},
  \bibinfo{person}{{Jun Yang}}, \bibinfo{person}{P.~S. {Yu}}, {and}
  \bibinfo{person}{J.~X. {Yu}}.} \bibinfo{year}{2006}\natexlab{}.
\newblock \showarticletitle{Dual Labeling: Answering Graph Reachability Queries
  in Constant Time}. In \bibinfo{booktitle}{\emph{Proc. 22nd Int. Conf. on Data
  Engineering}}. \bibinfo{pages}{75--75}.
\newblock
\urldef\tempurl%
\url{https://doi.org/10.1109/ICDE.2006.53}
\showDOI{\tempurl}


\bibitem[Hanauer et~al\mbox{.}(2020)]%
        {10.1137/1.9781611976007.9}
\bibfield{author}{\bibinfo{person}{Kathrin Hanauer}, \bibinfo{person}{Monika
  Henzinger}, {and} \bibinfo{person}{Christian Schulz}.}
  \bibinfo{year}{2020}\natexlab{}.
\newblock \showarticletitle{Fully Dynamic Single-Source Reachability in
  Practice: An Experimental Study}. In \bibinfo{booktitle}{\emph{Proc. 22nd
  Symp. on Algorithm Eng. and Experiments}}. \bibinfo{pages}{106--119}.
\newblock
\urldef\tempurl%
\url{https://doi.org/10.1137/1.9781611976007.9}
\showDOI{\tempurl}


\bibitem[Hanauer et~al\mbox{.}(2022)]%
        {10.1145/3556540}
\bibfield{author}{\bibinfo{person}{Kathrin Hanauer}, \bibinfo{person}{Christian
  Schulz}, {and} \bibinfo{person}{Jonathan Trummer}.}
  \bibinfo{year}{2022}\natexlab{}.
\newblock \showarticletitle{O’Reach: Even Faster Reachability in Large
  Graphs}.
\newblock \bibinfo{journal}{\emph{ACM J. Exp. Algorithmics}}
  \bibinfo{volume}{27}, Article \bibinfo{articleno}{4.2}
  (\bibinfo{year}{2022}), \bibinfo{numpages}{27}~pages.
\newblock
\showISSN{1084-6654}
\urldef\tempurl%
\url{https://doi.org/10.1145/3556540}
\showDOI{\tempurl}


\bibitem[Henzinger and King(1995)]%
        {492668}
\bibfield{author}{\bibinfo{person}{M.R. Henzinger} {and} \bibinfo{person}{V.
  King}.} \bibinfo{year}{1995}\natexlab{}.
\newblock \showarticletitle{Fully dynamic biconnectivity and transitive
  closure}. In \bibinfo{booktitle}{\emph{Proc. 36th Annual Symp. on Foundations
  of Computer Science}}. \bibinfo{pages}{664--672}.
\newblock
\urldef\tempurl%
\url{https://doi.org/10.1109/SFCS.1995.492668}
\showDOI{\tempurl}


\bibitem[Jagadish(1990)]%
        {10.1145/99935.99944}
\bibfield{author}{\bibinfo{person}{H.~V. Jagadish}.}
  \bibinfo{year}{1990}\natexlab{}.
\newblock \showarticletitle{A Compression Technique to Materialize Transitive
  Closure}.
\newblock \bibinfo{journal}{\emph{ACM Trans. Database Syst.}}
  \bibinfo{volume}{15}, \bibinfo{number}{4} (\bibinfo{year}{1990}),
  \bibinfo{pages}{558–598}.
\newblock
\showISSN{0362-5915}
\urldef\tempurl%
\url{https://doi.org/10.1145/99935.99944}
\showDOI{\tempurl}


\bibitem[Jin and Salihoglu(2022)]%
        {10.14778/3510397.3510400}
\bibfield{author}{\bibinfo{person}{Guodong Jin} {and} \bibinfo{person}{Semih
  Salihoglu}.} \bibinfo{year}{2022}\natexlab{}.
\newblock \showarticletitle{Making RDBMSs efficient on graph workloads through
  predefined joins}.
\newblock \bibinfo{journal}{\emph{Proc. VLDB Endowment}} \bibinfo{volume}{15},
  \bibinfo{number}{5} (\bibinfo{date}{Jan.} \bibinfo{year}{2022}),
  \bibinfo{pages}{1011–1023}.
\newblock
\urldef\tempurl%
\url{https://doi.org/10.14778/3510397.3510400}
\showDOI{\tempurl}


\bibitem[Jin et~al\mbox{.}(2010)]%
        {10.1145/1807167.1807183}
\bibfield{author}{\bibinfo{person}{Ruoming Jin}, \bibinfo{person}{Hui Hong},
  \bibinfo{person}{Haixun Wang}, \bibinfo{person}{Ning Ruan}, {and}
  \bibinfo{person}{Yang Xiang}.} \bibinfo{year}{2010}\natexlab{}.
\newblock \showarticletitle{Computing Label-Constraint Reachability in Graph
  Databases}. In \bibinfo{booktitle}{\emph{Proc. ACM SIGMOD Int. Conf. on
  Management of Data}}. \bibinfo{pages}{123–134}.
\newblock
\showISBNx{9781450300322}
\urldef\tempurl%
\url{https://doi.org/10.1145/1807167.1807183}
\showDOI{\tempurl}


\bibitem[Jin et~al\mbox{.}(2020)]%
        {10.1145/3392717.3392745}
\bibfield{author}{\bibinfo{person}{Ruoming Jin}, \bibinfo{person}{Zhen Peng},
  \bibinfo{person}{Wendell Wu}, \bibinfo{person}{Feodor Dragan},
  \bibinfo{person}{Gagan Agrawal}, {and} \bibinfo{person}{Bin Ren}.}
  \bibinfo{year}{2020}\natexlab{}.
\newblock \showarticletitle{Parallelizing Pruned Landmark Labeling: Dealing
  with Dependencies in Graph Algorithms}. In \bibinfo{booktitle}{\emph{Proc.
  34th Annual Int. Conf. on Supercomputing}}. Article \bibinfo{articleno}{11},
  \bibinfo{numpages}{13}~pages.
\newblock
\showISBNx{9781450379830}
\urldef\tempurl%
\url{https://doi.org/10.1145/3392717.3392745}
\showDOI{\tempurl}


\bibitem[Jin et~al\mbox{.}(2012)]%
        {10.1145/2213836.2213856}
\bibfield{author}{\bibinfo{person}{Ruoming Jin}, \bibinfo{person}{Ning Ruan},
  \bibinfo{person}{Saikat Dey}, {and} \bibinfo{person}{Jeffrey~Yu Xu}.}
  \bibinfo{year}{2012}\natexlab{}.
\newblock \showarticletitle{SCARAB: Scaling Reachability Computation on Large
  Graphs}. In \bibinfo{booktitle}{\emph{Proc. ACM SIGMOD Int. Conf. on
  Management of Data}}. \bibinfo{pages}{169–180}.
\newblock
\showISBNx{9781450312479}
\urldef\tempurl%
\url{https://doi.org/10.1145/2213836.2213856}
\showDOI{\tempurl}


\bibitem[Jin et~al\mbox{.}(2011)]%
        {10.1145/1929934.1929941}
\bibfield{author}{\bibinfo{person}{Ruoming Jin}, \bibinfo{person}{Ning Ruan},
  \bibinfo{person}{Yang Xiang}, {and} \bibinfo{person}{Haixun Wang}.}
  \bibinfo{year}{2011}\natexlab{}.
\newblock \showarticletitle{Path-Tree: An Efficient Reachability Indexing
  Scheme for Large Directed Graphs}.
\newblock \bibinfo{journal}{\emph{ACM Trans. Database Syst.}}
  \bibinfo{volume}{36}, \bibinfo{number}{1}, Article \bibinfo{articleno}{7}
  (\bibinfo{year}{2011}), \bibinfo{numpages}{44}~pages.
\newblock
\showISSN{0362-5915}
\urldef\tempurl%
\url{https://doi.org/10.1145/1929934.1929941}
\showDOI{\tempurl}


\bibitem[Jin and Wang(2013)]%
        {10.14778/2556549.2556578}
\bibfield{author}{\bibinfo{person}{Ruoming Jin} {and} \bibinfo{person}{Guan
  Wang}.} \bibinfo{year}{2013}\natexlab{}.
\newblock \showarticletitle{Simple, Fast, and Scalable Reachability Oracle}.
\newblock \bibinfo{journal}{\emph{Proc. VLDB Endowment}} \bibinfo{volume}{6},
  \bibinfo{number}{14} (\bibinfo{year}{2013}), \bibinfo{pages}{1978–1989}.
\newblock
\showISSN{2150-8097}
\urldef\tempurl%
\url{https://doi.org/10.14778/2556549.2556578}
\showDOI{\tempurl}


\bibitem[Jin et~al\mbox{.}(2009)]%
        {10.1145/1559845.1559930}
\bibfield{author}{\bibinfo{person}{Ruoming Jin}, \bibinfo{person}{Yang Xiang},
  \bibinfo{person}{Ning Ruan}, {and} \bibinfo{person}{David Fuhry}.}
  \bibinfo{year}{2009}\natexlab{}.
\newblock \showarticletitle{{3-HOP}: A High-Compression Indexing Scheme for
  Reachability Query}. In \bibinfo{booktitle}{\emph{Proc. ACM SIGMOD Int. Conf.
  on Management of Data}}. \bibinfo{pages}{813–826}.
\newblock
\showISBNx{9781605585512}
\urldef\tempurl%
\url{https://doi.org/10.1145/1559845.1559930}
\showDOI{\tempurl}


\bibitem[Jin et~al\mbox{.}(2008)]%
        {10.1145/1376616.1376677}
\bibfield{author}{\bibinfo{person}{Ruoming Jin}, \bibinfo{person}{Yang Xiang},
  \bibinfo{person}{Ning Ruan}, {and} \bibinfo{person}{Haixun Wang}.}
  \bibinfo{year}{2008}\natexlab{}.
\newblock \showarticletitle{Efficiently Answering Reachability Queries on Very
  Large Directed Graphs}. In \bibinfo{booktitle}{\emph{Proc. ACM SIGMOD Int.
  Conf. on Management of Data}}. \bibinfo{pages}{595–608}.
\newblock
\showISBNx{9781605581026}
\urldef\tempurl%
\url{https://doi.org/10.1145/1376616.1376677}
\showDOI{\tempurl}


\bibitem[Koutrouli et~al\mbox{.}(2020)]%
        {koutrouli2020guide}
\bibfield{author}{\bibinfo{person}{Mikaela Koutrouli},
  \bibinfo{person}{Evangelos Karatzas}, \bibinfo{person}{David Paez-Espino},
  {and} \bibinfo{person}{Georgios~A Pavlopoulos}.}
  \bibinfo{year}{2020}\natexlab{}.
\newblock \showarticletitle{A Guide to Conquer the Biological Network Era Using
  Graph Theory}.
\newblock \bibinfo{journal}{\emph{Front. Bioeng. Biotechnol.}}
  \bibinfo{volume}{8} (\bibinfo{year}{2020}), \bibinfo{pages}{34}.
\newblock
\showISSN{2296-4185}
\urldef\tempurl%
\url{https://doi.org/10.3389/fbioe.2020.00034}
\showDOI{\tempurl}


\bibitem[Kuijpers et~al\mbox{.}(2021)]%
        {Kuijpers2021PathII}
\bibfield{author}{\bibinfo{person}{Jochem Kuijpers}, \bibinfo{person}{George
  Fletcher}, \bibinfo{person}{Tobias Lindaaker}, {and} \bibinfo{person}{Nikolay
  Yakovets}.} \bibinfo{year}{2021}\natexlab{}.
\newblock \showarticletitle{Path Indexing in the Cypher Query Pipeline}. In
  \bibinfo{booktitle}{\emph{Proc. 24th Int. Conf. on Extending Database
  Technology}}. \bibinfo{pages}{582--587}.
\newblock
\urldef\tempurl%
\url{https://openproceedings.org/2021/conf/edbt/p156.pdf}
\showURL{%
\tempurl}


\bibitem[Li et~al\mbox{.}(2019)]%
        {10.1145/3299869.3319877}
\bibfield{author}{\bibinfo{person}{Wentao Li}, \bibinfo{person}{Miao Qiao},
  \bibinfo{person}{Lu Qin}, \bibinfo{person}{Ying Zhang},
  \bibinfo{person}{Lijun Chang}, {and} \bibinfo{person}{Xuemin Lin}.}
  \bibinfo{year}{2019}\natexlab{}.
\newblock \showarticletitle{Scaling Distance Labeling on Small-World Networks}.
  In \bibinfo{booktitle}{\emph{Proc. ACM SIGMOD Int. Conf. on Management of
  Data}}. \bibinfo{pages}{1060–1077}.
\newblock
\showISBNx{9781450356435}
\urldef\tempurl%
\url{https://doi.org/10.1145/3299869.3319877}
\showDOI{\tempurl}


\bibitem[Libkin and Vrgo\v{c}(2012)]%
        {10.1145/2274576.2274585}
\bibfield{author}{\bibinfo{person}{Leonid Libkin} {and}
  \bibinfo{person}{Domagoj Vrgo\v{c}}.} \bibinfo{year}{2012}\natexlab{}.
\newblock \showarticletitle{Regular Path Queries on Graphs with Data}. In
  \bibinfo{booktitle}{\emph{Proc. 15th Int. Conf. on Database Theory}} (Berlin,
  Germany). \bibinfo{pages}{74–85}.
\newblock
\urldef\tempurl%
\url{https://doi.org/10.1145/2274576.2274585}
\showDOI{\tempurl}


\bibitem[Livi and Rizzi(2013)]%
        {10.1007/s10044-012-0284-8}
\bibfield{author}{\bibinfo{person}{Lorenzo Livi} {and}
  \bibinfo{person}{Antonello Rizzi}.} \bibinfo{year}{2013}\natexlab{}.
\newblock \showarticletitle{The Graph Matching Problem}.
\newblock \bibinfo{journal}{\emph{Pattern Anal. Appl.}} \bibinfo{volume}{16},
  \bibinfo{number}{3} (\bibinfo{date}{aug} \bibinfo{year}{2013}),
  \bibinfo{pages}{253–283}.
\newblock
\showISSN{1433-7541}
\urldef\tempurl%
\url{https://doi.org/10.1007/s10044-012-0284-8}
\showDOI{\tempurl}


\bibitem[Lyu et~al\mbox{.}(2021)]%
        {10.1007/978-3-030-73197-7_52}
\bibfield{author}{\bibinfo{person}{Qiuyi Lyu}, \bibinfo{person}{Yuchen Li},
  \bibinfo{person}{Bingsheng He}, {and} \bibinfo{person}{Bin Gong}.}
  \bibinfo{year}{2021}\natexlab{}.
\newblock \showarticletitle{DBL: Efficient Reachability Queries on Dynamic
  Graphs}. In \bibinfo{booktitle}{\emph{Proc. 26th Int. Conf. on Database
  Systems for Advanced Applications}}. \bibinfo{pages}{761--777}.
\newblock
\showISBNx{978-3-030-73197-7}
\urldef\tempurl%
\url{https://doi.org/10.1007/978-3-030-73197-7_52}
\showURL{%
\tempurl}


\bibitem[Mathur(2021)]%
        {financialg}
\bibfield{author}{\bibinfo{person}{Nav Mathur}.}
  \bibinfo{year}{2021}\natexlab{}.
\newblock \bibinfo{title}{White Paper: Neo4j for Financial Services}.
\newblock
  \bibinfo{howpublished}{\url{https://neo4j.com/whitepapers/financial-services-neo4j}}.
\newblock


\bibitem[Merz and Sanders(2014)]%
        {merz2014preach}
\bibfield{author}{\bibinfo{person}{Florian Merz} {and} \bibinfo{person}{Peter
  Sanders}.} \bibinfo{year}{2014}\natexlab{}.
\newblock \showarticletitle{PReaCH: A fast lightweight reachability index using
  pruning and contraction hierarchies}. In \bibinfo{booktitle}{\emph{In Proc.
  22th European Symp. on Algorithms}}. \bibinfo{pages}{701--712}.
\newblock
\urldef\tempurl%
\url{https://doi.org/10.1007/978-3-662-44777-2_58}
\showURL{%
\tempurl}


\bibitem[Mhedhbi and Salihoglu(2019)]%
        {10.14778/3342263.3342643}
\bibfield{author}{\bibinfo{person}{Amine Mhedhbi} {and} \bibinfo{person}{Semih
  Salihoglu}.} \bibinfo{year}{2019}\natexlab{}.
\newblock \showarticletitle{Optimizing subgraph queries by combining binary and
  worst-case optimal joins}.
\newblock \bibinfo{journal}{\emph{Proc. VLDB Endowment}} \bibinfo{volume}{12},
  \bibinfo{number}{11} (\bibinfo{date}{July} \bibinfo{year}{2019}),
  \bibinfo{pages}{1692–1704}.
\newblock
\showISSN{2150-8097}


\bibitem[Neumann and Weikum(2008)]%
        {10.14778/1453856.1453927}
\bibfield{author}{\bibinfo{person}{Thomas Neumann} {and}
  \bibinfo{person}{Gerhard Weikum}.} \bibinfo{year}{2008}\natexlab{}.
\newblock \showarticletitle{RDF-3X: A RISC-Style Engine for RDF}.
\newblock \bibinfo{journal}{\emph{Proc. VLDB Endowment}} \bibinfo{volume}{1},
  \bibinfo{number}{1} (\bibinfo{year}{2008}), \bibinfo{pages}{647–659}.
\newblock
\showISSN{2150-8097}
\urldef\tempurl%
\url{https://doi.org/10.14778/1453856.1453927}
\showDOI{\tempurl}


\bibitem[Newman(2010)]%
        {Newman2010}
\bibfield{author}{\bibinfo{person}{Mark Newman}.}
  \bibinfo{year}{2010}\natexlab{}.
\newblock \bibinfo{booktitle}{\emph{{Networks: An Introduction}}}.
\newblock \bibinfo{publisher}{Oxford University Press}.
\newblock
\showISBNx{9780199206650}
\urldef\tempurl%
\url{https://doi.org/10.1093/acprof:oso/9780199206650.001.0001}
\showDOI{\tempurl}


\bibitem[Pacaci et~al\mbox{.}(2020)]%
        {10.1145/3318464.3389733}
\bibfield{author}{\bibinfo{person}{Anil Pacaci}, \bibinfo{person}{Angela
  Bonifati}, {and} \bibinfo{person}{M.~Tamer \"{O}zsu}.}
  \bibinfo{year}{2020}\natexlab{}.
\newblock \showarticletitle{Regular Path Query Evaluation on Streaming Graphs}.
  In \bibinfo{booktitle}{\emph{Proc. ACM SIGMOD Int. Conf. on Management of
  Data}}. \bibinfo{pages}{1415–1430}.
\newblock
\showISBNx{9781450367356}
\urldef\tempurl%
\url{https://doi.org/10.1145/3318464.3389733}
\showDOI{\tempurl}


\bibitem[Pacaci et~al\mbox{.}(2022)]%
        {9835463}
\bibfield{author}{\bibinfo{person}{Anil Pacaci}, \bibinfo{person}{Angela
  Bonifati}, {and} \bibinfo{person}{M.~Tamer Özsu}.}
  \bibinfo{year}{2022}\natexlab{}.
\newblock \showarticletitle{Evaluating Complex Queries on Streaming Graphs}. In
  \bibinfo{booktitle}{\emph{Proc. 38th Int. Conf. on Data Engineering}}.
  \bibinfo{pages}{272--285}.
\newblock
\urldef\tempurl%
\url{https://doi.org/10.1109/ICDE53745.2022.00025}
\showDOI{\tempurl}


\bibitem[Peng et~al\mbox{.}(2020)]%
        {10.14778/3380750.3380753}
\bibfield{author}{\bibinfo{person}{You Peng}, \bibinfo{person}{Ying Zhang},
  \bibinfo{person}{Xuemin Lin}, \bibinfo{person}{Lu Qin}, {and}
  \bibinfo{person}{Wenjie Zhang}.} \bibinfo{year}{2020}\natexlab{}.
\newblock \showarticletitle{Answering Billion-Scale Label-Constrained
  Reachability Queries within Microsecond}.
\newblock \bibinfo{journal}{\emph{Proc. VLDB Endowment}} \bibinfo{volume}{13},
  \bibinfo{number}{6} (\bibinfo{year}{2020}), \bibinfo{pages}{812–825}.
\newblock
\showISSN{2150-8097}
\urldef\tempurl%
\url{https://doi.org/10.14778/3380750.3380753}
\showDOI{\tempurl}


\bibitem[Quer and Calabrese(2020)]%
        {computation8040103}
\bibfield{author}{\bibinfo{person}{Stefano Quer} {and} \bibinfo{person}{Andrea
  Calabrese}.} \bibinfo{year}{2020}\natexlab{}.
\newblock \showarticletitle{Graph Reachability on Parallel Many-Core
  Architectures}.
\newblock \bibinfo{journal}{\emph{Computation}} \bibinfo{volume}{8},
  \bibinfo{number}{4} (\bibinfo{year}{2020}).
\newblock


\bibitem[Riesen et~al\mbox{.}(2010)]%
        {Riesen2010}
\bibfield{author}{\bibinfo{person}{Kaspar Riesen}, \bibinfo{person}{Xiaoyi
  Jiang}, {and} \bibinfo{person}{Horst Bunke}.}
  \bibinfo{year}{2010}\natexlab{}.
\newblock \showarticletitle{Exact and Inexact Graph Matching: Methodology and
  Applications}.
\newblock In \bibinfo{booktitle}{\emph{Managing and Mining Graph Data}}.
  Vol.~\bibinfo{volume}{40}. \bibinfo{pages}{217--247}.
\newblock
\showISBNx{978-1-4419-6045-0}
\urldef\tempurl%
\url{https://doi.org/10.1007/978-1-4419-6045-0_7}
\showDOI{\tempurl}


\bibitem[Roditty(2013)]%
        {10.5555/2627817.2627899}
\bibfield{author}{\bibinfo{person}{Liam Roditty}.}
  \bibinfo{year}{2013}\natexlab{}.
\newblock \showarticletitle{Decremental Maintenance of Strongly Connected
  Components}. In \bibinfo{booktitle}{\emph{Proc. 24th Annual ACM-SIAM Symp. on
  Discrete Algorithms}}. \bibinfo{pages}{1143–1150}.
\newblock
\showISBNx{9781611972511}
\urldef\tempurl%
\url{https://doi.org/10.1137/1.9781611973105.82}
\showURL{%
\tempurl}


\bibitem[Roditty and Zwick(2004)]%
        {10.1145/1007352.1007387}
\bibfield{author}{\bibinfo{person}{Liam Roditty} {and} \bibinfo{person}{Uri
  Zwick}.} \bibinfo{year}{2004}\natexlab{}.
\newblock \showarticletitle{A Fully Dynamic Reachability Algorithm for Directed
  Graphs with an Almost Linear Update Time}. In \bibinfo{booktitle}{\emph{Proc.
  36th Annual ACM Symp. on Theory of Computing}}. \bibinfo{pages}{184–191}.
\newblock
\showISBNx{1581138520}
\urldef\tempurl%
\url{https://doi.org/10.1145/1007352.1007387}
\showDOI{\tempurl}


\bibitem[Sahu et~al\mbox{.}(2017)]%
        {10.1145/3186728.3164139}
\bibfield{author}{\bibinfo{person}{Siddhartha Sahu}, \bibinfo{person}{Amine
  Mhedhbi}, \bibinfo{person}{Semih Salihoglu}, \bibinfo{person}{Jimmy Lin},
  {and} \bibinfo{person}{M.~Tamer \"{O}zsu}.} \bibinfo{year}{2017}\natexlab{}.
\newblock \showarticletitle{The Ubiquity of Large Graphs and Surprising
  Challenges of Graph Processing}.
\newblock \bibinfo{journal}{\emph{Proc. VLDB Endowment}} \bibinfo{volume}{11},
  \bibinfo{number}{4} (\bibinfo{year}{2017}), \bibinfo{pages}{420–431}.
\newblock
\showISSN{2150-8097}
\urldef\tempurl%
\url{https://doi.org/10.1145/3186728.3164139}
\showDOI{\tempurl}


\bibitem[Sahu et~al\mbox{.}(2020)]%
        {sahu2020ubiquity}
\bibfield{author}{\bibinfo{person}{Siddhartha Sahu}, \bibinfo{person}{Amine
  Mhedhbi}, \bibinfo{person}{Semih Salihoglu}, \bibinfo{person}{Jimmy Lin},
  {and} \bibinfo{person}{M~Tamer {\"O}zsu}.} \bibinfo{year}{2020}\natexlab{}.
\newblock \showarticletitle{The ubiquity of large graphs and surprising
  challenges of graph processing: extended survey}.
\newblock \bibinfo{journal}{\emph{VLDB J.}} \bibinfo{volume}{29},
  \bibinfo{number}{2} (\bibinfo{year}{2020}), \bibinfo{pages}{595--618}.
\newblock
\urldef\tempurl%
\url{https://doi.org/10.1007/s00778-019-00548-x}
\showURL{%
\tempurl}


\bibitem[Sakr et~al\mbox{.}(2021)]%
        {SakrBVIAAAABBDV21}
\bibfield{author}{\bibinfo{person}{Sherif Sakr}, \bibinfo{person}{Angela
  Bonifati}, \bibinfo{person}{Hannes Voigt}, \bibinfo{person}{Alexandru Iosup},
  \bibinfo{person}{Khaled Ammar}, \bibinfo{person}{Renzo Angles},
  \bibinfo{person}{Walid~G. Aref}, \bibinfo{person}{Marcelo Arenas},
  \bibinfo{person}{Maciej Besta}, \bibinfo{person}{Peter~A. Boncz},
  \bibinfo{person}{Khuzaima Daudjee}, \bibinfo{person}{Emanuele~Della Valle},
  \bibinfo{person}{Stefania Dumbrava}, \bibinfo{person}{Olaf Hartig},
  \bibinfo{person}{Bernhard Haslhofer}, \bibinfo{person}{Tim Hegeman},
  \bibinfo{person}{Jan Hidders}, \bibinfo{person}{Katja Hose},
  \bibinfo{person}{Adriana Iamnitchi}, \bibinfo{person}{Vasiliki Kalavri},
  \bibinfo{person}{Hugo Kapp}, \bibinfo{person}{Wim Martens},
  \bibinfo{person}{M.~Tamer {\"{O}}zsu}, \bibinfo{person}{Eric Peukert},
  \bibinfo{person}{Stefan Plantikow}, \bibinfo{person}{Mohamed Ragab},
  \bibinfo{person}{Matei Ripeanu}, \bibinfo{person}{Semih Salihoglu},
  \bibinfo{person}{Christian Schulz}, \bibinfo{person}{Petra Selmer},
  \bibinfo{person}{Juan~F. Sequeda}, \bibinfo{person}{Joshua Shinavier},
  \bibinfo{person}{G{\'{a}}bor Sz{\'{a}}rnyas}, \bibinfo{person}{Riccardo
  Tommasini}, \bibinfo{person}{Antonino Tumeo}, \bibinfo{person}{Alexandru
  Uta}, \bibinfo{person}{Ana~Lucia Varbanescu}, \bibinfo{person}{Hsiang{-}Yun
  Wu}, \bibinfo{person}{Nikolay Yakovets}, \bibinfo{person}{Da Yan}, {and}
  \bibinfo{person}{Eiko Yoneki}.} \bibinfo{year}{2021}\natexlab{}.
\newblock \showarticletitle{The future is big graphs: A community view on graph
  processing systems}.
\newblock \bibinfo{journal}{\emph{Commun. ACM}} \bibinfo{volume}{64},
  \bibinfo{number}{9} (\bibinfo{year}{2021}), \bibinfo{pages}{62--71}.
\newblock
\urldef\tempurl%
\url{https://doi.org/10.1145/3434642}
\showURL{%
\tempurl}


\bibitem[Schenkel et~al\mbox{.}(2005)]%
        {1410143}
\bibfield{author}{\bibinfo{person}{R. Schenkel}, \bibinfo{person}{A. Theobald},
  {and} \bibinfo{person}{G. Weikum}.} \bibinfo{year}{2005}\natexlab{}.
\newblock \showarticletitle{Efficient creation and incremental maintenance of
  the HOPI index for complex XML document collections}. In
  \bibinfo{booktitle}{\emph{Proc. 21st Int. Conf. on Data Engineering}}.
  \bibinfo{pages}{360--371}.
\newblock
\urldef\tempurl%
\url{https://doi.org/10.1109/ICDE.2005.57}
\showDOI{\tempurl}


\bibitem[Sengupta et~al\mbox{.}(2019)]%
        {8731546}
\bibfield{author}{\bibinfo{person}{Neha Sengupta}, \bibinfo{person}{Amitabha
  Bagchi}, \bibinfo{person}{Maya Ramanath}, {and} \bibinfo{person}{Srikanta
  Bedathur}.} \bibinfo{year}{2019}\natexlab{}.
\newblock \showarticletitle{ARROW: Approximating Reachability Using Random
  Walks Over Web-Scale Graphs}. In \bibinfo{booktitle}{\emph{Proc. 35th Int.
  Conf. on Data Engineering}}. \bibinfo{pages}{470--481}.
\newblock
\urldef\tempurl%
\url{https://doi.org/10.1109/ICDE.2019.00049}
\showDOI{\tempurl}


\bibitem[{Seufert} et~al\mbox{.}(2013)]%
        {6544893}
\bibfield{author}{\bibinfo{person}{S. {Seufert}}, \bibinfo{person}{A. {Anand}},
  \bibinfo{person}{S. {Bedathur}}, {and} \bibinfo{person}{G. {Weikum}}.}
  \bibinfo{year}{2013}\natexlab{}.
\newblock \showarticletitle{FERRARI: Flexible and efficient reachability range
  assignment for graph indexing}. In \bibinfo{booktitle}{\emph{Proc. 29th Int.
  Conf. on Data Engineering}}. \bibinfo{pages}{1009--1020}.
\newblock
\urldef\tempurl%
\url{https://doi.org/10.1109/ICDE.2013.6544893}
\showDOI{\tempurl}


\bibitem[Shi et~al\mbox{.}(2018)]%
        {10.1145/3128571}
\bibfield{author}{\bibinfo{person}{Xuanhua Shi}, \bibinfo{person}{Zhigao
  Zheng}, \bibinfo{person}{Yongluan Zhou}, \bibinfo{person}{Hai Jin},
  \bibinfo{person}{Ligang He}, \bibinfo{person}{Bo Liu}, {and}
  \bibinfo{person}{Qiang-Sheng Hua}.} \bibinfo{year}{2018}\natexlab{}.
\newblock \showarticletitle{Graph Processing on GPUs: A Survey}.
\newblock \bibinfo{journal}{\emph{ACM Comput. Surv.}} \bibinfo{volume}{50},
  \bibinfo{number}{6}, Article \bibinfo{articleno}{81} (\bibinfo{date}{Jan.}
  \bibinfo{year}{2018}), \bibinfo{numpages}{35}~pages.
\newblock


\bibitem[Shun and Blelloch(2013)]%
        {10.1145/2442516.2442530}
\bibfield{author}{\bibinfo{person}{Julian Shun} {and} \bibinfo{person}{Guy~E.
  Blelloch}.} \bibinfo{year}{2013}\natexlab{}.
\newblock \showarticletitle{Ligra: a lightweight graph processing framework for
  shared memory}. In \bibinfo{booktitle}{\emph{Proc. 18th ACM SIGPLAN Symp. on
  Principles and Practice of Parallel Programming}}
  \emph{(\bibinfo{series}{PPoPP '13})}. \bibinfo{pages}{135–146}.
\newblock


\bibitem[Steve and Andy(2013)]%
        {sparql}
\bibfield{author}{\bibinfo{person}{Harris Steve} {and}
  \bibinfo{person}{Seaborne Andy}.} \bibinfo{year}{2013}\natexlab{}.
\newblock \bibinfo{title}{SPARQL 1.1 Query Language}.
\newblock
  \bibinfo{howpublished}{\url{https://www.w3.org/TR/2013/REC-sparql11-overview-20130321/}}.
\newblock


\bibitem[{Su} et~al\mbox{.}(2017)]%
        {7750623}
\bibfield{author}{\bibinfo{person}{J. {Su}}, \bibinfo{person}{Q. {Zhu}},
  \bibinfo{person}{H. {Wei}}, {and} \bibinfo{person}{J.~X. {Yu}}.}
  \bibinfo{year}{2017}\natexlab{}.
\newblock \showarticletitle{Reachability Querying: Can It Be Even Faster?}
\newblock \bibinfo{journal}{\emph{IEEE Trans. Knowl. and Data Eng.}}
  \bibinfo{volume}{29}, \bibinfo{number}{3} (\bibinfo{year}{2017}),
  \bibinfo{pages}{683--697}.
\newblock
\urldef\tempurl%
\url{https://doi.org/10.1109/TKDE.2016.2631160}
\showDOI{\tempurl}


\bibitem[Sumrall et~al\mbox{.}(2017)]%
        {10.1007/978-3-319-58943-5_43}
\bibfield{author}{\bibinfo{person}{Jonathan~M. Sumrall},
  \bibinfo{person}{George H.~L. Fletcher}, \bibinfo{person}{Alexandra
  Poulovassilis}, \bibinfo{person}{Johan Svensson}, \bibinfo{person}{Magnus
  Vejlstrup}, \bibinfo{person}{Chris Vest}, {and} \bibinfo{person}{Jim
  Webber}.} \bibinfo{year}{2017}\natexlab{}.
\newblock \showarticletitle{Investigations on Path Indexing for Graph
  Databases}. In \bibinfo{booktitle}{\emph{Euro-Par 2016: Parallel Processing
  Workshops}}. \bibinfo{pages}{532--544}.
\newblock
\showISBNx{978-3-319-58943-5}


\bibitem[Sundaram et~al\mbox{.}(2015)]%
        {10.14778/2809974.2809983}
\bibfield{author}{\bibinfo{person}{Narayanan Sundaram},
  \bibinfo{person}{Nadathur Satish}, \bibinfo{person}{Md~Mostofa~Ali Patwary},
  \bibinfo{person}{Subramanya~R. Dulloor}, \bibinfo{person}{Michael~J.
  Anderson}, \bibinfo{person}{Satya~Gautam Vadlamudi},
  \bibinfo{person}{Dipankar Das}, {and} \bibinfo{person}{Pradeep Dubey}.}
  \bibinfo{year}{2015}\natexlab{}.
\newblock \showarticletitle{GraphMat: high performance graph analytics made
  productive}.
\newblock \bibinfo{journal}{\emph{Proc. VLDB Endowment}} \bibinfo{volume}{8},
  \bibinfo{number}{11} (\bibinfo{date}{July} \bibinfo{year}{2015}),
  \bibinfo{pages}{1214–1225}.
\newblock


\bibitem[Tarjan(1972)]%
        {tarjan1972depth}
\bibfield{author}{\bibinfo{person}{Robert Tarjan}.}
  \bibinfo{year}{1972}\natexlab{}.
\newblock \showarticletitle{Depth-first search and linear graph algorithms}.
\newblock \bibinfo{journal}{\emph{SIAM J. on Comput.}} \bibinfo{volume}{1},
  \bibinfo{number}{2} (\bibinfo{year}{1972}), \bibinfo{pages}{146--160}.
\newblock
\urldef\tempurl%
\url{https://doi.org/10.1137/0201010}
\showURL{%
\tempurl}


\bibitem[Team(2016)]%
        {openCypher}
\bibfield{author}{\bibinfo{person}{The~Neo4j Team}.}
  \bibinfo{year}{2016}\natexlab{}.
\newblock \bibinfo{title}{openCypher}.
\newblock \bibinfo{howpublished}{\url{http://www.opencypher.org}}.
\newblock


\bibitem[ten Wolde et~al\mbox{.}(2023)]%
        {Wolde2023DuckPGQEP}
\bibfield{author}{\bibinfo{person}{Daniel ten Wolde}, \bibinfo{person}{Tavneet
  Singh}, \bibinfo{person}{G{\'a}bor Sz{\'a}rnyas}, {and}
  \bibinfo{person}{Peter~A. Boncz}.} \bibinfo{year}{2023}\natexlab{}.
\newblock \showarticletitle{DuckPGQ: Efficient Property Graph Queries in an
  analytical RDBMS}. In \bibinfo{booktitle}{\emph{Proc. 13th Biennial Conf. on
  Innovative Data Systems Research}}.
\newblock
\urldef\tempurl%
\url{https://www.cidrdb.org/cidr2023/papers/p66-wolde.pdf}
\showURL{%
\tempurl}


\bibitem[Tri\ss{}l and Leser(2007)]%
        {10.1145/1247480.1247573}
\bibfield{author}{\bibinfo{person}{Silke Tri\ss{}l} {and} \bibinfo{person}{Ulf
  Leser}.} \bibinfo{year}{2007}\natexlab{}.
\newblock \showarticletitle{Fast and Practical Indexing and Querying of Very
  Large Graphs}. In \bibinfo{booktitle}{\emph{Proc. ACM SIGMOD Int. Conf. on
  Management of Data}}. \bibinfo{pages}{845–856}.
\newblock
\showISBNx{9781595936868}
\urldef\tempurl%
\url{https://doi.org/10.1145/1247480.1247573}
\showDOI{\tempurl}


\bibitem[Valstar et~al\mbox{.}(2017)]%
        {10.1145/3035918.3035955}
\bibfield{author}{\bibinfo{person}{Lucien~D.J. Valstar},
  \bibinfo{person}{George~H.L. Fletcher}, {and} \bibinfo{person}{Yuichi
  Yoshida}.} \bibinfo{year}{2017}\natexlab{}.
\newblock \showarticletitle{Landmark Indexing for Evaluation of
  Label-Constrained Reachability Queries}. In \bibinfo{booktitle}{\emph{Proc.
  ACM SIGMOD Int. Conf. on Management of Data}}. \bibinfo{pages}{345–358}.
\newblock
\showISBNx{9781450341974}
\urldef\tempurl%
\url{https://doi.org/10.1145/3035918.3035955}
\showDOI{\tempurl}


\bibitem[Veloso et~al\mbox{.}(2014)]%
        {Veloso2014ReachabilityQI}
\bibfield{author}{\bibinfo{person}{Ren{\^{e}}~Rodrigues Veloso},
  \bibinfo{person}{Lo{\"{\i}}c Cerf}, \bibinfo{person}{Wagner~Meira Jr.}, {and}
  \bibinfo{person}{Mohammed~J. Zaki}.} \bibinfo{year}{2014}\natexlab{}.
\newblock \showarticletitle{Reachability Queries in Very Large Graphs: A Fast
  Refined Online Search Approach}. In \bibinfo{booktitle}{\emph{Proc. 17th Int.
  Conf. on Extending Database Technology}}. \bibinfo{pages}{511--522}.
\newblock
\urldef\tempurl%
\url{https://openproceedings.org/EDBT/2014/paper_166.pdf}
\showURL{%
\tempurl}


\bibitem[Wang et~al\mbox{.}(2023b)]%
        {10.1145/3589259}
\bibfield{author}{\bibinfo{person}{Letong Wang}, \bibinfo{person}{Xiaojun
  Dong}, \bibinfo{person}{Yan Gu}, {and} \bibinfo{person}{Yihan Sun}.}
  \bibinfo{year}{2023}\natexlab{b}.
\newblock \showarticletitle{Parallel Strong Connectivity Based on Faster
  Reachability}.
\newblock \bibinfo{journal}{\emph{Proc. ACM Manag. Data}} \bibinfo{volume}{1},
  \bibinfo{number}{2}, Article \bibinfo{articleno}{114} (\bibinfo{date}{June}
  \bibinfo{year}{2023}), \bibinfo{numpages}{29}~pages.
\newblock
\urldef\tempurl%
\url{https://doi.org/10.1145/3589259}
\showDOI{\tempurl}


\bibitem[Wang et~al\mbox{.}(2023a)]%
        {10184762}
\bibfield{author}{\bibinfo{person}{Qiange Wang}, \bibinfo{person}{Xin Ai},
  \bibinfo{person}{Yanfeng Zhang}, \bibinfo{person}{Jing Chen}, {and}
  \bibinfo{person}{Ge Yu}.} \bibinfo{year}{2023}\natexlab{a}.
\newblock \showarticletitle{HyTGraph: GPU-Accelerated Graph Processing with
  Hybrid Transfer Management}. In \bibinfo{booktitle}{\emph{Proc. 39th Int.
  Conf. on Data Engineering}}. \bibinfo{pages}{558--571}.
\newblock


\bibitem[Wei et~al\mbox{.}(2014)]%
        {10.14778/2732977.2732992}
\bibfield{author}{\bibinfo{person}{Hao Wei}, \bibinfo{person}{Jeffrey~Xu Yu},
  \bibinfo{person}{Can Lu}, {and} \bibinfo{person}{Ruoming Jin}.}
  \bibinfo{year}{2014}\natexlab{}.
\newblock \showarticletitle{Reachability Querying: An Independent Permutation
  Labeling Approach}.
\newblock \bibinfo{journal}{\emph{Proc. VLDB Endowment}} \bibinfo{volume}{7},
  \bibinfo{number}{12} (\bibinfo{year}{2014}), \bibinfo{pages}{1191–1202}.
\newblock
\showISSN{2150-8097}
\urldef\tempurl%
\url{https://doi.org/10.14778/2732977.2732992}
\showDOI{\tempurl}


\bibitem[Wei et~al\mbox{.}(2018)]%
        {10.1007/s00778-017-0468-3}
\bibfield{author}{\bibinfo{person}{Hao Wei}, \bibinfo{person}{Jeffrey~Xu Yu},
  \bibinfo{person}{Can Lu}, {and} \bibinfo{person}{Ruoming Jin}.}
  \bibinfo{year}{2018}\natexlab{}.
\newblock \showarticletitle{Reachability Querying: An Independent Permutation
  Labeling Approach}.
\newblock \bibinfo{journal}{\emph{VLDB J.}} \bibinfo{volume}{27},
  \bibinfo{number}{1} (\bibinfo{year}{2018}), \bibinfo{pages}{1–26}.
\newblock
\showISSN{1066-8888}
\urldef\tempurl%
\url{https://doi.org/10.1007/s00778-017-0468-3}
\showDOI{\tempurl}


\bibitem[Weller(2014)]%
        {weller2014optimal}
\bibfield{author}{\bibinfo{person}{Mathias Weller}.}
  \bibinfo{year}{2014}\natexlab{}.
\newblock \bibinfo{title}{Optimal Hub Labeling is NP-complete}.
\newblock
\newblock
\showeprint[arxiv]{1407.8373}~[cs.CC]


\bibitem[Wood(2012)]%
        {10.1145/2206869.2206879}
\bibfield{author}{\bibinfo{person}{Peter~T. Wood}.}
  \bibinfo{year}{2012}\natexlab{}.
\newblock \showarticletitle{Query Languages for Graph Databases}.
\newblock \bibinfo{journal}{\emph{ACM SIGMOD Rec.}} \bibinfo{volume}{41},
  \bibinfo{number}{1} (\bibinfo{year}{2012}), \bibinfo{pages}{50–60}.
\newblock
\showISSN{0163-5808}
\urldef\tempurl%
\url{https://doi.org/10.1145/2206869.2206879}
\showDOI{\tempurl}


\bibitem[Xu et~al\mbox{.}(2011)]%
        {10.1145/2063576.2063807}
\bibfield{author}{\bibinfo{person}{Kun Xu}, \bibinfo{person}{Lei Zou},
  \bibinfo{person}{Jeffery~Xu Yu}, \bibinfo{person}{Lei Chen},
  \bibinfo{person}{Yanghua Xiao}, {and} \bibinfo{person}{Dongyan Zhao}.}
  \bibinfo{year}{2011}\natexlab{}.
\newblock \showarticletitle{Answering Label-Constraint Reachability in Large
  Graphs}. In \bibinfo{booktitle}{\emph{Proc. 20th ACM Int. Conf. on
  Information and Knowledge Management}}. \bibinfo{pages}{1595–1600}.
\newblock
\showISBNx{9781450307178}
\urldef\tempurl%
\url{https://doi.org/10.1145/2063576.2063807}
\showDOI{\tempurl}


\bibitem[Yan et~al\mbox{.}(2016)]%
        {10.1145/2911996.2912035}
\bibfield{author}{\bibinfo{person}{Junchi Yan}, \bibinfo{person}{Xu-Cheng Yin},
  \bibinfo{person}{Weiyao Lin}, \bibinfo{person}{Cheng Deng},
  \bibinfo{person}{Hongyuan Zha}, {and} \bibinfo{person}{Xiaokang Yang}.}
  \bibinfo{year}{2016}\natexlab{}.
\newblock \showarticletitle{A Short Survey of Recent Advances in Graph
  Matching}. In \bibinfo{booktitle}{\emph{Proc. of the 2016 ACM on Int. Conf.
  on Multimed. Retr.}} \bibinfo{pages}{167–174}.
\newblock
\showISBNx{9781450343596}
\urldef\tempurl%
\url{https://doi.org/10.1145/2911996.2912035}
\showDOI{\tempurl}


\bibitem[Yang et~al\mbox{.}(2019)]%
        {10.14778/3364324.3364329}
\bibfield{author}{\bibinfo{person}{Bohua Yang}, \bibinfo{person}{Dong Wen},
  \bibinfo{person}{Lu Qin}, \bibinfo{person}{Ying Zhang}, \bibinfo{person}{Xubo
  Wang}, {and} \bibinfo{person}{Xuemin Lin}.} \bibinfo{year}{2019}\natexlab{}.
\newblock \showarticletitle{Fully Dynamic Depth-First Search in Directed
  Graphs}.
\newblock \bibinfo{journal}{\emph{Proc. VLDB Endowment}} \bibinfo{volume}{13},
  \bibinfo{number}{2} (\bibinfo{date}{oct} \bibinfo{year}{2019}),
  \bibinfo{pages}{142–154}.
\newblock
\showISSN{2150-8097}
\urldef\tempurl%
\url{https://doi.org/10.14778/3364324.3364329}
\showDOI{\tempurl}


\bibitem[Yano et~al\mbox{.}(2013)]%
        {10.1145/2505515.2505724}
\bibfield{author}{\bibinfo{person}{Yosuke Yano}, \bibinfo{person}{Takuya
  Akiba}, \bibinfo{person}{Yoichi Iwata}, {and} \bibinfo{person}{Yuichi
  Yoshida}.} \bibinfo{year}{2013}\natexlab{}.
\newblock \showarticletitle{Fast and Scalable Reachability Queries on Graphs by
  Pruned Labeling with Landmarks and Paths}. In \bibinfo{booktitle}{\emph{Proc.
  22nd ACM Int. Conf. on Information and Knowledge Management}}.
  \bibinfo{pages}{1601–1606}.
\newblock
\showISBNx{9781450322638}
\urldef\tempurl%
\url{https://doi.org/10.1145/2505515.2505724}
\showDOI{\tempurl}


\bibitem[Yildirim et~al\mbox{.}(2010)]%
        {10.14778/1920841.1920879}
\bibfield{author}{\bibinfo{person}{Hilmi Yildirim}, \bibinfo{person}{Vineet
  Chaoji}, {and} \bibinfo{person}{Mohammed~J. Zaki}.}
  \bibinfo{year}{2010}\natexlab{}.
\newblock \showarticletitle{GRAIL: Scalable Reachability Index for Large
  Graphs}.
\newblock \bibinfo{journal}{\emph{Proc. VLDB Endowment}} \bibinfo{volume}{3},
  \bibinfo{number}{1–2} (\bibinfo{year}{2010}), \bibinfo{pages}{276–284}.
\newblock
\showISSN{2150-8097}
\urldef\tempurl%
\url{https://doi.org/10.14778/1920841.1920879}
\showDOI{\tempurl}


\bibitem[Yildirim et~al\mbox{.}(2013)]%
        {Yildirim2013DAGGERAS}
\bibfield{author}{\bibinfo{person}{Hilmi Yildirim}, \bibinfo{person}{Vineet
  Chaoji}, {and} \bibinfo{person}{Mohammed~J. Zaki}.}
  \bibinfo{year}{2013}\natexlab{}.
\newblock \bibinfo{title}{DAGGER: A Scalable Index for Reachability Queries in
  Large Dynamic Graphs}.
\newblock
\newblock
\showeprint[arxiv]{1301.0977}


\bibitem[Yu and Cheng(2010)]%
        {yu2010graph}
\bibfield{author}{\bibinfo{person}{Jeffrey~Xu Yu} {and}
  \bibinfo{person}{Jiefeng Cheng}.} \bibinfo{year}{2010}\natexlab{}.
\newblock \showarticletitle{Graph Reachability Queries: A Survey}.
\newblock In \bibinfo{booktitle}{\emph{Managing and Mining Graph Data}}.
  Vol.~\bibinfo{volume}{40}. \bibinfo{pages}{181--215}.
\newblock
\showISBNx{978-1-4419-6045-0}
\urldef\tempurl%
\url{https://doi.org/10.1007/978-1-4419-6045-0_6}
\showDOI{\tempurl}


\bibitem[Yue et~al\mbox{.}(2023)]%
        {ICDE55515.2023.00172}
\bibfield{author}{\bibinfo{person}{Pang Yue}, \bibinfo{person}{Zou Lei}, {and}
  \bibinfo{person}{Liu Yu}.} \bibinfo{year}{2023}\natexlab{}.
\newblock \showarticletitle{IFCA: Index-Free Community-Aware Reachability
  Processing Over Large Dynamic Graphs}. In \bibinfo{booktitle}{\emph{Proc.
  39th Int. Conf. on Data Engineering}}. \bibinfo{pages}{2211--2225}.
\newblock
\urldef\tempurl%
\url{https://doi.org/10.1109/ICDE55515.2023.00172}
\showDOI{\tempurl}


\bibitem[Zeng et~al\mbox{.}(2022)]%
        {10.1007/978-3-031-00129-1_37}
\bibfield{author}{\bibinfo{person}{Li Zeng}, \bibinfo{person}{Jinhua Zhou},
  \bibinfo{person}{Shijun Qin}, \bibinfo{person}{Haoran Cai},
  \bibinfo{person}{Rongqian Zhao}, {and} \bibinfo{person}{Xin Chen}.}
  \bibinfo{year}{2022}\natexlab{}.
\newblock \showarticletitle{SQLG+: Efficient k-hop Query Processing on RDBMS}.
  In \bibinfo{booktitle}{\emph{Proc. 27th Int. Conf. on Database Systems for
  Advanced Applications}}. \bibinfo{pages}{430--442}.
\newblock
\showISBNx{978-3-031-00129-1}


\bibitem[Zhang et~al\mbox{.}(2023c)]%
        {10.1109/ICDE55515.2023.00013}
\bibfield{author}{\bibinfo{person}{Chao Zhang}, \bibinfo{person}{Angela
  Bonifati}, \bibinfo{person}{Hugo Kapp}, \bibinfo{person}{Vlad~Ioan Haprian},
  {and} \bibinfo{person}{Jean-Pierre Lozi}.} \bibinfo{year}{2023}\natexlab{c}.
\newblock \showarticletitle{A Reachability Index for Recursive
  Label-Concatenated Graph Queries}. In \bibinfo{booktitle}{\emph{Proc. 39th
  Int. Conf. on Data Engineering}}. \bibinfo{pages}{66--80}.
\newblock
\urldef\tempurl%
\url{https://doi.org/10.1109/ICDE55515.2023.00013}
\showDOI{\tempurl}


\bibitem[Zhang et~al\mbox{.}(2023a)]%
        {10.1145/3555041.3589408}
\bibfield{author}{\bibinfo{person}{Chao Zhang}, \bibinfo{person}{Angela
  Bonifati}, {and} \bibinfo{person}{M.~Tamer \"{O}zsu}.}
  \bibinfo{year}{2023}\natexlab{a}.
\newblock \showarticletitle{An Overview of Reachability Indexes on Graphs}. In
  \bibinfo{booktitle}{\emph{Companion of ACM SIGMOD Int. Conf. on Management of
  Data}}. \bibinfo{pages}{61–68}.
\newblock
\showISBNx{9781450395076}
\urldef\tempurl%
\url{https://doi.org/10.1145/3555041.3589408}
\showDOI{\tempurl}


\bibitem[Zhang et~al\mbox{.}(2024a)]%
        {10.14778/3675034.3675040}
\bibfield{author}{\bibinfo{person}{Chao Zhang}, \bibinfo{person}{Angela
  Bonifati}, {and} \bibinfo{person}{M.~Tamer \"{O}zsu}.}
  \bibinfo{year}{2024}\natexlab{a}.
\newblock \showarticletitle{Incremental Sliding Window Connectivity over
  Streaming Graphs}.
\newblock \bibinfo{journal}{\emph{Proc. VLDB Endowment}} \bibinfo{volume}{17},
  \bibinfo{number}{10} (\bibinfo{date}{June} \bibinfo{year}{2024}),
  \bibinfo{pages}{2473–2486}.
\newblock


\bibitem[Zhang et~al\mbox{.}(2023b)]%
        {zhang2023indexingtechniquesgraphreachability}
\bibfield{author}{\bibinfo{person}{Chao Zhang}, \bibinfo{person}{Angela
  Bonifati}, {and} \bibinfo{person}{M.~Tamer Özsu}.}
  \bibinfo{year}{2023}\natexlab{b}.
\newblock \bibinfo{title}{Indexing Techniques for Graph Reachability Queries}.
\newblock
\newblock
\showeprint[arxiv]{2311.03542}~[cs.DB]
\urldef\tempurl%
\url{https://arxiv.org/abs/2311.03542}
\showURL{%
\tempurl}


\bibitem[Zhang et~al\mbox{.}(2025)]%
        {zhang2025lowlatencyslidingwindowconnectivity}
\bibfield{author}{\bibinfo{person}{Chao Zhang}, \bibinfo{person}{Angela
  Bonifati}, {and} \bibinfo{person}{Tamer Özsu}.}
  \bibinfo{year}{2025}\natexlab{}.
\newblock \bibinfo{title}{Low-Latency Sliding Window Connectivity}.
\newblock
\newblock
\showeprint[arxiv]{2410.00884}~[cs.DB]
\urldef\tempurl%
\url{https://arxiv.org/abs/2410.00884}
\showURL{%
\tempurl}


\bibitem[Zhang et~al\mbox{.}(2024b)]%
        {10.1145/3639260}
\bibfield{author}{\bibinfo{person}{Siyuan Zhang}, \bibinfo{person}{Zhenying
  He}, \bibinfo{person}{Yinan Jing}, \bibinfo{person}{Kai Zhang}, {and}
  \bibinfo{person}{X.~Sean Wang}.} \bibinfo{year}{2024}\natexlab{b}.
\newblock \showarticletitle{MWP: Multi-Window Parallel Evaluation of Regular
  Path Queries on Streaming Graphs}.
\newblock \bibinfo{journal}{\emph{Proc. ACM Manag. Data}} \bibinfo{volume}{2},
  \bibinfo{number}{1}, Article \bibinfo{articleno}{5} (\bibinfo{date}{March}
  \bibinfo{year}{2024}), \bibinfo{numpages}{26}~pages.
\newblock


\bibitem[Zhou et~al\mbox{.}(2023)]%
        {zhou2021fast}
\bibfield{author}{\bibinfo{person}{Junfeng Zhou}, \bibinfo{person}{Jeffrey~Xu
  Yu}, \bibinfo{person}{Yaxian Qiu}, \bibinfo{person}{Xian Tang},
  \bibinfo{person}{Ziyang Chen}, {and} \bibinfo{person}{Ming Du}.}
  \bibinfo{year}{2023}\natexlab{}.
\newblock \showarticletitle{Fast Reachability Queries Answering Based on
  $\mathsf{RCN}$RCN Reduction}.
\newblock \bibinfo{journal}{\emph{IEEE Trans. Knowl. and Data Eng.}}
  \bibinfo{volume}{35}, \bibinfo{number}{3} (\bibinfo{year}{2023}),
  \bibinfo{pages}{2590--2609}.
\newblock
\urldef\tempurl%
\url{https://doi.org/10.1109/TKDE.2021.3108433}
\showDOI{\tempurl}


\bibitem[Zhou et~al\mbox{.}(2017)]%
        {zhou2017dag}
\bibfield{author}{\bibinfo{person}{Junfeng Zhou}, \bibinfo{person}{Shijie
  Zhou}, \bibinfo{person}{Jeffrey~Xu Yu}, \bibinfo{person}{Hao Wei},
  \bibinfo{person}{Ziyang Chen}, {and} \bibinfo{person}{Xian Tang}.}
  \bibinfo{year}{2017}\natexlab{}.
\newblock \showarticletitle{DAG Reduction: Fast Answering Reachability
  Queries}. In \bibinfo{booktitle}{\emph{Proc. ACM SIGMOD Int. Conf. on
  Management of Data}}. \bibinfo{pages}{375–390}.
\newblock
\showISBNx{9781450341974}
\urldef\tempurl%
\url{https://doi.org/10.1145/3035918.3035927}
\showDOI{\tempurl}


\bibitem[Zhu et~al\mbox{.}(2014)]%
        {10.1145/2588555.2612181}
\bibfield{author}{\bibinfo{person}{Andy~Diwen Zhu}, \bibinfo{person}{Wenqing
  Lin}, \bibinfo{person}{Sibo Wang}, {and} \bibinfo{person}{Xiaokui Xiao}.}
  \bibinfo{year}{2014}\natexlab{}.
\newblock \showarticletitle{Reachability Queries on Large Dynamic Graphs: A
  Total Order Approach}. In \bibinfo{booktitle}{\emph{Proc. ACM SIGMOD Int.
  Conf. on Management of Data}}. \bibinfo{pages}{1323–1334}.
\newblock
\showISBNx{9781450323765}
\urldef\tempurl%
\url{https://doi.org/10.1145/2588555.2612181}
\showDOI{\tempurl}


\bibitem[Zou et~al\mbox{.}(2014)]%
        {ZOU201447}
\bibfield{author}{\bibinfo{person}{Lei Zou}, \bibinfo{person}{Kun Xu},
  \bibinfo{person}{Jeffrey~Xu Yu}, \bibinfo{person}{Lei Chen},
  \bibinfo{person}{Yanghua Xiao}, {and} \bibinfo{person}{Dongyan Zhao}.}
  \bibinfo{year}{2014}\natexlab{}.
\newblock \showarticletitle{Efficient Processing of Label-Constraint
  Reachability Queries in Large Graphs}.
\newblock \bibinfo{journal}{\emph{Inf. Syst.}}  \bibinfo{volume}{40}
  (\bibinfo{date}{March} \bibinfo{year}{2014}), \bibinfo{pages}{47–66}.
\newblock
\showISSN{0306-4379}
\urldef\tempurl%
\url{https://doi.org/10.1016/j.is.2013.10.003}
\showDOI{\tempurl}


\bibitem[Özsu(2016)]%
        {tamer16rdfsurvey}
\bibfield{author}{\bibinfo{person}{M.~Tamer Özsu}.}
  \bibinfo{year}{2016}\natexlab{}.
\newblock \showarticletitle{A survey of RDF data management systems}.
\newblock \bibinfo{journal}{\emph{Frontiers of Computer Science}}
  \bibinfo{volume}{10} (\bibinfo{date}{April} \bibinfo{year}{2016}),
  \bibinfo{pages}{418--432}.
\newblock
\urldef\tempurl%
\url{https://doi.org/10.1007/s11704-016-5554-y}
\showDOI{\tempurl}


\end{thebibliography}
